\documentclass[11pt, reqno]{article}
\pdfoutput=1

\usepackage{jheppub}
\usepackage{amssymb}
\usepackage{amsmath}
\usepackage{mathtools}
\usepackage[usenames,dvipsnames]{xcolor}
\usepackage{epsfig}
\usepackage{dcolumn}
\usepackage{esvect}
\usepackage{upgreek}
\usepackage{setspace}
\usepackage{enumitem}
\usepackage{array,multirow,bigdelim,arydshln}
\usepackage{appendix}
\usepackage{xparse}
\usepackage{empheq}
\usepackage{dsfont}
\usepackage[utf8]{inputenc}
\usepackage{color}

\hypersetup{
	colorlinks,
	urlcolor=Maroon,
	linkcolor=Maroon,
	citecolor=Maroon
}

\makeatletter
\def\@fpheader{}
\makeatother

\allowdisplaybreaks

\def \nn {\nonumber}

\def \be {\begin{equation}}
\def \ee {\end{equation}}
\def \vol {\text{vol}\,}
\def \slc {\text{SL}(2, \mathbb{C})}

\def \I {\mathbb{I}}

\def \N {\mathcal{N}}
\def \cp {\mathbb{CP}^1}
\def \C {\mathbb{C}}
\def \glc {\text{GL}(1, \mathbb{C})}
\def \doth#1{\hat{#1}}

\newcommand{\bea}{\begin{eqnarray}}
\newcommand{\eea}{\end{eqnarray}}

\def\one{\mbox{1 \kern-.59em {\rm l}}}

\def\a{\alpha}

\def\d{\delta}  \def\D{\Delta}  
\def\e{\epsilon}

\def\l{\lambda} 
\def\m{\mu}

\def\r{\rho}
\def\s{\sigma}  
\def\t{\tau}

\NewDocumentCommand{\eulerian}{omm}
{%
\genfrac<>{0pt}{}{#2}{#3}%
\IfValueT{#1}{_{\!#1}}%
}

\title{The S Matrix of 6D Super Yang--Mills and\\ Maximal Supergravity from Rational Maps}

\author[a]{Freddy Cachazo,}\emailAdd{fcachazo@pitp.ca}
\author[a,b,c]{Alfredo Guevara,}\emailAdd{aguevara@pitp.ca}
\author[d]{Matthew Heydeman,}\emailAdd{mheydema@caltech.edu}
\author[a,b]{\\Sebastian Mizera,}\emailAdd{smizera@pitp.ca}
\author[d]{John H. Schwarz,}\emailAdd{jhs@theory.caltech.edu}
\author[d,e]{and Congkao Wen}\emailAdd{c.wen@qmul.ac.uk}

\affiliation[a]{Perimeter Institute for Theoretical Physics, Waterloo, ON N2L 2Y5, Canada}
\affiliation[b]{Department of Physics \& Astronomy, University of Waterloo, Waterloo, ON N2L 3G1, Canada}
\affiliation[c]{CECs Valdivia \& Departamento de F\'isica, Universidad de Concepci\'on, Casilla 160-C,\\ Concepci\'on, Chile}
\affiliation[d]{Walter Burke Institute for Theoretical Physics, California Institute of Technology 452-48,\\ Pasadena, CA 91125, USA}
\affiliation[e]{Centre for Research in String Theory,
School of Physics \& Astronomy, \\ Queen Mary University of London,
Mile End Road, London E1 4NS, UK}

\abstract{We present new formulas for $n$-particle tree-level scattering amplitudes of six-dimensional $\N=(1,1)$ super Yang--Mills (SYM) and $\N=(2,2)$ supergravity (SUGRA). They are written as integrals over the moduli space of certain rational maps localized on the $(n-3)!$ solutions of the scattering equations. Due to the properties of spinor-helicity variables in six dimensions, the even-$n$ and odd-$n$ formulas are quite different and have to be treated separately. We first propose a manifestly supersymmetric expression for the even-$n$ amplitudes of $\N=(1,1)$ SYM theory and perform various consistency checks. By considering soft-gluon limits of the even-$n$ amplitudes, we deduce the form of the rational maps and the integrand for $n$ odd. The odd-$n$ formulas obtained in this way have a new redundancy that is intertwined with the usual $\slc$ invariance on the Riemann sphere. We also propose an alternative form of the formulas,  analogous to the Witten--RSV formulation, and explore its relationship with the symplectic (or Lagrangian) Grassmannian. Since the amplitudes are formulated in a way that manifests double-copy properties, formulas for the six-dimensional $\N=(2,2)$ SUGRA amplitudes follow. These six-dimensional results allow us to deduce new formulas for five-dimensional SYM and SUGRA amplitudes, as well as massive amplitudes of four-dimensional $\N=4$ SYM on the Coulomb branch.}

\preprint{CALT-TH-2018-019,\; QMUL-PH-18-07}
\arxivnumber{1805.11111}

\begin{document}

\maketitle
\addtocontents{toc}{\protect\setcounter{tocdepth}{2}}
\numberwithin{equation}{section}
\setcounter{page}{2}

\section{Introduction}

Scattering amplitudes have been the subject of great interest especially since the introduction of Witten's twistor string theory in 2003 \cite{Witten:2003nn}. Witten proposed a formulation of the complete tree-level S matrix of four-dimensional (4D) ${\cal N}=4$ super Yang--Mills theory (SYM) based on an integral over the moduli space of maps from $n$-punctured spheres into twistor space. The moduli space contains components characterized by the degree of the maps and by the number of disconnected curves in the target space. Soon after, Roiban, Spradlin, and Volovich (RSV) conjectured and gave evidence that by integrating over only maps to {\it connected} curves the complete tree-level S matrix could be recovered \cite{Roiban:2004yf}. The Witten--RSV formula then expresses $n$-particle amplitudes as integrals over the moduli space of maps from $n$-punctured spheres into connected curves in twistor space. The formula can then be translated into momentum space. The key ingredients, in modern terminology, are rational maps from ${\mathbb{CP}^1}$ into the null cone in spinor coordinates:
\be
z \;\to\; \rho_\alpha(z)\tilde\rho_{\dot\alpha}(z),
\ee
with $\rho_\a(z)$ and $\tilde{\rho}_{\dot{\a}} (z)$ polynomials of degree $d$ and $\tilde{d}$ respectively, such that $d+\tilde d = n-2$.

Extending such worldsheet formulations to other theories then became a natural open problem.  For example, formulas based on rational maps into twistor space for 4D ${\cal N}=8$ supergravity (SUGRA) were developed in 2012 by Geyer, Skinner, Mason and one of the authors in a series of works \cite{Cachazo:2012da,Cachazo:2012kg,Cachazo:2012pz}. These developments gave more impetus to the search for similar phenomena in other theories and perhaps other spacetime dimensions.

One of the main obstacles to extending the formalism to higher dimensions was the heavy use of spinor-helicity variables in 4D. This obstruction was removed in 2009 when Cheung and O'Connell introduced the 6D spinor-helicity formalism \cite{Cheung:2009dc}. However, straightforward extensions of connected formulas were not found, hinting that new ingredients were needed in 6D. Effective theories in 6D are very interesting for a variety of reasons. On a practical side, besides the interest in their own right, computing 6D SYM formulas would allow, for instance, via dimensional reduction, for a unification of 4D helicity sectors for massless amplitudes \cite{Huang:2011um} as well as for obtaining amplitudes along the Coulomb branch of ${\cal N}=4$ SYM, which contains massive particles such as W bosons \cite{Bern:2010qa}.

Unifying different helicity sectors of 4D ${\cal N}=4$ SYM into a generalization of the Witten--RSV formula in 6D proved to be a difficult problem resisting a solution. This motivated He, Yuan and one of the authors to look for alternative formulations, leading to what is now known as the scattering equations and the CHY formulation \cite{Cachazo:2013gna,Cachazo:2013hca,Cachazo:2013iea}. This formulation opened up worldsheet-like constructions for a large variety of theories in any number of dimensions at the expense of giving up on fermions and hence supersymmetry. The search for a conformal field theory that reproduced the CHY formulas led to the discovery of ambitwistor strings \cite{Mason:2013sva,Geyer:2014fka,Casali:2015vta} whose development allowed computations beyond tree-level \cite{Geyer:2015bja,Geyer:2015jch,Cachazo:2015aol,Geyer:2016wjx,Geyer:2018xwu} (for a recent review see \cite{Geyer:2016nsh}).

In a recent development, three of the authors found connected formulas for the effective theories living on a D5-brane and a M5-brane in 10D and 11D Minkowski spacetime, respectively \cite{Heydeman:2017yww}. These are 6D theories with $\mathcal{N}=(1,1)$ and $(2,0)$ supersymmetry, respectively. The former is the supersymmetric version of Born--Infeld theory, and the latter describes analogous interactions for a supermultiplet containing an Abelian self-dual tensor.

One key feature of theories describing spontaneously broken symmetries, such as the brane theories, is the vanishing of all odd-multiplicity amplitudes. This allows the introduction of polynomial maps:
\be\label{mapsE}
z \;\to\; \rho^A_a(z)\rho^B_b(z)\epsilon^{ab},
\ee
with ${\rm deg }\,\rho^A_a(z)=n/2-1$, such that the total degree of the maps is $n-2$. Here $A$ is a 6D spinor index while $a$ is a ``global" little-group index transforming in ${\text{SL}}(2,\mathbb{C})$. ``Global" means that it does not refer to a specific particle. The formulas found in \cite{Heydeman:2017yww} using these maps are very compact and manifest the symmetries of the theories.

It is well known that scattering amplitudes can make symmetries manifest that the corresponding Lagrangian does not. 
A striking and unexpected example is dual superconformal invariance of ${\cal N}=4$ SYM, which combines with the standard super-conformal invariance to generate an infinite-dimensional structure known as the Yangian of ${\rm PSU}(2,2|4)$ \cite{Drummond:2010km}. In 6D, the M5-brane theory provides an even more fundamental example. The self-dual condition on the three-form field strength causes difficulties in writing down a manifestly Lorentz invariant action for the two-form gauge field \cite{Perry:1996mk, Aganagic:1997zq}. In contrast, the formulas found in \cite{Heydeman:2017yww} for the complete tree-level S matrix are manifestly Lorentz invariant.
These examples highlight the importance of finding explicit formulas for the complete tree-level S matrix, as they can provide new insights into known symmetries of theories or even the discovery of unexpected ones.

The 6D formulas presented in \cite{Heydeman:2017yww} are built using two half-integrands, usually called \emph{left} and \emph{right} integrands, ${\cal I}_{L/R}$, in the sense of the CHY formulation. In the case of the D5- and M5-brane theories, the right integrand carries the supersymmetric information while the left one is purely bosonic. Amusingly, the only difference in the choice of right integrands is $ \N = (1,1)$ or $\N = (2,0)$ supersymmetry, and the left half-integrands agree in both theories:
\be
{\cal I}^{\rm D5}_L ={\cal I}^{\rm M5}_L = \left({\rm Pf}'A_n\right)^2.
\ee
Here $A_n$ is an antisymmetric $n\times n$ matrix whose reduced Pfaffian has made an appearance in CHY formulas for the non-linear sigma model, special Galileon and Born--Infeld theories \cite{Cachazo:2014xea,Cachazo:2016njl}. The entries of $A_n$ are given by:
\begin{align}\label{even-matrix-A}
[A_n]_{ij} = \begin{cases}
\dfrac{p_{i} \cdot p_{j}}{\sigma_{i} - \sigma_{j}} &\quad\text{if}\quad i\neq j, \\
\quad\; 0\phantom{\dfrac{1}{1}} &\quad\text{if}\quad i=j,
\end{cases}   \qquad\qquad\text{for}\qquad i,j=1,2,\ldots,n.
\end{align}
Here the puncture associated to the $i$th particle is located at $z=\sigma_i$. The right integrands in both theories also contain a single power of the reduced Pfaffian of $A_n$ in addition to the supersymmetry information.

In this paper we continue the exploration of worldsheet formulas in 6D and provide explicit such formulas for the complete tree-level S matrix of ${\cal N}=(1,1)$ SYM with $\text{U}(N)$ gauge group, the effective theory on $N$ coincident D5-branes. The S matrix of this theory has been studied previously in \cite{Dennen:2009vk,Bern:2010qa,Huang:2010rn,Dennen:2010dh,Brandhuber:2010mm,Czech:2011dk,Plefka:2014fta}.

A proposal for amplitudes with an even number of particles is naturally obtained \cite{Congkao-UCLA-talk} by noticing that the CHY formulation for Yang--Mills partial amplitudes of gluons can be obtained from that of the Born--Infeld theory by the replacement
\be
{\cal I}^{\rm BI}_L = \left({\rm Pf}'A_n\right)^2 \quad\longrightarrow\quad {\cal I}^{\text{YM}}_L = {\rm PT}(12\cdots n),
\ee
where ${\rm PT}(12\cdots n)$ is the famous Parke--Taylor factor \cite{Parke:1986gb,Nair:1988bq,Witten:2003nn}. Applying the same substitution to the D5-brane formula we get a formula for ${\cal N} = (1,1)$ SYM amplitudes with even multiplicity, and we provide strong evidence for its validity.

The formula for odd-point amplitudes proves to be a more difficult task, since the maps given in \eqref{mapsE} do not have an obvious generalization to odd multiplicity. In all previously known formulations of Yang--Mills amplitudes, soft limits have provided a way of generating $(n{-}1)$-particle amplitudes from $n$-particle ones since the leading singular behavior is controlled by Weinberg's soft theorem \cite{Weinberg:1965nx}. However, in all such cases the measure over the moduli space of maps has had the same structure for $n{-}1$ and $n$ particles.

In 6D the soft limit needs additional technical considerations, in part due to the  $\text{SL}(2,\mathbb{C})$ redundancy of the maps, inherited from the little group. (In 4D the redundancy was only $\text{GL}(1,\mathbb{C})$.) The $\text{SL}(2,\mathbb{C})$ structure introduces new degrees of freedom in the computation of the soft limit. In contrast to the 4D case, these degrees of freedom in 6D turn out to be inherently intertwined with the M\"obius group $\text{SL}(2,\mathbb{C})$ acting on $\mathbb{CP}^1$.

One of the main results of this paper is to uncover a fascinating structure that appears in the definition of the maps for an odd number of particles. In a nutshell, we find that the maps for $n$ odd can be defined by
\be
\rho^A_a(z) = \sum_{k=0}^{(n-3)/2}\rho^A_{k,a}\,z^k + w^A \xi_a\, z^{(n-1)/2},
\ee
while the moduli space is obtained by modding out by a novel redundancy, which we call T-shift invariance. It acts on the maps in addition to the two ${\rm SL}(2,\mathbb{C})$'s. In fact, the new T-shift action emerges due to the non-commutativity between such groups. This fact becomes apparent already from the soft-limit perspective as mentioned above. We start the exploration of the corresponding algebra and find that when the coefficients of the maps are partially fixed, the remaining redundancies take a form of a semi-direct product ${\rm SL}(2,\mathbb{C})\ltimes \mathbb{C}^2$. We introduce a formula for the integration measure for the space of maps of odd multiplicity as well as its supersymmetric extension. It can be used both for super Yang--Mills and supergravity theories.

Finally, we derive an explicit integrand for the $\mathcal{N}=(1,1)$ SYM odd-multiplicity amplitudes. The ingredients are a Parke--Taylor factor for the left-integrand and a generalization of the $A_n$ matrix whose reduced Pfaffian enters in the right-integrand together with the supersymmetric part of the measure. The new matrix and its reduced Pfaffian behave as quarter-integrands, again in standard CHY terminology. This means that it provides a new building block that can be used to construct potentially consistent theories by mixing it with other quarter-integrands. The new matrix ${\widehat A}_n$ is given by
\begin{align}
[\widehat{A}_n]_{ij} = \begin{cases}
\dfrac{p_{i} \cdot p_{j}}{\sigma_{i} - \sigma_{j}} &\quad\text{if}\quad i\neq j, \\
\quad\; 0\phantom{\dfrac{1}{1}} &\quad\text{if}\quad i=j,
\end{cases}   \qquad\qquad\text{for}\qquad i,j=1,2,\ldots,n,\star,
\end{align}
where the final column and row feature a new null vector $p_\star$ of the form:
\be
p_{\star}^{AB} = \frac{ 2\, q^{[A} \langle \rho^{B]}(\sigma_\star)\,  \rho^{C}(\sigma_\star)  \rangle\tilde{q}_C }{ q^D  [\tilde{\rho}_D(\sigma_\star )\, \tilde{\xi}] \langle \rho^E(\sigma_\star)\, \xi \rangle  \tilde{q}_E},
\ee
where $q$ and $\tilde{q}$ are reference spinors and $\sigma_\star$ is a reference puncture that can take an arbitrary value.

The connected formula for odd multiplicity is derived using the soft limit of the corresponding even-multiplicity result. We also obtain it by assuming the supersymmetric quarter-integrand and matching the rest by comparing to the CHY formula for $n-1$ scalars and one gluon. The same strategy of examining component amplitudes can also be used for even multiplicity using the same assumptions.

In addition to constraints which connect the external momenta to the product of maps as in (\ref{mapsE}), we also find a linear form of the constraints which leads to an alternative expression for the amplitudes. This form is the direct analog of the original Witten--RSV formula and connects the maps directly to the external 6D spinor-helicity variables. We further recast the linearized form of maps in the form of the so-called Veronese maps, and explore their relations to the symplectic (or Lagrangian) Grassmannian.

Having explicit integrands for the complete 6D ${\cal N} = (1,1)$ SYM tree amplitudes allows the construction of the 6D ${\cal N} = (2,2)$ SUGRA integrand by the standard replacement of the left-integrand Parke--Taylor factor by a copy of the right integrand, which contains the necessary new supersymmetric information.

We end with various applications to other theories in four, five, and six dimensions. These include mixed superamplitudes of 6D ${\cal N} = (1,1)$ SYM coupled to a single D5-brane, 5D SYM and SUGRA, and also 4D scattering amplitudes involving massive particles of ${\cal N}=4$ SYM on the Coulomb branch of its moduli space. The formulas for 5D theories take forms very similar to those of 6D, but with additional constraints on the rational maps to incorporate 5D massless kinematics. In order to describe the massive amplitudes of ${\cal N}=4$ SYM on the  Coulomb branch, we utilize the spinor-helicity formalism recently developed for massive particles in 4D \cite{Arkani-Hamed:2017jhn}, which in fact can naturally be viewed as a dimensional reduction of 6D massless helicity spinors.  We also would like to emphasize that, although it is a straightforward reduction of our 6D formula, this is the first time that a connected formula has been proposed for 4D ${\cal N}=4$ SYM away from the massless point of the moduli space. 

This paper is organized as follows. In Section~\ref{sec:rational-maps}, we review the general construction of rational maps from $\mathbb{CP}^1$ to the null cone in general spacetime dimensions. We also review 4D constructions and then 6D maps for an even number of particles. Sections \ref{sec:even-pt} and \ref{sec:odd-pt} are devoted to $\mathcal{N}=(1,1)$ SYM amplitudes in 6D. Section \ref{sec:even-pt} deals with an even number of particles while Section \ref{sec:odd-pt} contains the main results of this work by presenting formulas and consistency checks for odd multiplicity. In Section~\ref{sec:LinearConstraints} we discuss a linear form of the scattering maps in 6D and its relationship to the symplectic Grassmannian. Extensions and applications are presented in Section~\ref{sec:applications}. We conclude and give a discussion of future directions in Section~\ref{sec:outlook}. In Appendix \ref{app:symmetry-algebra} we present the algebra of the new T-shift, and in Appendix~\ref{app:soft-limits} we give details of the soft-limit calculations.

\section{\label{sec:rational-maps}Rational Maps and Connected Formulas}

This section begins by reviewing rational maps and the CHY formulas for an arbitrary space-time dimension. It then discusses the specialization to 4D and the Witten--RSV formulas. Finally, it gives an overview of the form of the even-$n$ rational maps in 6D, whose generalizations will be the subject of later sections.

\subsection{\label{arbitrarydim}Arbitrary Dimension} 

Let us consider scattering of $n$ massless particles in an arbitrary space-time dimension. To each particle, labeled by the index $i$, we associate a puncture at $z=\sigma_i$ on the Riemann sphere, $\cp$, whose local coordinate is $z$. We then introduce polynomial maps, $p^\mu (z)$, of degree $n-2$. They are constructed such that the momentum $p_i^\mu$ associated to the $i$th particle is given by:
\be\label{map eqs}
p_i^\mu = \frac{1}{2\pi i}\oint_{|z - \sigma_i|=\varepsilon} \frac{p^\mu(z)}{\prod_{j=1}^n (z-\sigma_j)}\, dz,
\ee
which means that $p^\mu(z)$ can be written as a polynomial in $z$:
\be\label{scattering-map}
p^\mu(z) = \sum_{i=1}^n p_i^\mu \prod_{j \neq i} (z-\sigma_j).
\ee
Here we take all momenta to be incoming, so that momentum conservation is given by $\sum_{i=1}^n p_i^\mu = 0$. We call $p^\mu(z)$ the \emph{scattering map}.

In order to relate the positions of the punctures $\sigma_i$ to the kinematics, the additional condition that the scattering map is null, i.e., $p^2(z) = 0$ for all $z$, is imposed. Since $p^2(z)$ is of degree $2n-4$ and it is already required to vanish at $n$ points, $\sigma_i$, requiring $p^\mu(z)$ to be null gives $n-3$ additional constraints. Using \eqref{scattering-map} these constraints can be identified by considering the combination
\be\label{maps-to-se}
\frac{p^2(z)}{\prod_{i=1}^n (z-\sigma_i)^2} = \sum_{i,j=1}^{n} \frac{p_{i} \cdot p_{j}}{(z-\sigma_i)(z-\sigma_j)} = 0.
\ee
The expression \eqref{maps-to-se} does not have any double poles, since the punctures are distinct and all of the momenta are null, $p^2_i =0$. Requiring that residues on all the poles vanish implies:
\be\label{scattering-equations}
E_i := \sum_{j \neq i} \frac{p_{i} \cdot p_{j}}{\sigma_{ij}} = 0 \qquad\text{for all } i,
\ee
where $\sigma_{ij} = \sigma_i - \sigma_j$. These are the so-called \emph{scattering equations} \cite{Cachazo:2013gna}. Due to the above counting, only $n-3$ of them are independent. In fact, $\sum_i \sigma_i^\ell E_i$ automatically vanishes for $\ell=0,1,2$ as a consequence of the mass-shell and momentum-conservation conditions. Using the $\slc$ symmetry of the scattering equations to fix three of the $\sigma_i$ coordinates, there are $(n-3)!$ solutions of the scattering equations for the remaining $\sigma_i$'s for generic kinematics \cite{Cachazo:2013gna}.

The scattering equations connect the moduli space of $n$-punctured Riemann spheres to the external kinematic data. Tree-level $n$-particle scattering amplitudes of massless theories can be computed using the Cachazo--He--Yuan (CHY) formula, which takes the form \cite{Cachazo:2013hca}:
\be\label{CHY-integral}
\mathcal{A}^{\text{theory}}_n = \int d\mu_n\, \mathcal{I}_L^{\text{theory}}\, \mathcal{I}_R^{\text{theory}}.
\ee
$\mathcal{I}_L^{\text{theory}}$ and $\mathcal{I}_R^{\text{theory}}$ are left- and right-integrand factors, respectively, and they depend on the theory under consideration. Their precise form is not important for now, other than that they carry weight $-2$ under an $\slc$ transformation for each puncture, i.e., $\mathcal{I}^{\text{theory}}_{L/R} \to \prod_{i=1}^n (C\sigma_i + D)^2\, \mathcal{I}^{\text{theory}}_{L/R}$ when $\sigma_i \to (A\sigma_i + B)/(C\sigma_i + D)$ and $AD-BC=1$. Correctly identifying the separation into left- and right-integrands is important for making the double-copy properties of amplitudes manifest. 

Let us now review the CHY measure:
\be\label{CHY-measure}
 d\mu_n = \delta^D({\textstyle\sum_{i=1}^n} p^{\mu}_i)\left(\prod_{i=1}^n \delta(p_i^2)\right)\,  \frac{\prod_{i=1}^n d\sigma_i}{\vol \slc} {\prod_i}^{\prime} \delta(E_i).
\ee
This is a distribution involving momentum conservation and null conditions for the external momenta. The factor $\vol \slc$ denotes the fact that it is necessary to quotient by the $\slc$ redundancy on the Riemann surface by fixing the positions of three of the punctures, specifically $i=p,q,r$. Similarly, the prime means that the corresponding three scattering equations are redundant and should be removed. Fixing these redundancies leads to
\begin{align} \label{eq:CHY-measure-1}
\int d\mu_n &= \delta^D({\textstyle\sum_{i=1}^n} p^{\mu}_i)\left( \prod_{i=1}^n \delta(p_i^2)\right)\,\int (\sigma_{pq}\sigma_{qr}\sigma_{rp})^2 \prod_{i \neq p,q,r}\!\!\left( d\sigma_i \, \delta(E_i) \right) \nn \\
&= \delta^D({\textstyle\sum_{i=1}^n} p^{\mu}_i)\left(\prod_{i=1}^n \delta(p_i^2)\right)\sum_{s=1}^{(n-3)!} \frac{(\sigma_{pq}\sigma_{qr}\sigma_{rp})^2}{\det \left[ \frac{\partial{E_i}}{\partial\sigma_j} \right]} \Bigg|_{\sigma_i = \sigma_i^{(s)}},
\end{align}
which can be shown to be independent of the choice of labels $p,q,r$. The delta functions fully localize the measure on the $(n-3)!$ solutions $\{\sigma_i^{(s)}\}$ of the scattering equations. The measure transforms with $\slc$-weight $4$ in each puncture, so that the CHY integral \eqref{CHY-integral} is $\slc$-invariant. 

Finally, one of the advantages of the CHY formulation is that soft limits can be derived from a simple application of the residue theorem \cite{Cachazo:2013gna}. Under the soft limit of an $(n+1)$-point amplitude with the last particle soft, i.e., $\tau\rightarrow 0$ where $p_{n+1}=\tau \hat{p}_{n+1}$, the measure behaves as
\be\label{soft-behavior}
\int d\mu_{n+1} = \delta(p_{n+1}^2)\int d\mu_{n}\, \frac{1}{2\pi i}\oint_{|\hat{E}_{n+1}|=\varepsilon} \frac{d\sigma_{n+1}}{E_{n+1}}+ O(\tau^0).
\ee 
Here we have rewritten the scattering equation $\hat{E}_{n+1} = 0$ as a residue integral. Note that $E_{n+1} = \tau \hat{E}_{n+1}$ is proportional to $\tau$, and thus the displayed term is dominant. Therefore the scattering equation associated to the last particle completely decouples in the limit $\tau \to 0$. For each of the $(n-3)!$ solutions of the remaining scattering equations, the contour $\{|\hat{E}_{n+1}| = \varepsilon \}$ localizes on $n-2$ solutions \cite{Cachazo:2013gna}. 

\subsection{Four Dimensions: Unification of Sectors} \label{sec:4d_rational}

Since the scattering equations are valid in an arbitrary dimension, they do not capture aspects specific to certain dimensions, such as fermions or supersymmetry. In order to do so, it is convenient to express the scattering maps using the spinor-helicity variables appropriate to a given dimension. We start with the well-understood case of 4D. Various aspects of specifying CHY formulations to 4D have also been discussed in \cite{He:2016vfi, Zhang:2016rzb}. 

The momentum four-vector of a massless particle in 4D Lorentzian spacetime
can be written in terms of a pair of two-component bosonic spinors, $\l^\a$ and $\tilde\l^{\dot\a}$,
which transform as ${\bf 2}$ and ${\bf \overline 2}$ representations of 
the $\text{SL}(2,\C) = \text{Spin}(3,1)$ Lorentz group
\begin{equation}
p^{\a\dot\a} = \sigma_{\mu}^{\alpha\dot{\alpha}} p^{\mu} =\l^\a \tilde\l^{\dot\a} \qquad\qquad \alpha=1,2,\quad\dot{\alpha} = \dot{1},\dot{2}.
\end{equation}
For physical momenta, $\l$ and $\pm\tilde\l$ are complex conjugates. However, when
considering analytic continuations, it is convenient to treat them as independent.
The little group for a massless particle\footnote{In this work we only consider massless particles that transform trivially under translations of the full little group of Euclidean motions in $D-2$ dimensions.} in 4D is $\text{U}(1)$. Its complexification is $ \text{GL}(1, \C)$. $\l$ and
$\tilde\l$ transform oppositely under this group so that the momentum is
invariant. In discussing $n$-particle scattering amplitudes, we label the particles by an index $i=1,2,\ldots,n$. It is important to understand that there is a distinct little group associated to each of the $n$ particles. Thus,
the little group $\text{GL}(1,\C)$ transforms the spinors as $\lambda_i \rightarrow t_i \lambda_i$ and $\tilde{\lambda}_i \rightarrow t_i^{-1} \tilde{\lambda}_i$, leaving only three independent degrees of freedom for the momentum. Lorentz-invariant spinor products are given by: $\langle \lambda_i \lambda_j \rangle = \varepsilon_{\alpha\beta} \lambda_i^{\alpha} \lambda_j^{\beta}$ and $[ \tilde{\lambda}_i \tilde{\lambda}_j ] = \varepsilon_{\dot{\alpha}\dot{\beta}} \tilde{\lambda}_i^{\dot{\alpha}} \tilde{\lambda}_j^{\dot{\beta}}$. It is sometimes convenient to simplify further and write $\langle ij \rangle$ or $[ij]$. Given a scattering amplitude, expressed in terms of spinor-helicity variables, one can deduce the helicity of the $i$th particle by determining the power of $t_i$ by which the amplitude transforms. For example, the most general Parke--Taylor (PT) formula for maximally helicity violating (MHV) amplitudes in 4D YM theory is as follows \cite{Parke:1986gb}: if gluons $i$ and $j$ have negative helicity, while the other $n-2$ gluons have positive helicity, then the (color-stripped) amplitude is
\begin{equation}
A^{\text{YM}}_n (1^+ 2^+ \cdots i^- \cdots j^- \cdots n^+) = \frac{{\langle i j\rangle}^4}{\langle 1 2 \rangle \langle 2 3 \rangle
\cdots \langle n 1\rangle}.
\end{equation}

Since the scattering map $p^\mu(z)$ in \eqref{scattering-map} is required to be null for all $z$, it can also be expressed in a factorized form in terms of spinors:
\be\label{4d-scattering-maps}
p^{\alpha \dot{\alpha}}(z) = \rho^\alpha (z) \tilde{\rho}^{\dot{\alpha}}(z).
\ee
The roots of $p^{\alpha \dot{\alpha}}(z)$ can be distributed among the polynomials $\rho(z)$, $\tilde{\rho}(z)$ in different ways, such that their degrees add up to $n-2$. When $\deg \rho(z) = d$ and $\deg \tilde{\rho}(z) = \tilde d =n-d-2$, the maps are said to belong to the $d$th sector. We parametrize the polynomials as:
\be
\rho^\alpha(z) = \sum_{k=0}^{d} \rho^{\alpha}_k z^k, \qquad \tilde{\rho}^{\dot{\alpha}}(z) = \sum_{k=0}^{\tilde d} \tilde{\rho}^{\dot{\alpha}}_k z^k.
\ee
The spinorial maps \eqref{4d-scattering-maps} carry the same information as the scattering equations, and therefore they can be used to redefine the measure. Here it is natural to introduce a measure for each sector as:
\be\label{4d-measure}
\int d\mu^{\text{4D}}_{n,d} = \int\frac{\prod_{i=1}^n d\sigma_i \prod_{k=0}^{d} d^2 \rho_k \prod_{k=0}^{\tilde d} d^2 \tilde{\rho}_{k}}{\vol\, \slc \times \glc} \frac{1}{R(\rho) R(\tilde{\rho})} \prod_{i=1}^{n} \delta^4 \left( p_i^{\alpha\dot{\alpha}} - \frac{\rho^\alpha(\sigma_i) \tilde{\rho}^{\dot{\alpha}}(\sigma_i)}{\prod_{j\neq i}\sigma_{ij}}\right).
\ee
These measures contain an extra $\glc$ redundancy, analogous to the little group symmetries of the momenta, which allows fixing one coefficient of $\rho(z)$ or $\tilde{\rho}(z)$. $R(\rho)$ denotes the resultant $R(\rho^1(z),\rho^2(z),z)$ and similarly for $R(\tilde{\rho})$ \cite{gelfand2009discriminants,Cachazo:2013zc}. The physical reason resultants appear in the denominator can be understood by finding the points in the moduli space of maps where they vanish. A resultant of any two polynomials, say $\rho_1(z)$ and $\rho_2(z)$, vanishes if and only if the two polynomials have a common root $z^*$. If such a $z^*$ exists then the map takes it to the tip of the momentum-space null cone, i.e., to the strict soft-momentum region. This is a reflection of the fact that in four (and lower) dimensions IR divergences are important in theories of massless particles. The measure is giving the baseline for the IR behavior while integrands can change it depending on the theory. As reviewed below, the gauge theory and gravity integrands contain $(R(\rho)R(\tilde\rho))^{\texttt{s}}$, where $\texttt{s}=1$ for YM and $\texttt{s}=2$ for gravity, which coincides with the spins of the particles. Combined with the factor in the measure one has $(R(\rho)R(\tilde\rho))^{\texttt{s}-1}$, which indicates that the IR behavior improves as one goes from a scalar theory, with $\texttt{s}=0$, to gravity \cite{Cachazo:2016sdc}.

Summing over all sectors gives the original CHY measure:
\be\label{measure-sum}
\int d\mu_n = \sum_{d=1}^{n-3} \int d\mu^{\text{4D}}_{n,d}.
\ee
This separation works straightforwardly for theories where the integrand only depends on $\sigma_i$'s and not on the maps. One such theory is the bi-adjoint scalar whose amplitudes are given by 
\be\label{wx}
m(\alpha |\beta) = \int d\mu_n\, {\rm PT}(\alpha )\, {\rm PT}(\beta ) = \sum_{d=1}^{n-3} m_{n,d}(\alpha |\beta) ,
\ee
where ${\rm PT}(\alpha )$ is the Parke--Taylor factor. The definition 
for the identity permutation is
\bea \label{eq:PTfactor}
{\rm PT}(12\cdots n) = {1 \over  \sigma_{12} \sigma_{23} \cdots \sigma_{n1} }.
\eea
In general $\a$ denotes a permutation of the indices $1,2,\ldots,n$. The quantities $m_{n,d}(\alpha |\beta)$ are the ``scalar blocks" defined in \cite{Cachazo:2016sdc}. In the $d$th sector the number of solutions is given by the Eulerian number $\eulerian{n-3}{d-1}$, as conjectured in \cite{Spradlin:2009qr} and proved in \cite{Cachazo:2013iaa}. Upon summation \eqref{measure-sum} gives all $\sum_{d=1}^{n-3} \eulerian{n-3}{d-1} = (n-3)!$ solutions  of the scattering equations. Note that momentum conservation and the factorization conditions that ensure masslessness are built into the measure \eqref{4d-measure}.

An alternative version of the above constraints, which is closer to the original Witten--RSV formulas, can be obtained by integrating-in auxiliary variables $t_i$ and $\tilde t_i$
\begin{align}\label{t-variables}
\delta^4 \left( p_i^{\alpha\dot{\alpha}} - \frac{\rho^\alpha(\sigma_i) \tilde{\rho}^{\dot{\alpha}}(\sigma_i)}{\prod_{j\neq i}\sigma_{ij}}\right) &= \delta(p_i^2)\, \int dt_i\, d\tilde{t}_i\, \delta\left(t_i \tilde{t}_i - \frac{1}{\prod_{j\neq i} \sigma_{ij}} \right)\\
&\qquad\qquad \times \delta^2\big(\lambda^\alpha_i - t_i \,\rho^\alpha(\sigma_i)\big) \delta^2 \big(\tilde{\lambda}^{\dot{\alpha}}_i - \tilde{t}_i \, \tilde{\rho}^{\dot{\alpha}}(\sigma_i)\big) .\nn
\end{align}
This formulation helps to linearize the constraints and make the little-group properties of theories with spin, such as Yang--Mills theory, more manifest. 

The on-shell tree amplitudes of ${\cal N}=4$ SYM theory in 4D are usually written as a sum over sectors
\be\label{symm}
\mathcal{A}^{\mathcal{N}=4 \text{ SYM}}_n= \sum_{d=1}^{n-3} \mathcal{A}_{n,d}^{\mathcal{N}=4 \text{ SYM}}.
\ee
The $d$th sector has $n-2-2d$ units of ``helicity violation'': $d \to n-2-d$ corresponds to reversing the helicities. Partial amplitudes in each sector are given by
\be\label{sect}
\mathcal{A}_{n,d}^{\mathcal{N}=4 \text{ SYM}}(\alpha) = \int d\mu_{n,d}^{\text{4D}}\; \text{PT}(\alpha) \, \left( R(\rho)R(\tilde\rho) \int d\Omega_{\text{F},d}^{(4)} \right) ,
\ee
where $d\Omega_{\text{F},d}^{(4)}$ denotes integrations over fermionic analogs of the maps $\rho(z)$ and $\tilde\rho(z)$ implementing the $\N=4$ supersymmetry, whose precise form can be found in \cite{Cachazo:2013iaa}.

Due to the fact that the little group is $\text{Spin}(4)$ in 6D, it is expected that the SYM amplitudes in 6D should not separate into helicity sectors. Dimensional reduction to 4D would naturally lead to a formulation with unification of sectors. This may appear somewhat puzzling as \eqref{symm} and \eqref{sect} seem to combine the measure in a given sector with an integrand that is specific to that sector. This puzzle is resolved by noticing that 
\be\label{detp}
R(\rho) = {\det} '\, \Phi_d, \qquad R(\tilde\rho) = {\det}'\, \tilde\Phi_{\tilde d},
\ee
where $[\Phi_d]_{ij} :=\langle ij\rangle/(t_it_j\sigma_{ij})$ and $[\tilde\Phi_{\tilde d}]_{ij} :=[ij]/(\tilde t_i\tilde t_j\sigma_{ij})$ for $i\neq j$. The diagonal components are more complicated and depend on $d$ and $\tilde d$ \cite{Cachazo:2012kg,Cachazo:2013zc}. The corresponding reduced determinants are computed using submatrices of size $d\times d$ and $\tilde d\times \tilde{d}$, respectively. One of the main properties of these reduced determinants is that they vanish when evaluated on solutions in sectors that differ from their defining degree, i.e., 
\be\label{obs}
\int d\mu_{n,d}^{\text{4D}}\; {\det}'\, \Phi_{d'} \, {\det}'\, \tilde\Phi_{\tilde{d}'} = \delta_{d,d'} \int d\mu_{n,d}^{\text{4D}} \; {\det}'\, \Phi_d \, {\det}'\, \tilde\Phi_{\tilde d}.
\ee
Using this it is possible to write the complete amplitude in terms of factors that can be uplifted to 6D and unified!
\be\label{unify}
\mathcal{A}^{\mathcal{N}=4 \text{ SYM}}_n (\alpha) =  \int \left(\,\sum_{d=1}^{n-3} d\mu^{\text{4D}}_{n,d}\right) \; \text{PT}(\alpha) \left( \sum_{d'=1}^{n-3} {\det}'\, \Phi_{d'}\, {\det}'\, \tilde\Phi_{\tilde{d}'} \int d\Omega_{\text{F},d'}^{(4)} \right).
\ee
Finally, it is worth mentioning that \eqref{obs} can be used to write unified 4D ${\cal N}=8$ SUGRA amplitudes, via the double copy, as
\be\label{sugraUni}
\mathcal{M}^{\mathcal{N}=8 \text{ SUGRA}}_n  =\!  \int \!\left(\sum_{d=1}^{n-3} d\mu^{\text{4D}}_{n,d}\right) \!\left( \sum_{d'=1}^{n-3} {\det}'\, \Phi_{d'}\, {\det}'\, \tilde\Phi_{\tilde{d}'} \!\int \!\! d\Omega_{\text{F},d'}^{(4)} \right) \!\left( \sum_{d'=1}^{n-3} {\det}'\, \Phi_{d'}\, {\det}'\, \tilde\Phi_{\tilde{d}'} \!\int \!\! d\hat{\Omega}_{\text{F},d'}^{(4)} \right) \!.\nn
\ee

\subsection{\label{sec:even-point}Six Dimensions: Even Multiplicity}

We now turn to a review of scattering maps in 6D.  It turns out that the 6D spinor-helicity formalism requires separate treatments for amplitudes with an even and an odd number of particles. In this subsection we review the construction for an even number of particles, as was recently introduced in the context of M5- and D5-brane scattering amplitudes \cite{Heydeman:2017yww}. (These theories only have non-vanishing amplitudes for $n$ even.) A formula for odd multiplicity, which is required for Yang--Mills theories, is one of the main results of this paper and it is given in Section \ref{sec:odd-pt}.

The little group for massless particles in 6D is $\text{Spin}(4) \sim \text{SU}(2) \times \text{SU}(2)$. We use indices without hats when referring to representations of the first $\text{SU}(2)$ or its $\slc$ complexification and ones with hats when referring to the second $\text{SU}(2)$ or its $\slc$ complexification. Momenta of massless particles are parametrized in terms of 6D spinor-helicity variables $\l_i^{A,a}$ by \cite{Cheung:2009dc}:
\be
p_i^{AB} = \sigma^{AB}_{\mu} p^{\mu}_i = \langle \lambda_i^A \lambda_i^B \rangle,\qquad\qquad A,B=1,2,3,4,
\ee
where $\sigma_\mu^{AB}$ are six antisymmetric $4\times 4$ matrices, which form an invariant tensor of $\text{Spin}(5,1)$. The angle bracket denotes a contraction of the little-group indices:
\be
\langle \lambda_i^A \lambda_i^B \rangle = \epsilon_{a b} \lambda_i^{A,a} \lambda_i^{B,b} = \lambda_i^{A+} \lambda_i^{B-} -\, \lambda_i^{A-} \lambda_i^{B+}, \qquad a,b=+,-.
\ee
$\e_{ab}$ is an invariant tensor of the $\text{SU}(2)$ little group, as well
as its $\slc$ complexification. The on-shell condition, $p_i^2=0$, is equivalent to the vanishing of the Pfaffian of $p_i^{AB}$. The little group transforms the spinors as $\lambda_i^{A,a} \rightarrow (L_i)^a_b \lambda_i^{A,b}$, where $L_i \in \slc$, leaving only five independent degrees of freedom for the spinors, appropriate for a massless particle in six dimensions. The momenta can be equally well described by conjugate spinors $\tilde{\lambda}_{i,A,\doth{a}}$ :
\be
p_{i,AB} = \frac{1}{2} \epsilon_{ABCD}\, p_i^{CD} = [ \tilde{\lambda}_{i,A} \tilde{\lambda}_{i,B} ],
\ee
where
\be
[ \tilde{\lambda}_{i,A} \tilde{\lambda}_{i,B} ] = \epsilon^{\doth{a} \doth{b}} \tilde{\lambda}_{i,A,\doth{a}} \tilde{\lambda}_{i,B,\doth{b}} = \tilde{\lambda}_{i,A,\doth{+}} \tilde{\lambda}_{i,B,\doth{-}} -\, \tilde{\lambda}_{i,A,\doth{-}} \tilde{\lambda}_{i,B,\doth{+}}, \qquad \doth{a},\doth{b}=\doth{+},\doth{-}.
\ee
These conjugate spinors belong to the second (inequivalent) four-dimensional representation of the $\text{Spin}(5,1) \sim \text{SU}^*(4)$ Lorentz group, and they transform under the right-handed little group. Using the invariant tensors of $\text{SU}^*(4)$, Lorentz invariants can be constructed as follows:
\be
\langle \lambda_i^a \lambda_j^b \lambda_k^c \lambda_l^d \rangle = \epsilon_{ABCD} \lambda_i^{A,a} \lambda_j^{B,b} \lambda_k^{C,c} \lambda_l^{D,d}, 
\ee
\be [ \tilde{\lambda}_{i,\doth{a}} \tilde{\lambda}_{j,\doth{b}} \tilde{\lambda}_{k,\doth{c}} \tilde{\lambda}_{l,\doth{d}} ] = \epsilon^{ABCD} \tilde{\lambda}_{i,A,\doth{a}} \tilde{\lambda}_{j,B,\doth{b}} \tilde{\lambda}_{k,C,\doth{c}} \tilde{\lambda}_{l,D,\doth{d}},
\ee
\be
\langle \lambda_{i}^{a} | \tilde{\lambda}_{j,\doth{b}} ] = \lambda_{i}^{A,a} \tilde{\lambda}_{j,A,\doth{b}} = [ \tilde{\lambda}_{j,\doth{b}} | \lambda_{i}^{a}  \rangle.
\ee
The $\l$ and $\tilde\l$ variables are not independent. They are related by the condition 
\be
\langle \lambda_i^a | \tilde{\lambda}_{i,\doth{a}} ] = 0,
\ee
for all $a$ and $\doth{a}$. We also have
\be
\epsilon_{ABCD}\, p_i^{AB} p_j^{CD} = 2 \, p_{i,AB}\, p_j^{AB} = 8\, p_i \cdot p_j .
\ee

Using the notation given above, the scattering maps can be written in terms of 6D spinor-helicity variables:
\be\label{evenmap}
p^{AB}(z) = \langle \rho^{A}(z) \rho^B (z) \rangle. 
\ee
In the following we take the spinorial maps $\rho^{A,a}(z)$, for $a\in \{+,-\}$, to be polynomials of the same degree. In contrast to 4D, we can also consider non-polynomial forms of the maps (such that \eqref{evenmap} is still a polynomial), see discussion at the end of Section \ref{oddmeasure}. Note that this choice is consistent with the action of the group denoted $\text{SL}(2,\mathbb{C})_{\rho}$. This is the same abstract group as the little group, but it does not refer to a specific particle. Let us now focus on the construction for $n$ even. In this case the degree of the polynomials is $m=\frac{n}{2}-1$. Thus they can be expanded as:
\be
\rho^{A,a}(z) = \sum_{k=0}^{m} \rho_{k}^{A,a} z^k.
\ee
With these maps the polynomial constructed in \eqref{evenmap} is null and has the correct degree $n-2$. By the arguments reviewed in Section \ref{arbitrarydim} we conclude that the equations constructed from $\rho^{A,a}(z)$,
\begin{equation}\label{eq:6dmapeqs}
p^{AB}_i = \frac{\langle \rho^{A}(\sigma_i)\, \rho^{B}(\sigma_i) \rangle }{\prod_{j\neq i} \sigma_{ij}}\,,
\end{equation}
imply the scattering equations for $\{\s_i\}$. However, the converse, i.e., that any solution of the scattering equations is a solution to \eqref{eq:6dmapeqs} is not guaranteed. This was checked numerically in \cite{Heydeman:2017yww} for even multiplicity up to  $n=8$ particles. In this work we give an inductive proof of this fact in Appendix \ref{app:softeven}, obtained by considering consecutive soft limits of the maps. Using this fact together with the counting of delta functions we then argue that the following measure
\be\label{6d-measure}
\int d\mu_{n \text{ even}}^{\text{6D}} = \int \frac{\prod_{i=1}^n d\sigma_i\, \prod_{k=0}^{m} d^8 \rho_k}{\vol( \slc_\sigma \times \slc_\rho)} \frac{1}{V_n^2}\prod_{i=1}^n \delta^6 \left( p^{AB}_i - \frac{\langle \rho^{A}(\sigma_i)\, \rho^{B}(\sigma_i) \rangle }{\prod_{j\neq i} \sigma_{ij}}\right)
\ee
is equivalent to the CHY measure given in \eqref{CHY-measure}, after integrating out the $\rho$ moduli. Also, it has momentum conservation and null conditions built-in. The formula contains the Vandermonde factor 
\be
V_n = \prod_{1 \leq i<j \leq n} \sigma_{ij}.
\ee
which is needed to match the $\slc_\sigma$ weight of \eqref{CHY-measure}. In order to avoid confusion, we use the notation $\slc_\sigma$ for the M\"obius group acting on the Riemann sphere. Just as the $\slc_\s$ symmetry can be used to fix three of the $\s$ coordinates, the $\slc_\rho$ symmetry can be used to fix three of the coefficients of the polynomial maps $\rho^{A,a}(z)$. This form of the measure imposes $6n$ constraints on $5n -6$ integration variables, leaving a total of $n+6$ delta functions which account for the $n$ on-shell conditions and the six momentum conservation conditions. Fixing the values of $\s_1$, $\s_2$, $\s_3$ and of $\rho_0^{1,+}$, $\rho_0^{1,-}$, $\rho_0^{2,+}$, the gauge-fixed form of the measure becomes:
\be
\int d\m_{n \text{ even}}^{\text{6D}} =  \int\frac{J_{\rho}\, J_{\sigma}}{V_n^2} \left(\prod_{i=4}^{n}d\sigma_i\right) d\rho_{0}^{2,-} d^{2}\rho_{0}^{3} d^{2}\rho_{0}^{4} \left(\prod_{k=1}^{m}d^{8}\rho_{k}\right) \prod_{i=1}^{n} \delta^6 \left( p^{AB}_i - \frac{\langle \rho^{A}(\sigma_i) \,\rho^{B}(\sigma_i) \rangle }{\prod_{j\neq i} \sigma_{ij}}\right),\nn
\ee
where the Jacobians are\footnote{This Jacobian can be derived from the identity $\int d^6 p_0 \, \delta (p_0^2)= \int J_\rho \,  d\rho_0^{2,-}d\rho_0^{3,+}\,d\rho_0^{3,-}\,d\rho_0^{4,+}\,d\rho_0^{4,-}$, since the map component $p_0^{AB}=\langle \rho_0^A\, \rho_0^B\rangle $ is a null vector.}
\be\label{jacobians}
J_{\sigma} = \sigma_{12} \sigma_{23} \sigma_{31}, \qquad \qquad J_{\rho} = \rho_{0}^{1,+}\langle\rho_{0}^{1}\, \rho_{0}^{2}\rangle. 
\ee
It is convenient to use a short-hand notation for the bosonic delta functions:
\be\label{bosonic-delta}
\Delta_B =\prod_{i=1}^n \delta^6 \left( p^{AB}_i - \frac{\langle \rho^{A}(\sigma_i)\,\rho^{B}(\sigma_i) \rangle }{\prod_{j\neq i} \sigma_{ij}}\right) = \delta^6({\textstyle\sum_{i=1}^n} p^{AB}_i) \left(\prod_{i=1}^n \delta(p_i^2)\right) \hat{\Delta}_{B},
\ee
where $\hat{\Delta}_B$ is 
\be \label{bosonic-delta1}
\hat{\Delta}_{B}= \delta^4\left(p^{AB}_n - \frac{\langle \rho^{A}(\sigma_n)\,\rho^{B}(\sigma_n) \rangle }{\prod_{i\neq n} \sigma_{ni}} \right) \prod_{i=1}^{n-2} \delta^5 \left( p^{AB}_i - \frac{\langle \rho^{A}(\sigma_i)\,\rho^{B}(\sigma_i) \rangle }{\prod_{j\neq i} \sigma_{ij}}\right) \prod_{i=1}^n p_i^{12} \left(\frac{p_{n-1}^{24}}{p_{n-1}^{12}}-\frac{p_{n}^{24}}{p_{n}^{12}}\right).
\ee
Here the five dimensional delta functions are chosen such that $\{A,B\}\neq \{3,4\}$, whereas $\{A,B\}\neq \{3,4\},\{1,3\}$ for the four dimensional ones, and the additional factors are the Jacobian of taking out the momentum conservation and on-shell conditions \cite{Heydeman:2017yww}. Alternatively, a covariant extraction of the on-shell delta functions can be obtained by introducing auxiliary variables $M_i$ that linearize the constraints, analogous to the ones given in \eqref{t-variables}, as follows:
\begin{align}
\label{linearmeasure}
\delta^6 \left( p^{AB}_i - \frac{\langle \rho^{A}(\sigma_i)\, \rho^{B}(\sigma_i) \rangle }{\prod_{j\neq i} \sigma_{ij}}\right) &= \delta(p_i^2) \int d^4 M_i\, |M_i|^3 \, \delta\!\left( |M_i| - \prod_{j\neq i} \sigma_{ij}\right) \\
&\qquad\qquad\qquad\qquad \times \delta^8 \left( \rho^{A,a}(\sigma_i) - (M_i)^{a}_{b}\, \lambda^{A,b}_i \right),\nn
\end{align}
where $|M_i|$ denotes the determinant of the matrix $M_i$, and for some purpose it is more convenient to use this version of constraints. This form connects the maps directly to the external 6D spinors, and is a 6D version of the Witten--RSV constraints, which we explore in Section \ref{sec:LinearConstraints}.

\section{\label{sec:even-pt}$\mathbf{\N=(1,1)}$ Super Yang--Mills: Even Multiplicity}

In the following sections, we will propose a formula based on rational maps for the tree amplitudes of 6D maximal SYM theory, which has $\mathcal{N}=(1,1)$ non-chiral supersymmetry. This theory describes the non-abelian interactions of a vector, four scalars, and four spinors all of which are massless and belong to the adjoint representation of the gauge group. As usual, we will generally consider color-stripped SYM amplitudes. Some properties of these amplitudes have been discussed in \cite{Dennen:2009vk, Bern:2010qa, Brandhuber:2010mm, Plefka:2014fta} using 6D $\mathcal{N}=(1,1)$ superspace.

In addition to the usual spacetime and gauge symmetries of Yang--Mills theory, the $\mathcal{N}=(1,1)$ theory has a $\text{Spin}(4) \sim \text{SU}(2) \times \text{SU}(2)$ R symmetry group. The intuitive way to understand this is to note that this theory arises from dimensional reduction of 10D SYM theory, and the R symmetry corresponds to rotations in the four transverse directions. This group happens to be the same as the little group, which is just a peculiarity of this particular theory. From these and other considerations, one may argue that 6D $\mathcal{N}=(1,1)$ SYM with $\text{U}(N)$ gauge symmetry (in the perturbative regime with no theta term) describes the IR dynamics of $N$ coincident D5-branes in type IIB superstring theory \cite{Witten:1997kz}. In contrast to 4D $\mathcal{N} = 4$ SYM, the gauge coupling in six dimensions has inverse mass dimension, so this theory is non-renormalizable and not conformal. This is not an issue for the tree amplitudes that we consider in this work. Further dimensional reduction on a $T^2$ leads to 4D $\mathcal{N} = 4$ SYM, and this provides a consistency check of the results.

Six-dimensional $\mathcal{N}=(1,1)$ SYM is a theory with 16 supercharges. Its physical degrees of freedom form a 6D $\mathcal{N}=(1,1)$ supermultiplet consising of eight on-shell bosons and eight on-shell fermions. These may be organized according to their quantum numbers under the four $\text{SU}(2)$'s of the little group and R symmetry group. For example, the vectors belong to the representation $(\mathbf{2,2;1,1})$, which means that they are doublets of each of the little-group $\text{SU}(2)$'s and singlets of each of the R symmetry $\text{SU}(2)$'s. In this notation, the fermions belong to the representation $(\mathbf{1,2;2,1})+(\mathbf{2,1;1,2})$, and the scalars belong to the representation $(\mathbf{1,1;2,2})$. (Whether one writes $(\mathbf{1,2;2,1})+(\mathbf{2,1;1,2})$ or $(\mathbf{1,2;1,2})+(\mathbf{2,1;2,1})$ is a matter of convention.)

It is convenient to package all 16 of these particles into a single on-shell ``superparticle", 
by introducing four Grassmann numbers (per superparticle),
\begin{eqnarray}  \label{eq:spectrum11}
\Phi(\eta)
=  \phi^{1 \hat{1}} + \eta_a  \psi^{a \hat{1}} + \tilde{\eta}_{\hat{a}}  \hat{\psi}^{\hat{a}1} +  \eta_a \tilde{\eta}_{\hat{a}} A^{a\hat{a}} + (\eta)^2 \phi^{2 \hat{1}} + (\tilde{\eta})^2 \phi^{1 \hat{2}} 
+ \cdots  + (\eta)^2 (\tilde{\eta})^2 \phi^{2 \hat{2}} \,. 
\end{eqnarray}
Here $\eta_a$ and $\tilde{\eta}_{\hat{a}}$ are the four Grassmann numbers, and the $\text{SU}(2)$ indices $a$ and $\hat{a}$ are little-group indices as before. The explicit 1's and 2's in the spectrum described above are R symmetry indices. Since the superfield transforms as a little-group scalar, this formulation makes the little-group properties manifest, but it obscures the R symmetry. By means of an appropriate Grassmann Fourier transform one could make the R symmetry manifest, but then the little-group properties would be obscured as explained in \cite{Heydeman:2017yww}. The choice that has been made here turns out to be the more convenient one for the study of superamplitudes.

When discussing an $n$-particle amplitude the Grassmann coordinates carry an additional index $i$, labeling the $n$ particles, just like the spinor-helicity coordinates. Thus, the complete color-stripped on-shell $n$-particle tree amplitude will be a cyclically symmetric function of the $\l_i$'s and the $\eta_i$'s. The various component amplitudes correspond to the terms with the appropriate dependence on the Grassmann coordinates. Thus, the superamplitude is like a generating function in which the Grassmann coordinates play the role of fugacities. This is an on-shell analog of the use of superfields in the construction of Lagrangians. Fortunately, it exists in cases where the latter does not exist.

Often we will refer to the momenta $p^{AB}_i$ and supercharges $q_i^A$, $\tilde{q}_{iA}$ of the on-shell states. For $(1,1)$  supersymmetry, they can be expressed in terms of the Grassmann coordinates:
\begin{equation}
q_i^{A} = \epsilon^{ab} \lambda_{ia}^A \eta_{ib} = \langle \lambda_{i}^A\, \eta_i \rangle, \quad
\tilde q_{iA} = \epsilon^{\hat a \hat b} \tilde \lambda_{i A\hat a} \tilde \eta_{i \hat b}
= [\tilde \l_{i A} \, \tilde\eta_i] ,
\end{equation}
and the superamplitudes should be annihilated by the supercharges $Q^A = \sum_{i=1}^n q_i^{A}$ and $\tilde Q_A = \sum_{i=1}^n \tilde q_{i A}$. These symmetries will be manifest in the formulas that follow. However, there are eight more supercharges, involving derivatives with respect to the $\eta$ coordinates, which should also be conserved. Once one establishes the first eight supersymmetries and the R symmetry, these supersymmetries automatically follow. The explicit form of the derivatively realized supercharges is:
\begin{equation}
\bar{q}_i^{A} = \lambda_{ia}^A \frac{\partial}{\partial \eta_{ia}}, \quad
\tilde{\bar q}_{iA} = \tilde \l_{i A\hat a} \frac{\partial}{\partial \tilde\eta_{i\hat a}},
\end{equation}

In terms of these Grassmann variables, one may also write the generators of the $\text{SU}(2) \times \text{SU}(2)$ R symmetry group. One first notes that they obey the anti-commutation relations:
\begin{align}
\left \{ \eta_a , \frac{\partial}{\partial \eta^b} \right \} = \epsilon_{ab}, \quad \left \{ \tilde{\eta}_{\hat a} , \frac{\partial}{\partial \tilde{\eta}^{\hat b}} \right \} = \epsilon_{\hat{a}\hat{b}} \, .
\end{align}
In terms of these, the six generators of the R symmetry group may be defined as
\begin{align}
\label{eq:Rgenerators}
R^+ = \eta_a \eta^a, \quad R^- = \frac{\partial}{\partial \eta^a} \frac{\partial}{\partial \eta_a}, \quad R = \eta_a \frac{\partial}{\partial \eta_a} - 1, \\
\widetilde{R}^+ = \tilde{\eta}_{\hat a} \tilde{\eta}^{\hat a}, \quad \widetilde{R}^- = \frac{\partial}{\partial \tilde{\eta}^{\hat a}} \frac{\partial}{\partial \tilde{\eta}_{\hat a}}, \quad \widetilde{R} = \tilde{\eta}_{\hat a} \frac{\partial}{\partial \tilde{\eta}_{\hat a}} - 1\, ,
\end{align}
which have the standard raising and lowering commutation relations. These generate a global symmetry of $\mathcal{N}=(1,1)$ SYM. It is easy to see that linear generators $R$ and $\widetilde{R}$ annihilate amplitudes since they are homogeneous polynomials of degree $n$ in both $\eta$ and $\tilde{\eta}$. The non-linearly realized ones become more transparent in an alternative form of the constraints that we will discuss in Section~\ref{sec:LinearConstraints}. As explained earlier, this is due to the choice of parametrization of the non-chiral on-shell superspace.

As discussed in previous literature for tree-level amplitudes of 6D $\mathcal{N}=(1,1)$ SYM, the four-particle partial amplitude is particularly simple when expressed in terms of the supercharges: 
\begin{equation} \label{A4D5}
\mathcal{A}_4^{\N=(1,1) \text{ SYM}}(1234) = \d^6 \left(\sum_{i=1}^4 p_i^{AB} \right)\,
\frac{\d^4 \left(\sum_{i=1}^4 q_i^{A} \right)
\d^4 \left(\sum_{i=1}^4 \tilde q_{i,A} \right)}{s_{12} \, s_{23}}.
\end{equation}
Here and throughout this work one should view this expression as a superamplitude; the component amplitudes may be extracted by Grassmann integration. For example, in terms of the Lorentz invariant brackets the four-gluon amplitude is:
\begin{eqnarray}
\mathcal{A}_4( A_{a \hat a} A_{b \hat b} A_{c \hat c} A_{d \hat d} )
=\d^6 \left(\sum_{i=1}^4 p_i^{AB} \right) \frac{ \langle  1_{a} 2_{b} 3_{c} 4_{d}\rangle
[  1_{\hat a} 2_{\hat b} 3_{\hat c} 4_{\hat d}]}{s_{12}\, s_{23}}.
\end{eqnarray}
Using the formalism of rational maps for the 6D spinor-helicity variables, the main technical result of this section is a formula for the $n$-point generalization of the superamplitude when $n$ is even. The formula for odd $n$ will be given in Section~\ref{sec:odd-pt}.

\subsection{\label{sec:even_points}Connected Formula}

We propose that the connected formula for even-multiplicity 6D $\N=(1,1)$ SYM amplitudes is given by
\be \label{eq:(1,1)SYM-even}
\boxed{{\mathcal{A}}_{n\text{ even}}^{\N=(1,1)\text{ SYM}} (\alpha) = \int d\mu_{n \text{ even}}^{\text{6D}}\; {\rm PT}( \alpha ) \left(\,  {\rm Pf'} A_n \int d\Omega_F^{(1,1)} \right),}
\ee
where $d\mu_{n \text{ even}}^{\text{6D}}$ is the measure given in \eqref{6d-measure}, and we will shortly explain other ingredients that enter this formula. This formula is inspired by the D5-brane effective field theory scattering amplitudes written as a connected formula \cite{Heydeman:2017yww}, where the factor of $(\text{Pf}\,' A_n)^2$ has been replaced with $\text{PT}(\alpha)$ given in \eqref{eq:PTfactor}. This is a standard substitution in the CHY formalism for passing from a probe D-brane theory to a Yang--Mills theory. Since the only non-vanishing amplitudes of the D5-brane theory have even $n$, this only works for the even-point amplitudes of SYM.

As indicated explicitly in the expression (\ref{eq:(1,1)SYM-even}), the integrand of \eqref{eq:(1,1)SYM-even} factorizes into two half-integrands. Such a factorization of the integrand will be important later when we deduce the formulas for 6D SUGRA with $\mathcal{N} = (2,2)$ supersymmetry. The left half-integrand ${\rm PT}( \alpha )$ is the Parke--Taylor factor, where $\alpha$ is a permutation that denotes the color ordering of Yang--Mills partial amplitudes. The right half-integrand further splits into two quarter-integrands. The first of these is the reduced Pfaffian of the antisymmetric matrix $A_n$, whose entries are given by:
\begin{align}\label{A-matrix-def}
[A_n]_{ij} = \begin{cases}
\dfrac{p_{i} \cdot p_{j}}{\sigma_{ij}} &\quad\text{if}\quad i\neq j, \\
\quad 0\phantom{\dfrac{1}{1}} &\quad\text{if}\quad i=j,
\end{cases}   \qquad\qquad\text{for}\qquad i,j=1,2,\ldots,n.
\end{align}
Since this matrix has co-rank $2$, its Pfaffian vanishes. Instead, one defines the reduced Pfaffian:
\be\label{reduced-pfaffian}
\text{Pf}\,' A_n = \frac{(-1)^{p+q}}{\sigma_{pq}} \text{Pf}\, A_n^{[pq]}, 
\ee
where we have removed two rows and columns labeled by $p$ and $q$, and denoted the resulting reduced matrix by $A_n^{[pq]}$. The reduced Pfaffian is independent of the choice of $p$ and $q$ \cite{Cachazo:2014xea} and transforms under $\slc_\s$ in an appropriate way.

The remaining quarter integrand is the fermionic integration measure responsible for implementing the 6D $\N=(1,1)$ supersymmetry \cite{Heydeman:2017yww}, which we will review here. The formula is
\be\label{even-fermionic-measure}
d\Omega_F^{(1,1)} = V_n \left( \prod_{k=0}^{m}  d^2\chi_k\,  d^2\tilde{\chi}_k  \right)\, \D_F \widetilde{\D}_F,
\ee
where $m ={n \over 2} - 1$, as before. This measure contains the Vandermonde determinant $V_n$, as well as a fermionic measure and fermionic delta functions. The integration variables arise as the coefficients of the fermionic rational maps, which are defined by
\be
\chi^{a} (z)   = \sum^{m}_{k=0} \chi^{ a}_k \, z^k, \qquad \quad \tilde{\chi}^{\hat{a}} (z)   = \sum^{m}_{k=0} \tilde{\chi}^{ \hat{a}}_k \, z^k,
\ee
where $\chi^{ a}_k$ and $\tilde{\chi}^{ \hat{a}}_k$ are Grassmann variables. The fermionic delta functions, $\D_F$ and $\widetilde{\D}_F$ are given by:
\begin{align}\label{fermdelt}
\D_F &=  \prod^n_{i=1}  \delta^{4} \left(q^{A}_i -   { \langle \rho^A(\sigma_i)\, \chi (\sigma_i) \rangle \over \prod_{j \neq i} \sigma_{ij} }  \right) , \\
\widetilde{\D}_F &=  \prod^n_{i=1} \delta^{4} \left( \tilde{q}_{i,A} -   { [\tilde{\rho}_{A}(\sigma_i) \, \tilde{\chi} (\sigma_i) ] \over \prod_{j \neq i} \sigma_{ij} }  \right).
\end{align}
These delta functions are built from the external chiral and anti-chiral supercharges of each particle and are responsible for the $(1,1)$ supersymmetry in this formalism. Conservation of half of the 16 supercharges is made manifest by this expression. As in (\ref{A4D5}), the component amplitudes can be extracted by Grassmann integration of the appropriate $\eta_a$'s and $\tilde{\eta}_{\hat{a}}$'s, which enter via the supercharges.

Even though the maps $\tilde{\rho}_{A\hat{a}}(z)$ appear explicitly in $\widetilde{\D}_F$, just as in the construction of D5-brane amplitudes \cite{Heydeman:2017yww}, the integration measure does not include additional integrations associated to the maps $\tilde{\rho}_{A\hat{a}}(z)$. If it did, the formula, for instance, would have the wrong mass dimension to describe SYM amplitudes in 6D. Instead, the $\tilde\rho$ coefficients are fixed by the conjugate set of rational constraints
\bea
\label{eq:conjconstraints}
p_{i, AB} - { [\tilde{\rho}_{A }(\sigma_i) \, \tilde{\rho}_B(\sigma_i) ] \over \prod_{j \neq i} \sigma_{ij} } =0 \, ,
\eea
for all $i=1, 2, \ldots, n$. These equations are not enough to determine all of the $\tilde{\rho}^{\hat{a} }_{A,k}$'s. One needs to utilize $\slc_{\tilde{\rho}}$ to fix the remaining ones. The resulting amplitude is independent of choices that are made for the $\slc_{\tilde{\rho}}$ fixing because $\tilde{\rho}_{A,\hat{a}}(\sigma_i) \tilde{\chi}^{\hat{a}} (\sigma_i)$ and the fermionic measure $d^{2} \tilde{\chi}_k$ are $\slc_{\tilde{\rho}}$ invariant. The usual scattering amplitudes ${{A}}_n$ are obtained by removing the bosonic and fermionic on-shell conditions (``wave functions''), which appear as delta functions, namely,
\begin{equation} \label{eq:wave_function}
\mathcal{A}_n^{\N=(1,1) \text{ SYM}} = \delta^6 \left(\sum_{i=1}^n p_i\right) \left( \prod_{i=1}^n \delta(p_i^2) \delta^2(\tilde{\lambda}_{i,A,\hat{a}}\, q^A_i) \delta^2(\lambda_{i,b}^B\, \tilde{q}_{i,B})\right) A_n^{\N=(1,1) \text{ SYM}}.
\end{equation}

It is straightforward to show that this formula produces the correct four-point superamplitude of 6D $\N=(1,1)$ SYM, expressed in (\ref{A4D5}).
A quick way to see it is to utilize the relation between the D5-brane amplitudes and the amplitudes of 6D $\N=(1,1)$ SYM. As we discussed previously, they are related by the exchange of $({\rm Pf}^{\prime} A_n)^2$ with the Parke--Taylor factor $\text{PT}(\alpha)$. The four-point superamplitude for the D5-brane theory is given by \cite{Heydeman:2017yww}
\be \label{eq:D5_4pts}
A^{\text{D5-brane}}_4 = { \delta^4 \left( \sum_{i=1}^4 q^A_i \right) \delta^4 \left( \sum_{i=1}^4 \tilde{q}_{i,A} \right)  } \, .
\ee
From  the explicit solution of the four-point scattering equations for the $\s_i$'s, it is easy to check that the effect of changing from $(\text{Pf}\,' A_4)^2$ to $\text{PT}(1234)$, defined in (\ref{eq:PTfactor}), is to introduce an additional factor of $1/(s_{12}\,s_{23})$. Namely, on the support of the scattering equations, we have the following identity for the $\slc_\sigma$-invariant ratio, 
\bea
{\text{PT}(1234) \over (\text{Pf}\,' A_4)^2 } = {1 \over s_{12} \, s_{23} }\,.
\eea
Thus, combining this identity and the D5-brane formula (\ref{eq:D5_4pts}), we arrive at the result of the four-point of 6D $\mathcal{N}=(1,1)$ SYM (\ref{A4D5}). We have further checked numerically that the above formula reproduces the component amplitudes of scalars and gluons for $n=6,8$, obtained from Feynman diagram computations. 

\subsection{\label{sec:Integrand_CHY_Even}Comparison with CHY}

This section presents a consistency check of the integrand by comparing a special bosonic sector of the theory with a CHY formula of YM amplitudes valid in arbitrary spacetime dimensions. This comparison actually also gives a derivation of the integrand in (\ref{eq:(1,1)SYM-even}). We begin with the general form of the superamplitude, 
\be 
\mathcal{A}_{n\text{ even}}^{\N=(1,1)\text{ SYM}} (\alpha) = \int d\mu_{n \text{ even}}^{\text{6D}}\; \int d\Omega_F^{(1,1)} \times  \mathcal{J}_{n\,\rm even} \, ,
\ee
where the measures $d\mu_{n \text{ even}}^{\text{6D}}$ and $d\Omega_F^{(1,1)}$ take care of 6D massless kinematics and 6D $\N=(1,1)$ supersymmetry, respectively. The goal is then to determine the integrand $\mathcal{J}_{n\,\rm even}$. The strategy is to consider a particular component amplitude by performing fermionic integrations of the superamplitude $\mathcal{A}_{n\text{ even}}^{\N=(1,1)\text{ SYM}} (\alpha)$ such that our formula can be directly compared to the known CHY integrand, thereby determining $\mathcal{J}_{n\,\rm even}$. 

To make the fermionic integration as simple as possible, it is convenient to consider a specific all-scalar amplitude, for instance, 
\be
\mathcal{A}_n(\phi^{1\hat{1}}_1, \ldots, \phi^{1\hat{1}}_{\frac{n}{2}},  \phi^{2\hat{2}}_{\frac{n}{2}+1}, \ldots , \phi^{2\hat{2}}_{n}).
\ee
Half of the particles have been chosen to be the scalar of the top component of the superfield, while the other half are the scalar of the bottom component of the superfield. This equal division is required to obtain a non-zero amplitude, because the superamplitude is homogeneous of degree $n$ both in the $\eta$ and $\tilde{\eta}$ coordinates. Due to this convenient choice of the component amplitude, the fermionic integral over $\chi$'s and $\tilde{\chi}$'s can be done straightforwardly. Explicitly, for the component amplitude we are interested in, 
\be
\int d\Omega_F^{(1,1)} \;\Longrightarrow\; 
V_n\, J_{\rm w}   \int  \prod_{k=0}^m  d^2\chi_k\,  d^2\tilde{\chi}_k \, \D^{\rm proj}_F \widetilde{\D}^{\rm proj}_F ,
\ee
where we have taken out the fermionic wave functions as in (\ref{eq:wave_function}), which results in a Jacobian $J_{\rm w} = \prod_{i=1}^n \frac{1}{(p^{13}_i)^{2}}$ in the above expression. Furthermore, the fermionic delta functions are projected to the component amplitude of interest, 
\bea
\D^{\rm proj}_F &=&   \prod_{i \in Y} \prod_{A=1,3} \delta \left(   { \langle \rho^A(\sigma_i)\, \chi (\sigma_i) \rangle \over \prod_{j \neq i} \sigma_{ij} }  \right) \prod_{i \in \overline{Y} } p_i^{13} , \\
\widetilde{\D}^{\rm proj}_F &=& 
\prod_{i \in Y}  \prod_{A=2,4} \delta \left(  { [\tilde{\rho}_{A}(\sigma_i) \, \tilde{\chi} (\sigma_i) ] \over \prod_{j \neq i} \sigma_{ij} }  \right)
\prod_{i \in \overline{Y} } p_i^{13}.
\eea
Here $Y$ labels all the scalars $\phi^{1\hat{1}}$, namely $Y:=\{1, \ldots, \frac{n}{2} \}$, and $\overline{Y}$ labels the other type of scalars $\phi^{2\hat{2}}$, so $\overline{Y}:=\{\frac{n}{2}+1, \ldots, n\}$.

Carrying out the integrations over $d^2\chi_k$ and $d^2\tilde{\chi}_k$, we see that the maps $\rho^A_a(\sigma_i)$ combine nicely into $\langle \rho^A(\sigma_i)\, \rho^B(\sigma_i) \rangle$, which on the support of the rational map constraints becomes $p^{AB}_i \prod_{j \neq i} \sigma_{ij}$. Concretely, we have,
\bea
&& \int  \prod_{k=0}^m  d^2\chi_k\,  d^2\tilde{\chi}_k \prod_{i \in Y} \prod_{A=1,3} \delta \left(   { \langle \rho^A(\sigma_i)\, \chi (\sigma_i) \rangle \over \prod_{j \neq i} \sigma_{ij} }  \right)  \prod_{B=2,4} \delta \left(  { [\tilde{\rho}_{B}(\sigma_i) \, \tilde{\chi} (\sigma_i) ] \over \prod_{j \neq i} \sigma_{ij} }  \right) \cr
&& \qquad\qquad\qquad\qquad\qquad\qquad\qquad\qquad\qquad\qquad =\prod_{i \in Y} p_i^{13} p_{i, 24} \; \times \prod_{{i\in Y} {J\in \overline{Y}}} { 1 \over  \s^2_{iJ} } \, .
\eea
Collecting terms, we find that the wave-function Jacobian $J_{\rm w}$ cancels out completely, and we obtain the final result
\be \label{eq:JF}
V_n J_{\rm w}   \int  \prod_{k=0}^m  d^2\chi_k\,  d^2\tilde{\chi}_k \, \D^{\rm proj}_F \widetilde{\D}^{\rm proj}_F\; = V_n \prod_{{i\in Y} {J\in \overline{Y}}} { 1 \over  \s^2_{iJ} }:=J_F  \, ,
\ee
where we have defined the final result to be $J_F$. Therefore we have, 
\be \label{eq:evenpt_scalars}
\mathcal{A}_n (\phi^{1\hat{1}}_1, \ldots, \phi^{1\hat{1}}_{\frac{n}{2}},  \phi^{2\hat{2}}_{\frac{n}{2}+1}, \ldots , \phi^{2\hat{2}}_{n}) = 
\int d\mu_{n \text{ even}}^{\text{6D}}\;\, (J_F \times \mathcal{J}_{n\,\rm even}).
\ee
We are now ready to compare this result directly with CHY amplitude, which is given by
\be \label{eq:CHY_scalars}
\mathcal{A}_n (\phi^{1\hat{1}}_1, \ldots, \phi^{1\hat{1}}_{\frac{n}{2}},  \phi^{2\hat{2}}_{\frac{n}{2}+1}, \ldots , \phi^{2\hat{2}}_{n})=  \int d\mu_n\;   {\rm PT}(\alpha) \, 
{\rm Pf'} \Psi_n\big{|}_{\rm project} \, ,
\ee
and $d\mu_n = d\mu_{n \text{ even}}^{\text{6D}}$ if we restrict the CHY formula to 6D.

The notation $\Psi_n\big{|}_{\rm project}$ denotes projection of the matrix $\Psi_n$ of the CHY formulation to the specific scalar component amplitude we want via dimensional reduction. In particular, the ``polarization vectors'' should satisfy $\varepsilon_i \cdot \varepsilon_I =1$ if  $i \in Y$ and $I \in \overline{Y}$ or vice versa. If they belong to the same set, then we have $\varepsilon_i \cdot \varepsilon_j = \varepsilon_I \cdot \varepsilon_J = 0$. Furthermore, $p_i \cdot \varepsilon_j =0$ for all $i$ and $j$, i.e., both sets, since all of the vectors are dimensionally reduced to scalars. Let us now recall the definition of the matrix $\Psi_n$ that enters the CHY construction of YM amplitudes. It can be expressed as
\be
\Psi_n = \begin{pmatrix} 
A_n & -C_n^\intercal  \\
C_n & B_n
\end{pmatrix},
\ee
where $A_n$ is given in \eqref{A-matrix-def}, and $B_n$ and $C_n$ are $n\times n$ matrices defined as
\be
     \quad \quad
                  [B_n]_{ij}= \begin{cases}
                  \dfrac{\varepsilon_i \cdot \varepsilon_j}{\s_{ij}} \, \qquad &\text{if}\quad i\neq j, \\
                  0 \, \qquad &\text{if}\quad i=j.
            \end{cases} 
              \qquad\qquad
              [C_n]_{ij}=\begin{cases}
                  \dfrac{p_j \cdot \varepsilon_i}{\s_{ij}} \, \qquad &\text{if}\quad i\neq j, \\
          -  \displaystyle\sum_{l \neq i} \dfrac{p_l \cdot \varepsilon_i}{\s_{il}} \, \qquad & \text{if}\quad i=j.
                \end{cases}
\ee
Like $A_n$, the matrix $\Psi_n$ is also an antisymmetric matrix of co-rank $2$. Its non-vanishing reduced Pfaffian is defined as
\be
{\rm Pf'} \Psi_n = {(-1)^{p+q} \over \s_{pq}} {\rm Pf} \Psi^{[pq]}_n \,,
\ee
where $\Psi^{[pq]}_n$ denotes the matrix $\Psi_n$ with rows $p, q$ and columns $p, q$ removed. These should be chosen from the first $n$ rows and columns. Otherwise, the result is independent of the choice of $p, q$.  

For the specific choice of the component amplitude described above, $C_n =0$ and the reduced Pfaffian ${\rm Pf'} \Psi_n$ becomes
\be
{\rm Pf'} \Psi_n \big{|}_{\rm project} = {\rm Pf}' A_n \times {\rm Pf} \, B_n\big{|}_{\rm scalar}  ,
\ee
where the ``projected" matrix $B_n$ is 
\be
     \quad 
                  [B_n\big{|}_{\rm scalar}]_{iJ}= \begin{cases}
                  \dfrac{1}{\s_{iJ}} \, \qquad &\text{if}\quad i \in Y, \quad J \in \overline{Y}, \\
                  0 \, \qquad &\quad {\rm otherwise}.
            \end{cases} 
\ee
Using the above result, we find 
\be
{\rm Pf} \, B_n\big{|}_{\rm scalar} = {\rm det}  \left(\dfrac{1}{\s_{iJ}} \right) \, \qquad \text{where}\quad i \in Y, \quad J \in \overline{Y}.
\ee
Comparing (\ref{eq:evenpt_scalars}) with (\ref{eq:CHY_scalars}), we deduce that the even-point integrand should be given by
\be
\mathcal{J}_{n\,\rm even} (\alpha)= {\rm PT}(\alpha) \, {\rm Pf'} A_n {{\rm Pf}\, B_n\big{|}_{\rm scalar} \over  J_F} \, .
\ee
It is easy to prove that ${\rm Pf}\, B_n\big{|}_{\rm scalar}$ and $J_F$ are actually identical. In particular, one can see that they, as rational functions, have the same zeros and poles. So we obtain the desired result, $\mathcal{J}_{n\,\rm even}(\alpha) = {\rm PT}(\alpha) \, {\rm Pf'} A_n$.

\section{\label{sec:odd-pt}$\mathbf{\N=(1,1)}$ Super Yang--Mills: Odd Multiplicity}

This section presents the formula for $\N=(1,1)$ SYM amplitudes with odd multiplicity. This case is considerably subtler than the case of even $n$. It is perhaps the most novel aspect of the present work. Nevertheless, we will show that it can be written in a form entirely analogous to the even-point case:
\be \label{eq:(1,1)SYM-odd}
\boxed{\mathcal{A}_{n\text{ odd}}^{\N=(1,1)\text{ SYM}} (\alpha)  = \int d\mu_{n \text{ odd}}^{\text{6D}}\; {\rm PT}( \alpha ) \left( \,{\rm Pf'} \widehat{A}_n\, \int d\widehat{\Omega}_F^{(1,1)} \right).}
\ee
The following subsections describe the different ingredients in this expression.

Section \ref{sec:odd-rational-maps} starts by presenting the form of the rational maps for $n$ odd and studying the corresponding redundancies that enter in the integration measure. The explicit form of $d\mu_{n \text{ odd}}^{\text{6D}}$, given in \eqref{eq:prop}, is deduced by considering a soft limit of an amplitude with $n$ even. In particular, we deduce the existence of an emergent shift invariance acting on the rational maps. The discussion of how this new invariance interacts with the groups $\text{SL}(2,\mathbb{C})_{\sigma}$ and $\text{SL}(2,\mathbb{C})_{\rho}$ is relegated to Appendix~\ref{app:symmetry-algebra}. Appendix~\ref{app:soft-limits} presents the detailed derivation of the form of the maps, as well as the measure, from the soft limit of the even-point formula \eqref{eq:(1,1)SYM-even}.  

Section \ref{sec:odd-integrand} discusses the form of the integrand for odd $n$, which can also be derived by carefully examining the soft limit. The fermionic integration measure $d\widehat{\Omega}_F^{(1,1)}$ is given explicitly in \eqref{odd-fermionic-measure}. We show that the odd-$n$ analog of the $A_n$ matrix is an antisymmetric $(n+1) \times (n+1)$ matrix, which is denoted $\widehat{A}_n$. It is constructed from $(n+1)$ momenta: the original $n$ momenta of external particles and an additional special null vector, $p^{AB}_{\star}$, defined through an arbitrarily chosen puncture $\s_{\star}$. The formula for the matrix $\widehat{A}_n$ is given in \eqref{odd-matrix-A}, and $p^{AB}_{\star}$ in (\ref{eq:pstar}).

In Section \ref{sec:odd-consistency} we describe consistency checks of \eqref{eq:(1,1)SYM-odd}. This includes a comparison with the CHY formula in the bosonic sector, as was done for $n$ even in \ref{sec:Integrand_CHY_Even}. We also present a computation of the three-point superamplitude \cite{Dennen:2009vk} directly from the connected formula.

\subsection{\label{sec:odd-rational-maps}Rational Maps and the Measure}

Let us consider the definition of the scattering maps in 6D for the odd-point case $n=2m+1$:
\be \label{eq:mapss}
p_{i}^{AB}=\frac{p^{AB}(\sigma_{i})}{\prod_{j\neq i}\sigma_{ij}}.
\ee
This formula implies the scattering equations if $p^{AB}(z)$ is a polynomial of degree $n-2=2m-1$ such that the vector $p^{AB}(z)$ is massless for any value of $z$. The latter is realized by requiring
\be
p^{AB}(z)  =  \langle\rho^{A}(z)\rho^{B}(z)\rangle
  =  \rho^{A,+}(z)\rho^{B,-}(z)-\rho^{A,-}(z)\rho^{B,+}(z),
\ee
as in the case of even $n$. The polynomials $\rho^{A,+}(z)$ and $\rho^{A,-}(z)$ should have the same degree, since we want to maintain $\slc_\rho$ symmetry. This is achieved by choosing $\text{deg } \rho^{A,a}(z) =m$. However, this produces an undesired term of degree $2m = n-1$ in $p^{AB}(z)$. This term can be made to vanish by requiring that the coefficient of $z^m$ in $\rho^{A,a}(z)$ takes the special form
\be \label{eq:factor}
\rho_{m}^{A,a} = \omega^{A} \xi^a,
\ee
since then $\langle\rho_{m}^{A}\,\rho_{m}^{B}\rangle = 0$. This is the first new feature we encounter for odd $n$. The maps for $n=2m+1$ then become
\bea \label{eq:odd-pt-map}
\r^{A,a}(z) &=& \sum_{k=0}^{m-1} \r^{A,a}_{k}\,z^k + \omega^A \xi^a \,z^m , \\\label{eq:odd-pt-map2}
\tilde{\r}_A^{\hat{a}}(z) &=& \sum_{k=0}^{m-1} \tilde{\r}_{A\,k}^{\hat{a}} \,z^k + \tilde{\omega}_A \tilde{\xi}^{\hat a} \,z^m .
\eea

Note that the spinor $\xi^a$, which we also write as $|\xi\rangle$, involves a projective scale that can be absorbed into $\omega^{A}$, which is invariant under $\slc_\rho$. In other words, $\xi^a$ are homogeneous coordinates on $\mathbb{CP}^1$. For instance, this freedom can be used to set
\be
|\xi\text{\ensuremath{\rangle}}=\left(\begin{array}{c}
1\\
\xi
\end{array}\right).
\ee
In the following we use the symbol $\xi$ to denote both the two-component spinor and its only independent component. 

After plugging this form of the (chiral) maps into equations \eqref{eq:mapss} we find the expected $(n-3)!$ solutions. This is consistent since, as we show in Appendix \ref{app:soft-limits}, this version of the maps can be obtained directly from a soft limit of the even multiplicity ones. However, a counting argument quickly leads to the fact that we must fix an extra component of the maps when solving the equations: There are $5n-6$ independent equations for $5n+1$ variables, which implies the existence of seven redundancies. Six of them are of course the $\slc$'s present in the even case, but there is an emergent redundancy that we call \emph{T-shift} symmetry. It is the subject of the next section.

\subsubsection{Action of the T Shift}\label{sym}

Consider the following transformation on the polynomials
\begin{eqnarray}\label{shift}
\rho^{A}(z) \;\rightarrow\; \hat{\rho}^{A}(z) & = & (\mathbb{I}+z\,T)\rho^{A}(z). \label{eq:action}
\end{eqnarray}
Here $T$ is a $2\times2$ matrix labeled by little-group indices. In order
to preserve the bosonic delta functions, $\Delta_{B}$, we require that for any value of $z$ and for any polynomial $\rho^A (z)$:
\begin{eqnarray}
p^{AB}(z) & = & \hat{p}^{AB}(z)\\
& = & \langle(\mathbb{I}+zT)\rho^{A}(z)\,\:(\mathbb{I}+zT)\rho^{B}(z)\rangle\nn\\
& = & \langle\rho^{A}(z)\, \rho^{B}(z)\rangle+z\left(\langle T\rho^{A}(z) \; \rho^{B}(z)\rangle+\langle\rho^{A}(z) \; T\rho^{B}(z)\rangle\right) +z^{2}\langle T\rho^{A}(z)\; T\rho^{B}(z)\rangle.\nn
\end{eqnarray}
Thus we obtain the following conditions
\be
\label{eq:Tconditions}
T^\intercal \epsilon+\epsilon T = 0, \qquad\qquad T^\intercal \epsilon T =  0,
\ee
where $T^\intercal$ is the transpose of $T$ and $\epsilon$ is the $2\times2$
antisymmetric matrix. The first condition is equivalent to
\begin{equation}\label{Tr-T}
\text{Tr}\, T =0,
\end{equation}
which implies that $T^{2}\propto\mathbb{I}$. The second condition
then fixes
\begin{equation}\label{T2}
T^{2}=0.
\end{equation}
What is the meaning of the conditions \eqref{Tr-T} and \eqref{T2}? They guarantee that the transformation \eqref{shift} is a $z$-dependent $\slc_{\rho}$ transformation, hence preserving the polynomial map $p^{AB}(z)$. In other words, \eqref{Tr-T} and \eqref{T2} are equivalent to
\be
\det (\mathbb{I}+z\,T)=1\qquad \text{for any $z$.}
\ee
We discuss such transformations in more generality in the next sections. For now, let us further impose that $T$ preserves the
degree of the maps, i.e., $\text{deg } \hat{\rho}^{A,a}(z) = \text{deg } \rho^{A,a}(z) =m$,
that is
\be
T_{b}^{a}\rho_{m}^{A,b}=0,
\ee
where $\rho_{m}^{A,a}$ is the top coefficient. This means that the
kernel of $T$ consists of the four spinors $\rho_{m}^{A,a}$ with $A=1,2,3,4$. In general
this would force the $2\times 2$ matrix $T$ to vanish. However, for an odd number of particles
\be
\rho_{m}^{A,a}=\omega^{A}\xi^{a} \quad\implies\quad T_{b}^{a}\xi^{b}=0.
\ee
These two equations, together with condition \eqref{Tr-T}, fix three
of the four components of $T$. It is easy to see that the solution
is 
\be
T=\alpha|\xi\rangle\langle\xi|,
\ee
where $\alpha\in\mathbb{C}$ is a complex scale. Therefore we have found
a redundancy on the coefficients of the maps given by the transformation
\eqref{shift}. This is an inherent consequence of the description of the moduli space in terms of the polynomials \eqref{eq:odd-pt-map}. In fact, in Appendix \ref{app:symmetry-algebra} we show how $T$ is necessary from a purely group-theoretic point of view, when regarding the equivalent maps as representations of a bigger group, identified as $\slc \ltimes \mathbb{C}^2$. Finally, in Appendix \ref{app:soft-limits} we show how the soft limit of the even-multiplicity maps gives another interpretation of $T$ that is reminiscent of the little group of the soft particle.

Let us close this part of the section by noting that $T$ produces the following shift on the top component of the polynomial:
\be
\hat{\rho}_{m}^{A,a}  =  \rho_{m}^{A,a} + T\rho_{m-1}^{A,a} = \omega^{A}\xi^a+\alpha\langle\xi\,\rho_{m-1}^{A}\rangle \xi^a,
\ee
or equivalently,
\begin{equation}\label{omega-shift}
\hat{\omega}^{A}=\omega^{A}+\alpha\langle\xi \,\rho_{m-1}^{A}\rangle,
\end{equation} 
which will be useful in the next section.

\subsubsection{Measure \label{oddmeasure}}

Let us introduce the measure for $n=2m+1$, which can be obtained from the soft limit of the measure for $n=2m+2$. This leads to the correct choice of integration variables, and the integral localizes on the solutions of the scattering equations. Specifically, we consider an amplitude for $n+1=2m+2$ particles, the last one of which is chosen to be a gluon. In the soft limit of the gluon momentum, i.e., $p_{2m+2}=\tau\hat{p}_{2m+2}$ and $\tau \rightarrow 0$, the even-point measure takes the form
\be
\int d\mu_{2m+2}^{\text{6D}} = \delta(p_{n+1}^{2})\int d\mu_{2m+1}^{\text{6D}} \frac{1}{2\pi i} \oint_{|\hat{E}_{n+1}| = \varepsilon} \frac{d\sigma_{n+1}}{E_{n+1}}+O(\tau^{0}),\label{eq:main}
\ee
where the odd-point measure is given by:
\begin{equation}
\boxed{d\mu_{2m+1}^{\text{6D}} = \frac{\left(\prod_{i=1}^{n} d\sigma_i \prod_{k=0}^{m-1}d^{8}\rho_{k}\right)\, d^{4}\omega\, \langle \xi d\xi\rangle }{\vol \left( \slc_{\sigma} , \slc_{\rho} , \text{T} \right)}\, \frac{1}{V^{2}_n} \prod_{i=1}^{n}\delta^6 \left(p_{i}^{AB}-\frac{p^{AB}(\sigma_{i})}{\prod_{j\neq i} \sigma_{ij}}\right).} \label{eq:prop}
\end{equation}
This is derived in detail in Appendix \ref{app:even-to-odd}. The volume factor here implies modding out by the action of the two $\slc$ groups, as well as the T-shift. Furthermore,
\begin{equation}
E_{n+1}= \tau \hat{E}_{n+1} = \tau \sum^{n}_{i=1} \frac{\hat{p}_{n+1}\cdot p_i}{\sigma_{n+1,i}} =0
\end{equation}
corresponds to the scattering equation for the soft particle. The factor of $\tau$ in $E_{n+1}$ makes the first term in the expansion of the $(n+1)$-particle measure singular as $\tau \to 0$. As we explain below, the measure given here for $n=2m+1$ has the correct $\slc_\sigma$ scaling, which is degree $4n$. 

Let us now proceed to the explicit computation of the measure. Note that the redundancies can be used to fix seven of the $5n+1$ variables, leaving $5n-6$ integrations. This precisely matches the number of bosonic delta functions, which can be counted in the same way as in the even-point case. Therefore, as before, all the integration variables are localized by the delta functions. In order to carry out the computations, one needs use the seven symmetry generators to fix seven coordinates and obtain the corresponding Jacobian. The order in which this is done is also important, since $T$ does not commute with the other generators. In order to make contact with the even-point counterpart, let us first fix the T-shift symmetry. Because $T$ merely generates a shift in the coefficients of the polynomial, it can be seen that the measure in \eqref{eq:prop} is invariant.  Now, let us regard the symmetry parameter $\alpha$
as one of the integration variables in favor of fixing one of the
four components $\omega^{A}$. For instance, one can choose $\omega^{1}$
as fixed, and then integrate over the parameters $\{\alpha,\omega^{2},\omega^{3},\omega^{4}\}.$ It can be checked from \eqref{omega-shift} that this change of variables induces the Jacobian
\be
d^{4}\hat{\omega} = \langle\xi\,\rho_{m-1}^{1}\rangle\, d\alpha\, d\omega^{2}d\omega^{3}d\omega^{4}.\label{eq:omegameasure}
\ee
The other ingredients in the measure are invariant under this transformation, i.e.,
\begin{eqnarray}
\Delta_{B}(\hat{\rho},\sigma) & = & \Delta_{B}(\rho,\sigma),\\
\prod_{k=0}^{m-1}d^{8}\hat{\rho}_{k} & = & \prod_{k=0}^{m-1}d^{8}\rho_{k}.
\end{eqnarray}
The dependence on $\alpha$ can then be dropped, with the corresponding
integration formally canceling the volume factor for the T-shift in the denominator
of \eqref{eq:prop}. The measure in this partially-fixed form is now
\begin{equation}
d\mu_{2m+1}^{\text{6D}} = \frac{\left(\prod_{i=1}^{n} d\sigma_i \prod_{k=0}^{m-1}d^{8}\rho_{k}\right)\, d^{3}\omega \, \langle\xi\,\rho_{m-1}^{1}\rangle\langle\xi d\xi\rangle}{\vol \left( \slc_{\sigma} \times \slc_{\rho} \right)}\, \frac{1}{V^{2}_n} \prod_{i=1}^{n}\delta^6 \left(p_{i}^{AB}-\frac{p^{AB}(\sigma_{i})}{\prod_{j\neq i} \sigma_{ij}}\right) \label{eq:gfmeasure}.
\end{equation}
Note that the factor $d^{3}\omega\langle\xi \rho_{m-1}^{1}\rangle\langle\xi d\xi\rangle$
is invariant under the projective scaling of $\xi_{\alpha}$. By construction, it is also invariant under the action of the T shift, implying that $\omega^1$ may be set to any value. However, after making these choices Lorentz invariance is no longer manifest. 

Let us show explicitly how this measure has the correct $\text{SL}(2,\mathbb{C})_{\sigma}$-scaling under the transformation $\sigma \rightarrow t\sigma$ together with the scaling of the coefficients in the maps,
\be \label{eq:SL2scale}
\rho^{Aa}_{k}\rightarrow t^{m-k}\rho^{Aa}_{k}.
\ee
In particular, this implies that $\rho_{m}^{Aa}=\omega^{A}\xi^a$ is invariant. As is apparent from \eqref{eq:action}, the parameter $\alpha$ carries $\slc_{\sigma}$-scaling $-1$, as does the $T$ volume $\langle\xi \rho_{m-1}^{1}\rangle$ using (\ref{eq:SL2scale}). Since the projective scaling of $\xi$ is completely
independent from the $\slc_{\sigma}$ transformation, none of the components $\xi^{a}$ and $\omega^{A}$ transform. Now, we find
\begin{eqnarray}
\prod_{k=0}^{m-1}d^{8}\rho_{k}& \;\rightarrow\; & t^{n^{2}-1}\prod_{k=0}^{m-1}d^{8}\rho_{k},\\
\frac{1}{V^{2}_n} \prod_{i=1}^{n}d\sigma_i  & \;\rightarrow\; & t^{4n-n^{2}} \frac{1}{V^{2}_n} \prod_{i=1}^{n}d\sigma_i,  \\
\langle\xi\,\rho_{m-1}^{1}\rangle\langle\xi d\xi\rangle d^{3}\omega & \;\rightarrow\; &
t\,\langle\xi\,\rho_{m-1}^{1}\rangle\langle\xi d\xi\rangle d^{3}\omega,
\end{eqnarray}
leading to the scaling weight of $4n$ for the full measure as required. 

Having carried out these checks, we are now in position to give the final form of the measure for $n=2m+1$ in the same way as explained earlier for even $n$. For this, we eliminate the remaining $\text{SL}(2,\mathbb{C})_{\sigma}\times \text{SL}(2,\mathbb{C})_{\rho}$ symmetry by performing the standard fixing of $\sigma_{1},\sigma_2,\sigma_3$ and $\rho_{0}^{1,+},\rho_{0}^{1,-},\rho_{0}^{2,+}$. Note that we fixed the lowest coefficients $\rho^{A a}_{0}$, because they are not affected by the T-shift. Otherwise, this would interfere with the T-shift. Finally, we extract the mass shell and momentum conservation delta functions as in \eqref{bosonic-delta}. This leads to
\be
d\mu_{2m+1}^{\text{6D}} = \frac{J_{\rho}\,J_{\sigma}}{V_n^2} \left(\prod_{i=4}^{n}d\sigma_i \right)\, d\rho_{0}^{2,-} d^2\rho_{0}^{3}\, d^2\rho_{0}^{4} \left(\prod_{k=1}^{m-1}d^{8}\rho_{k}\right) \, d^{3}\omega\, d\xi\, \langle\xi\,\rho_{d-1}^{1}\rangle\, \hat{\Delta}_{B},
\ee
where the Jacobians are given in \eqref{jacobians}, and $\hat{\Delta}_{B}$ is given in \eqref{bosonic-delta1}.

\subsubsection{Transformations of the Maps \label{subsec:maptransf}}

Having checked the scaling of the measure, here we consider other  $\text{SL}(2,\mathbb{C})_{\sigma}$ transformations, as we will see that they lead to other interesting new features of the odd-point rational maps. In particular, let us consider the inversion $\s_i \rightarrow -1/\s_i$.\footnote{The minus sign is for convenience only. Sign reversal is already established as a consequence of the scaling symmetry.}  Under this inversion, the rational map transforms as, 
\bea
{\langle \r^A(\s_i) \, \r^B(\s_i) \rangle \over \prod_{j \neq i} \s_{ij} } \rightarrow {\langle \r'^A(\s_i) \, \r'^B(\s_i) \rangle \over ( \prod_{j=1}^n \s_j^{-1} )  (\prod_{j \neq i} \s_{ij}) } \, ,
\eea
and the new object $\r'^A(\s_i)$ entering the map is given by
\bea \label{eq:rho_prime}
\r'^A(\s_i) = (-1)^m \sum_{k=0}^m (-1)^k \r^A_{k,a} \, \s_i^{m-k-{1\over 2}} \, .
\eea
Note that this is actually not a polynomial due to the fact that $n$ is odd. To keep the rational-map constraints unchanged, we require that the coefficients transform as
\bea \label{eq:transformation_rho}
\r^A_{k,a} \rightarrow \r'^A_{k,a} = (-1)^{k}\, \r^A_{m-k,a} .
\eea
Then, up to an overall factor, the transformation exchanges the degree-$k$ coefficient with the degree $m-k$ coefficient just like in the case of even $n$. What is different from the even-point case is the non-polynomial property of $\r'^A(\s_i)$. Therefore, inversion turns the polynomial map into a non-polynomial one of the following form
\bea \label{eq:half_Integer_map}
{\r'^A_a(z) = \sum_{k=0}^{m} \r'^A_{k,a} \, z^{k-{1\over 2}} \, .}
\eea
Now the lowest-degree coefficient, $\r'^A_{0,a}$, which is proportional to the highest coefficient of the original map, has the special factorized form
\bea
\r'^A_{0,a} = \omega^A \, \xi_a   ,
\eea
where we have used \eqref{eq:factor}. Therefore the product $p^{AB}(z) = \langle \r^A(z)\, \, \r^B(z) \rangle$ remains a degree-$(n{-}2)$ polynomial. 

Although we only use polynomial maps throughout the paper, it is worth mentioning that the above non-polynomial form of the maps could be used equally well. This discussion makes it is clear that for odd multiplicity a general $\text{SL}(2,\mathbb{C})_{\sigma}$ transformation can take the original polynomial maps to a more complicated-looking, but equivalent, version of maps. This is a consequence of the fact that the seven generators of $\slc_{\sigma}$, $\slc_{\rho}$, ${T}$ do not close into a group, as can already be seen from the fact that 
\begin{equation}
\vol \left( \slc_{\sigma} , \slc_{\rho} , {T} \right)
\end{equation}
carries weight $-1$ under scaling of $\slc_{\sigma}$. Compositions of the action of these seven generators on the maps lead to the general transformations of the form
\begin{equation}\label{gen}
\rho^{A,a}(z)\rightarrow  (e^{\mathcal{T}(z)})^a_b\,\rho'^{A,b}(z),
\end{equation}
where $\mathcal{T}^a_b(z)$ is a traceless $2\times 2$ matrix depending on $z$. 

It is interesting to study the subalgebra that preserves the form of the polynomial maps, which for even multiplicity is just $\slc_{\sigma}\times \slc_{\rho}$.  In Appendix~\ref{app:symmetry-algebra} we obtain the corresponding algebra for odd multiplicity: We first show that the generators of $\slc_{\sigma}$ and $\slc_{\rho}$ do not commute in general and that a subset of these recombines into the algebra $\slc \ltimes \mathbb{C}^2$. This includes inversion and $\text{T}$-shift, but it requires a partially fixed $\text{SL}(2,\mathbb{C})_{\rho}$ frame.

\subsection{\label{sec:odd-integrand}Integrand from Soft Limits}

Here we apply the soft limit to the even-point integrand in order to obtain the odd-point version, with the soft factor included. The answer is composed of two pieces:
\be
\mathcal{I}^{\N=(1,1) \text{ SYM}}_{\text{odd}} = \int d\widehat{\Omega}_F^{(1,1)} \times \mathcal{J}_{\text{odd}}.
\ee
The fermionic measure $d\widehat{\Omega}_F^{(1,1)}$ can be obtained in a way similar to the bosonic one, and we relegate its derivation to Appendix~\ref{app:soft-2}. The result is
\be\label{odd-fermionic-measure}
\boxed{ d\widehat{\Omega}_F^{(1,1)} = V_n\, dg\,d\tilde{g}\prod_{k=0}^{m-1}d^2\chi_{k}\,d^2\tilde{\chi}_{k} \prod_{i=1}^{n}\delta^{4}\bigg(q_i^A - \frac{\langle \rho^A(\sigma_i) \, \chi (\sigma_i)\rangle}{\prod_{j\neq i}\sigma_{ij}} \bigg) \delta^{4}\bigg(\tilde{q}_{i,A} - \frac{[ \tilde{\rho}_A(\sigma_i) \, \tilde{\chi} (\sigma_i)]}{\prod_{j\neq i}\sigma_{ij}} \bigg).}
\ee
The fermionic maps are constructed such that $\langle \rho(z) \,\chi(z)\rangle$ and its conjugate are polynomials of degree $n-2$, and take a form similar to the bosonic map in (\ref{eq:odd-pt-map}). Specifically,
\begin{align}
\chi^{a}(z)& =\sum_{k=0}^{m-1} \chi^{a}_k\, z^k  + g \xi^a z^m,\label{fermaps1} \\
\tilde{\chi}^{\hat{a}}(z)& = \sum_{k=0}^{m-1} \tilde{\chi}^{\hat{a}}_k\, z^k + \tilde{g} \tilde{\xi}^{\hat{a}} z^m,\label{fermaps2}
\end{align}
where the $\chi$'s and $\tilde{\chi}$'s, as well as $g$ and $\tilde{g}$, are Grassmann coefficients. Note that the same spinors $\xi^a$ and $\tilde{\xi}^{\hat{a}}$ appear in the coefficients of $z^m$ for both the bosonic maps and the fermionic maps. This form ensures the coefficient of $z^{2m}$ in the product of maps vanishes, so that the product has the desired degree, $2m-1 = n-2$, for an odd number of particles.

With this parametrization of the maps, the first check is to show that the construction has the right Grassmann degree. As in the case of the even $n$, we need to remove the fermionic ``wave functions'' $\prod_{i=1}^n\delta^2(\tilde{\lambda}_{i,A,\hat{a}}\, q^A_i) \delta^2(\lambda_{i,b}^B \tilde{q}_B)$. This leaves an integrand with Grassmann degree of $4n$, as required. Having established the Grassmann degree of the integrand, let us next count the number of fermionic integrations. There are $4m$ $\chi$ and $\tilde{\chi}$ integrals and two $g$ and $\tilde{g}$ integrals, giving a total of $4m+2 = 2n$ integrations. The final amplitude thus has Grassmann degree $2n$. More precisely, just as for even $n$, it has degree $n$ in both the $\eta$'s and the $\tilde{\eta}$'s, which is what we expect for the superamplitudes of 6D $\N=(1,1)$ SYM in the representation (\ref{eq:spectrum11}). 

The factor $\mathcal{J}_{\text{odd}}$, which is purely bosonic, contains a contour integral in $\sigma_{n+1}$ that emerges from the soft limit of the measure \eqref{eq:main}. Therefore, it encodes all of the dependence on the soft particle. Using the identity permutation $\I_n$ and setting $\sigma_{n+1}=z$ for convenience, we show in Appendix~\ref{app:even-to-odd} that for a soft gluon
\be
S^{a\hat{a}} \mathcal{J}_{\text{odd}} = \text{PT}(\I_n)\, \frac{\sigma_{1n}}{2\pi i} \oint_{|\hat{\mathcal{E}}_{n+1}|=\varepsilon}\frac{dz}{\mathcal{E}_{n+1}}\times\frac{\text{Pf}'A_{n+1}}{(z-\sigma_{1})(z-\sigma_{n})}\,\frac{x^{a}}{\langle\xi\,\Xi\rangle}\,\frac{\tilde{x}^{\hat{a}}}{[\tilde{\xi}\,\tilde{\Xi}]}.
\ee
Let us explain the various terms appearing in this formula. First, the vanishing of $\mathcal{E}_{n+1} = \tau \hat{\mathcal{E}}_{n+1}= p(z) \cdot p_{n+1}$ is the rescaled scattering equation for the soft
$(n{+}1)$th particle (on the support of the hard scattering equations), such that $\mathcal{E}_{n+1} = E_{n+1} \prod_{i=1}^{n} (z{-}\sigma_i)$. In terms of the 6D spinor-helicity formalism, Weinberg's soft factor for a gluon is given by
\be\label{soft-factor}
S^{a\hat{a }}=\frac{[\tilde{\lambda}_{n+1}^{\hat{a}}|p_{1}\tilde{p}_{n}|\lambda_{n+1}^{a}\rangle}{s_{n+1,1}s_{n,n+1}} = \frac{\tilde{\lambda}_{n+1,A}^{\hat{a}}\,p_{1}^{AB}\,\tilde{p}_{n,BC }\,\lambda_{n+1}^{a,C}}{s_{n+1,1}s_{n,n+1}}.
\ee
The reduced Pfaffian can be expanded as
\be
\text{Pf}'A_{n+1} = \frac{(-1)^{n+1}}{\sigma_{1n}}\sum_{i=2}^{n-1}(-1)^{i}\frac{p_{n+1}\cdot p_{i}}{z-\sigma_{i}}\, \text{Pf}\, A^{[1,i,n,n+1]}_{n+1},
\ee
where $A^{[1,i,n,n+1]}_{n+1}$ denotes the matrix $A_{n+1}$ with rows and columns $1,i,n,n+1$ removed. This odd-point integrand by construction does not depend on $\tau$, the scaling parameter introduced to define the soft limit. It is also independent of the choice of polarization $(a,\hat{a})$ and the direction of the soft momentum $p_{n+1} = \t \, \hat p_{n+1}$. Recall that $\xi^a = (1,\xi)$ is determined from the hard scattering maps, while $\Xi^{a}=(\Xi^{+},\Xi^{-})$ and $x^{a}=(x,-1)$ are given by the following linear equations (for $z = \sigma_{n+1}$):
\be
\langle\Xi\,\rho^{A}(z)\rangle = \langle x\,\lambda_{n+1}^{A}\rangle, \qquad\qquad [\tilde{\Xi}\,\tilde{\rho}^{A}(z)]  =  [\tilde{x}\,\tilde{\lambda}_{n+1}^{A}].
\ee

Introducing a reference spinor $r^{A}$ and contracting the first of the preceding two equations with $\epsilon_{ABCD}\lambda_{n+1}^{B,a}\rho^{C,b}(z)r^{D}$, we obtain 
\be
x^{a}\langle\rho^{b}(z)|\tilde{p}_{n+1}|r\rangle = \Xi^{b}\langle\lambda_{n+1}^{a}|\tilde{p}(z)|r\rangle.
\ee
This can be used to make the $z$-dependence explicit in the integrand. Contracting with $\xi_{b}$ and repeating these
steps for the anti-chiral piece gives
\be
\frac{x^{a}}{\langle\xi\,\Xi\rangle}\,\frac{\tilde{x}^{\hat{a}}}{[\tilde{\xi}\,\tilde{\Xi}]} = \frac{\langle\lambda_{n+1}^{a} | \tilde{p}(z) | r\rangle \, [\tilde r|p(z)|\tilde{\lambda}_{n+1}^{\hat{a}}]}{\xi^{b}\langle\rho_{b}(z)|\tilde{p}_{n+1}|r\rangle \, [\tilde r|p_{n+1}|\tilde{\rho}_{\hat{b}}(z)]\tilde{\xi}^{\hat{b}}},
\ee
where $|r\rangle$ and $[\tilde r|$ are independent reference spinors. Hence
\be \label{beforedef}
{S^{a\hat{a}}} \mathcal{J}_{\text{odd}} =  \text{PT}(\I_n)\, \frac{\sigma_{1n}}{2\pi i} \oint_{|\hat{\mathcal{E}}_{n+1}|=\varepsilon}\frac{dz}{\mathcal{E}_{n+1}}\times\frac{\text{Pf}'A_{n+1}}{(z-\sigma_{1})(z-\sigma_{n})}\,\frac{\langle\lambda_{n+1}^{a}|\tilde{p}(z)|r\rangle[\tilde r|p(z)|\tilde{\lambda}_{n+1}^{\hat{a}}]}{\xi^{b}\langle\rho_{b}(z)|\tilde{p}_{n+1}|r\rangle[\tilde r|p_{n+1}|\tilde{\rho}_{\hat{b}}(z)]\tilde{\xi}^{\hat{b}}}.
\ee

In Section~\ref{sec:contour-deformation} we evaluate this integral via contour deformation. However, let us point out here the difficulties arising when trying to evaluate this integral. For a given solution of the hard punctures $\{\sigma_i\}_{i=1}^n$ the scattering equation $\hat{\mathcal{E}}_{n+1}=0$ is a polynomial equation of degree $n-2$ in $z$, which in general does not have closed-form solutions. In the CHY formalism the soft limit can be evaluated by deforming the contour and enclosing instead the hard punctures at $z=\sigma_i$. This is because the CHY integrand can be decomposed into Parke--Taylor factors, which altogether yield $1/z^2$ as the fall off at infinity. The argument can be straightforwardly repeated for the Witten--RSV formula in four dimensions, as we outline in Appendix~\ref{4dm}. In the case of \eqref{beforedef} we find the leading behavior at infinity to be exactly $1/z^2$. However, the new contour will also enclose the poles associated to the brackets in the denominator, which are given by the solutions of a polynomial equation of degree $({n-3})/{2}$. Since these contributions to the integral also turn out to be cumbersome to evaluate, in the next section we introduce a novel contour deformation that allows us to evaluate the integral without the need to compute these individual contributions.

\subsubsection{\label{sec:contour-deformation}Contour Deformation}

The soft factor $S^{a \hat a}$, given by \eqref{soft-factor}, is still contained in the integrand $\mathcal{J}_{\text{odd}}$ and introduces an apparent dependence on the soft momentum. In order to extract it and evaluate the contour integral at the same time, we perform a complex shift of the soft momentum $p_{n+1}$. More specifically, for a given solution of the hard data $\{\sigma_{i},\rho,\tilde{\rho}\}$, we perform a holomorphic shift in $|\lambda_{n+1}\rangle$ and use it to extract the odd-point integrand as a residue. 

First, consider a reference null six-vector $Q=|q_{a}\rangle\langle q^{a}|$. (The Lorentz indices are implicit.) Using the little-group symmetry, the spinors can be adjusted such that
\be \label{eq:rhoq}
\langle\rho_{a}(\sigma_{n})|\tilde{q}_{b}]=m\,\epsilon_{ab},
\ee
together with $\langle q_{a}|\tilde{q}_{b}]=0$. Here $m^2=2\,p(\sigma_n)\cdot Q $  is a mass scale that drops out at the end of the computation, so we set $m=1$ for convenience. Note that $\tilde{q}_{b,A}$ transforms under the antifundamental representation of the Lorentz group, $\text{SU}^*(4)$, but under the \textit{chiral} $\text{SL}(2,\mathbb{C})_\rho$. Now consider
a shift described by a complex variable $w$:
\begin{eqnarray}
|\lambda_{n+1}^{a}\rangle &  \;\rightarrow\; & |\lambda_{w}^{a}\rangle = |\rho_{n}^{a}\rangle + w\,|q^{a}\rangle\\
|\tilde{\lambda}_{n+1}^{\hat{+}}] & \;\rightarrow\; & |\tilde{\lambda}_{w}^{\hat{+}}] = |\tilde{\rho}_{n}]+ w\,C^{a}|\tilde{q}_{a}],
\end{eqnarray}
where $|\rho_{n}^{a}\rangle^{A}$ is shorthand for $\rho^{A,a}(\sigma_{n})$,
while 
\be
|\tilde{\rho}_{n}]_{A}=[\tilde{\xi}\,\tilde{\rho}_{A}(\sigma_{n})],
\ee
and the index $A$ has been suppressed in the preceding equations. Without loss of generality, we may make the deformation for a specific choice of the polarization, which we have chosen to be $\hat{a}=\hat{+}$ in the second line.
The only requirement for $|\tilde{\lambda}^{\hat+}_{n+1}]$ is that
\be
0 = \langle\lambda_{n+1}^{a}|\tilde{\lambda}_{n+1}^{\hat{+}}]=\langle\lambda_{w}^{a}|\tilde{\lambda}_{w}^{\hat{+}}]=\langle\rho_{n}^{a}|\tilde{q}_{b}]C^{b}+\langle q^{a}|\tilde{\rho}_{n}],\qquad\, a=+,- \, ,
\ee
and using \eqref{eq:rhoq} this implies $C^{a}=\langle q^{a}|\tilde{\rho}_{n}]$.

The shifted soft factor that we utilize is 
\be
S_{w}^{a\hat{+}}=\frac{\langle\lambda_{w}^{a}|\tilde{p}_{n}p_{1}|\tilde{\lambda}_{w}^{\hat{+}}]}{s_{w,1}s_{w,n}},
\ee
which has a simple pole at $w=0$.
After a short computation one can show that
\be
\frac{1}{2\pi i}\oint_{|w|=\varepsilon} dw \,S_{w}^{a\hat{+}}=\frac{C^{a}}{2\pi i} \oint_{|w|=\varepsilon}\frac{dw}{w}=C^{a}.
\ee
Thus, the odd-point integrand can be recast in the form
\be
\mathcal{J}_{\text{odd}} = \text{PT}(\I_n) \times \frac{1}{2\pi i \,C^a}\oint_{|w|=\varepsilon} dw \,I_{w}^{a},
\ee
with
\be
I_{w}^{a}= \frac{1}{2\pi i}\oint_{|p(z)\cdot p_{w}| = \varepsilon} \frac{dz}{p(z)\cdot p_{w}}\times\frac{\sigma_{1n}\,\text{Pf}'A_{n+1}}{(z-\sigma_{1})(z-\sigma_{n})}\,\frac{x^{a}}{\langle\xi\,\Xi\rangle}\,\frac{\tilde{x}^{\hat{+}}}{[\tilde{\xi}\,\tilde{\Xi}]}\,.
\ee

As $w\rightarrow0$, the soft momentum $p_{w}\rightarrow p(\sigma_{n})$,
and hence we expect $z\rightarrow\sigma_{n}$. In fact, we claim that
this solution is the only one contributing to the singularity in $w$.
Therefore we may redefine the contour as enclosing only the pole at $\sigma_{n}$, and
\begin{eqnarray}
I_{w}^{a} & = & \frac{1}{2\pi i} \oint_{|z-\sigma_n|=\varepsilon}\frac{dz}{p(z)\cdot p_{w}}\times\frac{\sigma_{1n}\,\text{Pf}'A_{n+1}}{(z-\sigma_{1})(z-\sigma_{n})}\,\frac{x^{a}}{\langle\xi\,\Xi\rangle}\,\frac{\tilde{x}^{\hat{+}}}{[\tilde{\xi}\,\tilde{\Xi}]}\\
 & = & \frac{\text{Pf}'A_{n+1}\vert_{z=\sigma_{n}}}{p(\sigma_{n})\cdot p_{\omega}}\,\frac{\langle\lambda_{w}^{a}|\tilde{p}(\sigma_{n})|r\rangle[\tilde{r}|p(\sigma_{n})|\tilde{\lambda}_{w}]}{\langle\rho_{_{n}}|\tilde{p}_{w}|r\rangle[\tilde{r}|p_{w}|\tilde{\rho}_{n}]}.
\end{eqnarray}
One can show that:
\begin{align}
p(\sigma_{n})\cdot p_{w} & = \frac{w^{2}}{2},\\
\text{Pf}'A_{n+1} \vert_{z=\sigma_n} & =\frac{\omega}{2} \frac{1}{\sigma_{n1}}\sum_{i=2}^{n-1}(-1)^{i}\frac{\langle q^{a}|\tilde{p}_{i}|\rho_{n,a}\rangle}{\sigma_{ni}}\text{Pf}\,A^{[1,i,n,n+1]}_{n+1} + \mathcal{O}(w^{2}),
\end{align}
where we have used the identity 
\be
\sum_{i=2}^{n-1}(-1)^{i}\frac{p_{n} \cdot p_{i}}{\sigma_{ni}}\text{Pf}\, A^{[1,i,n,n+1]}_{n+1}=0.
\ee
We also have that
\be
\frac{[\tilde{r}|p(\sigma_{n})|\tilde{\lambda}_{w}]}{[\tilde{r}|p_{w}|\tilde{\rho}_{n}]} =  \frac{ w\, [\tilde{r}|\rho_{n,a}\rangle\langle\rho_{n}^{a}|\tilde{q}_{b}]C^{b}}{w\, [\tilde{r}|\rho_{n,b}\rangle\, C^{b}}+\mathcal{O}(w) = 1 + \mathcal{O}(w).
\ee
Note that for the chiral piece we can set $|r\rangle$ such that $\tilde{p}(\sigma_{n})|r\rangle=|\tilde{\rho}_{n}]$. Then 
\be
\frac{\langle\lambda_{w}^{a}|\tilde{p}(\sigma_{n})|r\rangle}{\langle\rho_{{n}}|\tilde{p}_{w}|r\rangle} = \frac{w\,\langle q^{a}|\tilde{\rho}_{n}]}{ w\,\epsilon_{ABCD}\,\xi^{c}\,\rho_{n,c}^{A}\,\rho_{n,b}^{B}\,q^{C,b}\,r^{D}}+\mathcal{O}(w),
\ee
where the contraction in the denominator evaluates to
\be
\epsilon_{ABCD}\,\xi^{c}\,\rho_{n,c}^{A}\,\rho_{n,b}^{B}\,q^{C,b}\,r^{D}=\langle q|\tilde{p}(\sigma_{n})|r\rangle=\langle q|\tilde{\rho}_{n}],
\ee
with $|q\rangle^{A}:=\xi^{a}q_{a}^{A}$. Hence we obtain
\be
\lim_{w\rightarrow0}\frac{\langle\lambda_{w}^{a}|\tilde{p}(\sigma_{n})|r\rangle}{\langle\rho_{{n}}|\tilde{p}_{w}|r\rangle}=\frac{C^{a}}{\langle q|\tilde{\rho}_{n}]}.
\ee

Putting everything together we find
\begin{align}
\label{eq:Joddcontour}
\mathcal{J}_{\text{odd}} & =  \text{PT}(\I_n) \times \frac{1}{2\pi i\, C^a} \oint_{|w|=\varepsilon} \frac{dw}{w}\,\frac{C^{a}}{\sigma_{n1}}\sum_{i=2}^{n-1}(-1)^{i}\frac{\langle q^{a}|\tilde{p}_{i}|\rho_{n,a}\rangle}{\sigma_{ni}\langle q|\tilde{\rho}_{n}]} \text{Pf}\, A^{[1,i,n,n+1]}_{n+1}\nn\\
& =  \text{PT}(\I_n)\times\frac{\langle q^{a}|\tilde{X}_{(1,n)}|\rho_{n,a}\rangle}{\langle q|\tilde{\rho}_{n}]},
\end{align}
where for convenience we have defined the null vector
\be \label{eq:definition_X}
X_{(1,n)}^{AB} := \frac{1}{\sigma_{n1}}\sum_{i=2}^{n-1}(-1)^{i}\frac{p_{i}^{AB}}{\sigma_{ni}}\text{Pf}\,A^{[1,i,n,n+1]}_{n+1} = \frac{1}{2} \epsilon^{ABCD} \tilde{X}_{(1,n),CD}.
\ee
Despite using the notation $\text{Pf}\,A^{[1,i,n,n+1]}_{n+1}$, this Pfaffian is completely independent of the soft momentum and the associated puncture. As anticipated, the expression is independent of the scale of $q$, so we can remove
the normalization condition $2\,p(\sigma_{n})\cdot Q=1$, turning  $|q^{a}\rangle$ into a completely arbitrary spinor. Expanding the numerator of \eqref{eq:Joddcontour} in a basis given by
$\{\xi,\zeta\}$, where $\zeta$ is a reference spinor such that $\langle \xi \zeta \rangle = 1$, we find:
\be
\mathcal{J}_{\text{odd}} = \text{PT}(\I_n) \times\frac{\langle q|\tilde{X}_{(1,n)}|\pi_{n}\rangle-\langle w|\tilde{X}_{(1,n)}|\rho_{n}\rangle}{\langle q|\tilde{\rho}_{n}]},
\ee
where 
\be
|\pi_{n}\rangle^{A}=\langle\zeta\,\rho^{A}(\sigma_{n})\rangle
\ee
is the conjugate component of $|\rho_{n}\rangle$. Also,
\be
|w\rangle^{A}=\langle\zeta\,q^{A}\rangle.
\ee
In particular, the fact that the integrand is independent of $w$ implies the non-trivial
identity:
\be\label{identity}
\tilde{X}_{(1,n)}|\rho_{n}\rangle=0,
\ee
which yields the following form of the integrand
\be\label{chiralintegrand}
\mathcal{J}_{\text{odd}} = \text{PT}(\I_n) \times\frac{\langle q|\tilde{X}_{(1,n)}|\pi_{n}\rangle}{\langle q|\tilde{\rho}_{n}]} .
\ee

Using \eqref{identity} this expression can be recast in a non-chiral form. Let us introduce another reference spinor $|\tilde{q}]$ and consider
\be\label{odd-integrand-X}
\mathcal{J}_{\text{odd}} = \text{PT}(\I_n) \times\frac{\langle q|\tilde{X}_{(1,n)}\, p(\sigma_n)|\tilde{q}]}{\langle q|\tilde{\rho}_{n}]\langle \rho_{n}|\tilde{q}]}.
\ee
Note that $\tilde{p}(\sigma_n)_{AB} X_{(1,n)}^{BC} \ = - \tilde{X}_{(1,n)\,,AB}\, p(\sigma_n)^{BC} $. Finally, using the definition of $X_{(1,n)}$ in (\ref{eq:definition_X}) we recognize that the second factor in \eqref{odd-integrand-X} is in fact a reduced Pfaffian of an antisymmetric $(n+1)\times (n+1)$ matrix constructed out of $A_n$ with an additional column and row (labeled by $\star$) attached. We call this matrix $\widehat{A}_n$. Restoring the original integration variables, its entries are given by:
\begin{align}\label{odd-matrix-A}
\boxed{[\widehat{A}_n]_{ij} = \begin{cases}
\dfrac{p_{i} \cdot p_{j}}{\sigma_{ij}} &\quad\text{if}\quad i\neq j, \\
\quad 0\phantom{\dfrac{1}{1}} &\quad\text{if}\quad i=j,
\end{cases}   \qquad\qquad\text{for}\qquad i,j=1,2,\ldots,n,\star,}
\end{align}
where 
\be \label{eq:pstar}
p_{\star}^{AB} = \dfrac{ 2\, q^{[A} p^{B]C}(\sigma_\star) \tilde{q}_C }{ q^D  [\tilde{\rho}_D(\sigma_\star )\, \tilde{\xi}] \langle \rho^E(\sigma_\star)\, \xi \rangle  \tilde{q}_E}
\ee
is a reference null vector entering the final row and column, $q$ and $\tilde{q}$ are arbitrary spinors, and $\sigma_\star$ is a reference puncture that can be set to one of the punctures associated to removed rows and columns. In fact, we have numerical evidence that $\sigma_\star$ can be chosen completely arbitrarily without changing the result. Here, $q^{[A}p^{B]C}$ denotes the antisymmetrization $q^A p^{BC} - q^B p^{AC}$. The reduced Pfaffian is then defined analogously to \eqref{reduced-pfaffian}, with the restriction that the starred column and row are not removed. Independence of the choice of removed columns and rows follows from the analogous statement for $n$ even. It is straightforward to confirm that ${\text{Pf}}\,' \widehat{A}_n$ transforms as a quarter-integrand in the $\slc_\rho$-frame studied in Appendix~\ref{app:symmetry-algebra}, and that its mass dimension is $n{-}2$, as required. This completes the derivation of the odd-point formula \eqref{eq:(1,1)SYM-odd}. The reasoning was complicated, but the result is as simple as could be hoped for.

\subsection{\label{sec:odd-consistency}Consistency Checks}

We have checked numerically that the new formula \eqref{eq:(1,1)SYM-odd} correctly reproduces the 6D SYM amplitudes of gluons and scalars directly computed from Feynman diagrams, up to $n=7$. In this subsection we perform additional consistency checks of the formula. We begin by re-deriving the odd-point integrand $\mathcal{I}_{\rm odd}$  by comparing it with the corresponding CHY expression for a particular bosonic sector following a similar argument used earlier for the case of even $n$. We will then show analytically that the formula leads to the correct three-point super-amplitude of 6D SYM. It is worth noting that the three-point amplitudes in 6D YM are rather subtle due to the special kinematics first explained in \cite{Cheung:2009dc}. As we will see, our formula gives a natural parametrization of the special three-point kinematics. 

\subsubsection{Comparison with CHY} 

This section presents an alternative derivation of the integrand of the odd-point amplitudes. The method we will use here is similar to the one for the even-point case given in section \ref{sec:Integrand_CHY_Even}. It is based on comparison to known results of the CHY formulation of YM amplitudes in general spacetime dimensions. This method of derivation is independent of and very different from the soft-theorem derivation presented in the previous sections. Therefore it constitutes an additional consistency check. 

Let us begin with the general form of the odd-point amplitudes of 6D $\N=(1,1)$ SYM, 
\be \label{eq:odd_points}
\mathcal{A}^{\N=(1,1) \, {\rm SYM}}_{n} = \int d\mu_{n}^{\text{6D}}\,  d\widehat{\Omega}_F^{(1,1)} \times \mathcal{J}_{n\, \text{odd}},
\ee
for $n=2m+1$. 
Recall that the bosonic measure $d\mu_{n}^{\text{6D}}$ is defined in (\ref{eq:prop}), and the fermionic measure $d\widehat{\Omega}_F^{(1,1)}$ is given in (\ref{odd-fermionic-measure}), which is the part that is more relevant to the discussion here. The goal is to determine the integrand $\mathcal{J}_{n\,\text{odd}}$. As mentioned above, we will follow the same procedure as in the case of even $n$, namely comparison of our formula with the CHY formulation of amplitudes for adjoint scalars and gluons. To do so, we consider a particular component of the amplitude. Due to the fact that $n$ is odd and the scalars have to appear in pairs, it is not possible to choose all the particles to be scalars. The most convenient choice of the component amplitudes one with $n{-}1$ scalars and one gluon. Concretely, in the same notation as before, we choose to consider  
\be
\mathcal{A}_n (\phi^{1\hat{1}}_1, \ldots, \phi^{1\hat{1}}_m,  \phi^{2\hat{2}}_{m+1}, \ldots , \phi^{2\hat{2}}_{2m}, A^{a \hat{a}}_n ) \, ,
\ee
where $A^{a \hat{a}}_n$ is a gluon. 

As in Section \ref{sec:Integrand_CHY_Even}, we integrate out the fermionic variables so as to extract the desired component amplitude. The computation is similar to the one for even $n$, but slightly more complicated due to the appearance of $A^{a \hat{a}}_n$ in the middle term of the superfield. Projecting to this component amplitude, we obtain
\be
\int d\Omega_F^{(1,1)} \;\Longrightarrow\; 
V_n J_{\rm w}   \int  \prod_{k=0}^{m-1}  d^2\chi_k\,  d^2\tilde{\chi}_k \, dg\, d \tilde{g} \,  d \eta_n^a d \tilde{\eta}_n^{\hat{a}} \;\, \D^{\rm proj}_F \widetilde{\D}^{\rm proj}_F .
\ee
The factor $J_{\rm w} = \prod_{i=1}^n \frac{1}{(p^{13}_i)^{2}}$ arises from extracting the fermionic wave functions. The fermionic delta functions are given by   
\bea
\D^{\rm proj}_F &=& 
\prod_{A=1,3} \delta \left( q^A_n -  { \langle \rho^A(\sigma_n)\, \chi (\sigma_n) \rangle \over \prod_{j \neq n} \sigma_{nj} }  \right)    \prod_{i \in Y}  \delta \left(   { \langle \rho^A(\sigma_i)\, \chi (\sigma_i) \rangle \over \prod_{j \neq i} \sigma_{ij} }  \right) \prod_{i \in \overline{Y} } p_i^{13} , \\
\widetilde{\D}^{\rm proj}_F &=& 
  \prod_{A=2,4} \delta \left( \tilde{q}_{n A} -  { [\tilde{\rho}_{A}(\sigma_n) \, \tilde{\chi} (\sigma_n) ] \over \prod_{j \neq n} \sigma_{nj} }  \right) \prod_{i \in Y} \delta \left(  { [\tilde{\rho}_{A}(\sigma_i) \, \tilde{\chi} (\sigma_i) ] \over \prod_{j \neq i} \sigma_{ij} }  \right)
\prod_{i \in \overline{Y} } p_i^{13},
\eea
with $Y:=\{1, \ldots, m \}$ and $\overline{Y}:=\{m+1, \ldots, n-1\}$. Compared to the even-particle case, we have an additional contribution coming from the gluon $A^{a\hat{a}}_n$. Performing the fermionic integrations leads to the final result, 
\be
d\widehat{\Omega}_F^{(1,1)} \;\Longrightarrow\; (J_F)_{a \hat{a}} =  { \l_{n,a}^{[A} \langle \rho^{B]}(\s_n) \; \xi \rangle \over p_n^{AB} }
 { \tilde{\l}_{n, \hat{a}, [C} [ \tilde{\rho}_{D]}(\s_n) \; \tilde{\xi} ] \over p_{n, CD} } {V_n \over \prod_{j\neq n} \sigma_{nj}^2} \prod_{i \in Y, J \in \overline{Y}}  {1 \over  \s^2_{iJ} },
\ee  
where the square brackets denote anti-symmetrization on indices $A, B$ and $C, D$. Note that although the formula for $(J_F)_{a \hat{a}}$ exhibits explicit Lorentz indices $A, B$ and $C, D$, it is actually independent of the choice of these indices. Therefore, we have only made the dependence on the little-group indices $a$ and $\hat{a}$ explicit in $(J_F)_{a \hat{a}}$. They appear because the component amplitude contains a gluon $A^{a \hat{a}}_n$. 

Having extracted the component amplitude that we want, we can compare it to the corresponding result from the CHY formulation. From the comparison, we find that the odd-point integrand is given by
\be \label{eq:odd_integrand_CHY}
\mathcal{J}_{n\,\rm odd}(\alpha) = { {\rm Pf'} (\Psi_{\rm project})_{a \hat{a}}  \over (J_F)_{a \hat{a}} } \times  {\rm PT}(\alpha) \, .
\ee 
This ratio should be scalar, independent of the choice of the little-group indices $a$ and $\hat{a}$. As in the case of $n$ even, ${\rm Pf'} \Psi_{\rm project} $ is defined by the usual ${\rm Pf'} \Psi$, projected to the component amplitude 
under consideration. In the present case this means that the dot products of a pair of polarization vectors for scalars particles are the same as before, namely $\varepsilon_i \cdot \varepsilon_I =1$ if $i \in Y$ and $I \in \overline{Y}$, and otherwise they vanish. Furthermore $\varepsilon_i \cdot \varepsilon_n=0$, and $p_i \cdot \varepsilon_j =0$ if $j\neq n$. Using these rules, the original reduced Pfaffian ${\rm Pf'}\Psi$ simplifies to
\be\label{eq:Snp}
{\rm Pf'} (\Psi_{\rm project})_{a \hat{a}}
= 
{\rm det} (\D_m ) \sum_{i=1}^{n-2} (-1)^i\, {p_i \cdot (\varepsilon_n)_{a \hat{a}} \over \s_{in}}\, {\rm Pf} A_n^{[i,n-1,n]} \, ,
\ee
where the $m \times m$ matrix $\D_{m}$ has entries given by ${1\over \s_{iI}}$ for $i \in Y$ and $I \in \overline{Y}$. 

The ratio entering the integrand $\mathcal{J}_{n\,\rm odd}$ in (\ref{eq:odd_integrand_CHY}) can be dramatically simplified. To demonstrate this, note that as $\mathcal{J}_{n\,\rm odd}(\alpha)$ is a scalar, the following two tensors are proportional,
\be
 \l_{n,a}^{[A} \langle \rho^{B]}(\s_n) \; \xi \rangle  \tilde{\l}_{n, \hat{a}, [C} [ \tilde{\rho}_{D]}(\s_n)\; \tilde{\xi} ] \times R=p_n^{AB}\, p_{n, CD} \,\sum_{i=1}^{n-2} (-1)^i\, {p_i \cdot (\varepsilon_n)_{a \hat{a}} \over \s_{in}}\, {\rm Pf} A_n^{[i,n-1,n]},
\ee
where the proportionality factor $R$ is a scalar. After multiplying both sides of this equation with  $\l^{A,a}_n \tilde{\l}^{\hat{a}}_{n, C}$ and contracting indices $a$ and $\hat{a}$, we obtain
\begin{align}
\langle \rho^{A}(\s_n)\; \xi \rangle   [ \tilde{\rho}_{C}(\s_n)\; \tilde{\xi} ] \times R =& \sum_{i=1}^{n-2} (-1)^i\, {\l^{A,a}_n  \,  p_i \cdot (\varepsilon_n)_{a \hat {a}}  \,  \tilde{\l}^{\hat{a}}_{n, C} \over \s_{in}}\, {\rm Pf} A_n^{[i,n-1,n]} \, \nn \\
=& \sum_{i=1}^{n-2} \frac{(-1)^i}{\s_{in}} { p_n^{AB}\, p_{i, B D}\, \varrho^{DE}\, p_{n, EC} \over \varrho \cdot p_n }  {\rm Pf} A_n^{[i,n-1,n]},
\end{align}
where in the last line we used the spinor form of the polarization vector $(\varepsilon_n)_{a \hat {a}}$ \cite{Cheung:2009dc}, with $\varrho$  a reference vector. Collecting everything and plugging $R$ back into the integrand, we arrive at:
\be
\mathcal{J}_{n\,\rm odd}(\alpha) = {{\rm PT}(\alpha)  \over \s_{n-1,\,n} } \sum_{i=1}^{n-2}  \frac{(-1)^i}{\sigma_{in}}\, { p^{AB}(\sigma_n)\, p_{i, B D}\, \varrho^{DE}\, p_{n, EC} \over \varrho \cdot p_n \,   \r^A_n\,   \tilde{\r}_{n,C} }\,\, {\rm Pf} A_n^{[i,n-1,n]},
\ee
where we have also simplified the $\s$-dependent part, and defined
\be
\r^A_n := \langle \rho^{A}(\s_n)\, \, \xi \rangle  \, , 
\quad 
\tilde{\r}_{n,C} := [ \tilde{\rho}_{C}(\s_n)\, \, \tilde{\xi} ]  \, ,
\ee
as in the previous subsection. Furthermore, using the identity 
\be
\sum_{i =1}^{n-2} (-1)^i\, { p_i \cdot p_n  \over \s_{in}  }\, {\rm Pf} A_n^{[i,n-1,n]}  =0 \, , 
\ee
the summation in the expression of $\mathcal{J}_{n\,\rm odd}(\alpha)$ can be further simplified, leading to the final form of the integrand: 
\be
\mathcal{J}_{n\,\rm odd}(\alpha) = {{\rm PT}(\alpha)  \over \s_{n-1,\,n} }  \sum_{i=1}^{n-2}  (-1)^i\, { p^{AB}(\sigma_n)\, p_{i, B C}  \over \s_{in}  \,   \r^A_n\,   \tilde{\r}_{n,C} }\, {\rm Pf} A_n^{[i,n-1,n]} \, . 
\ee
This result is actually a Lorentz scalar, as it should be, even though it appears to depend on the explicit Lorentz spinor indices $A$ and $C$. The above expression agrees with \eqref{odd-integrand-X} after contraction with reference spinors in the numerator and denominator and choosing $\s_{\star}=\s_n$. In the derivation here, we have chosen particles $n$ as well as $n{-}1$ to be special. However, the final result should be independent of such a choice, and therefore we have a complete agreement with \eqref{odd-integrand-X}, the result obtained by using the soft theorem.  
\subsubsection{Three-point Amplitude} \label{sec:3pts}

Here we derive analytically the three-point amplitude from our odd-$n$ formula. As explained in \cite{Cheung:2009dc}, the three-point amplitude requires additional considerations such as an adequate parametrization of its special kinematics. Here we find that our formula naturally leads to such a parametrization together with the correct supersymmetric expression. Since the result, which is quite subtle, exists in the literature \cite{Dennen:2009vk}, it is nice to see that our formula reproduces the known result. It turns out that it is more convenient to use the linearized constraints introduced in (\ref{linearmeasure}). So we start with the following integral representation of the superamplitude:
\be \label{3ptamp}
\mathcal{A}_{3}^{\N=(1,1)\text{ SYM}}(123) = \int d\mu_{3}^{\text{6D}} \frac{\mathcal{J}_{3}}{(V_3)^3} \!\!\int d^2\chi_0 \, d^2\tilde{\chi}_0  \,dg\,d\tilde{g}\prod_{i=1}^{3}\delta^{2}(\eta_{i}^{b}M_{i,b}^{a}-\chi^{a}(\sigma_{i}))\delta^{2}(\tilde{\eta}_{i}^{\hat{b}}{\widetilde{M}}_{i,\hat{b}}^{\hat{a}}-\tilde{\chi}^{\hat{a}}(\sigma_{i})).
\ee
The fermionic delta functions in the above formula are the fermionic versions of the linear constraints, and we will discuss the $n$-point version of these constraints in Section~\ref{sec:LinearConstraints}. For now we take this as a given, and write the degree $1$ three-point maps as:
\begin{eqnarray}
\rho^{A,a}(z) & = & \rho_{0}^{A,a}+\omega^{A} \xi^{a}\, z,\nn \\
\chi^{a}(z) & = & \chi^{a}_{0}+g \, \xi^{a} \,z,
\end{eqnarray}
together with their conjugates $\tilde{\rho}_{A\hat{a}}(z)$
and $\tilde{\chi}_{\hat{a}}(z)$. Imposing the orthogonality condition $\rho^{A,a}(z)\tilde{\rho}_{A,\hat{a}}(z)=0$ we find:
\begin{align}
\rho_{0}^{A,a}\,\tilde{\rho}_{0,A,\hat{a}}&=0, \nn \\
\rho_{0}^{A,a}\,\tilde{\omega}_{A}\,\tilde{\xi}_{\hat{a}}+\xi^{a}\,\omega^{A}\,\tilde{\rho}_{0,A,\hat{a}}&=0,\nn \\
 \quad\omega^{A}\,\tilde{\omega}_{A}&=0.
\end{align}
The solution to the middle constraint is given by
\begin{eqnarray}
\rho_{0}^{A,a}\, \tilde{\omega}_{A} & = & t\,\xi^a,\nn \\
\omega^{A}\, \tilde{\rho}_{0,A,\hat{a}} & = & -t\,\tilde{\xi}_{\hat{a}},
\end{eqnarray}
for some scale $t$. Recall that the top component of each map, i.e.,
$\xi^{a}\omega^{A}$ and its conjugate, carries a $\glc$ freedom
which we previously used to fix $\xi^{+}=1$. For reasons that will
become apparent soon, here it is more convenient to use this scaling to fix $t=V_3$.
Using this and the previous equations we find the following relation:
\be\label{ort}
\rho^{A,a}(\sigma_{i})\tilde{\rho}_{A,\hat{a}}(\sigma_{j})=V_3\,\xi^{a}\tilde{\xi}_{\hat{a}}\,\sigma_{ij}.
\ee
Let us now evaluate the integrand in the representation of \eqref{chiralintegrand} and \eqref{eq:definition_X}:
\begin{eqnarray}
\mathcal{J}_{3}&=&\frac{1}{(V_3)^2 \sigma_{13}}\times \frac{\langle q|\tilde{p}(\sigma_{1})|\pi_{3}\rangle}{\langle q|\tilde{\rho}_{3}]}\nn \\
&=&\frac{1}{(V_3)^2  \sigma_{13}} \frac{q^{A}\tilde{\rho}_{A,\hat{a}}(\sigma_{1})\tilde{\rho}_{B}^{\hat{a}}(\sigma_{1})\rho_{a}^{B}(\sigma_{3})\zeta^{a}}{q^{A}\tilde{\rho}_{0,A}^{\hat{a}}\tilde{\xi}_{\hat{a}}}\nn\\
 & = & 1/V_3,
\end{eqnarray}
where we used $\tilde{X}_{(1,3)} = -\frac{\tilde{p}(\s_1)}{V_3 \s_{13}}$, $\text{Pf}\,A^{[1,2,3,4]}_{4} = 1$, $\langle\xi\zeta\rangle=1$ and $q^{A}\tilde{\rho}_{A}^{\hat{a}}(\sigma)\tilde{\xi}_{\hat{a}}=q^{A}\tilde{\rho}_{0,A}^{\hat{a}}\tilde{\xi}_{\hat{a}}$. For three points, the $\slc_{\sigma}$ symmetry completely fixes all three $\s$'s, and we have,
\be
\int d\mu_{3}^{\text{CHY}}=(V_3)^{2}.
\ee
Plugging this into \eqref{3ptamp} we are left with
\be
\mathcal{A}_{3}^{\N=(1,1) \text{ SYM}}(123) = F_{3}^{(1,0)} F_{3}^{(0,1)}\,,\qquad F_{3}^{(1,0)}=\frac{1}{V_3}\int d\chi^+_{0}d\chi^-_{0}\, dg\prod_{i=1}^{3}\delta^{2}(\eta_{i}^{b}M_{i,b}^{a}-\chi^{a}(\sigma_{i})),
\ee
together with its conjugate $F_3^{(0,1)}$. We find that now the three-point amplitude only involves fermionic integrals and factorizes into chiral and antichiral pieces. However, this form is not completely satisfactory as it still carries redundancies. In order to match this expression with the known ones \cite{Cheung:2009dc,Czech:2011dk}, we note that \eqref{ort} can be inverted as follows: Pick three labels $\{i,j,k\} = \{1,2,3\}$ for the external particles, then
\be
\langle \lambda_i^{a}| \tilde{\lambda}_j^{\hat{a}}]  =  V_3\,\frac{M_{i,b}^{a}\xi^{b}\widetilde{M}_{j\hat{b}}^{\hat{a}}\tilde{\xi}^{\hat{b}}}{|M_{i}|\,|\widetilde{M}_{j}|}\sigma_{ij}
  =  \epsilon_{ijk} \left(M_{i,b}^{a}\xi^{b}\right) \left(\widetilde{M}_{j,\hat{b}}^{\hat{a}}\tilde{\xi}^{\hat{b}}\right)\,, 
\ee
where $\epsilon_{ijk}$ is the sign of the permutation $(ijk)$, as usual. This allows us to read off the variables defined in \cite{Cheung:2009dc} for the special case of three-point kinematics. Since $\det \langle \lambda_i^{a}| \tilde{\lambda}_j^{\hat{a}}] = 0$, 
\be
u_{i}^a = M_{i,b}^a\,\xi^b\,,\qquad\tilde{u}_{i}^{\hat{a}} =\widetilde{M}_{i,\hat{b}}^{\hat{a}}\,\tilde{\xi}^{\hat{b}}.
\ee
It is easy to check that they satisfy $u_{i}^{a}\lambda_{i,a}^{A}=u_{j}^{a}\lambda_{j,a}^{A}$ for any $i,j$. Their duals, defined as
\be
w_{i}^a = \frac{M_{i,b}^a \zeta^b}{\sigma_{ij}\sigma_{ik}}\,,\qquad\tilde{w}_{i}^{\hat{a}}=\frac{\widetilde{M}_{i,\hat{b}}^{\hat{a}}\tilde{\zeta}^{\hat{b}}}{\sigma_{ij}\sigma_{ik}}\,,
\ee
satisfy $\langle u_{i}\, w_{i}\rangle=[\tilde{u}_{i}\,\tilde{w}_{i}]=1$. 
Since the maps are constructed such that momentum conservation
is guaranteed, the condition imposed in \cite{Cheung:2009dc},
\be
\sum_{i=1}^{3}\omega_{i}^{a}\lambda_{i,a}^{A} =  \zeta^{a}\sum_{i=1}^{3}\frac{\rho_{a}^{A}(\sigma_{i})}{| M_{i}|} = 0,
\ee
is also satisfied by virtue of the residue theorem. Furthermore, note that there are scaling and shifting redundancies in the definition of $u_i, \tilde{u}_i, w_i, \tilde{w}_i$ \cite{Cheung:2009dc}. In particular, these variables are defined up to a rescaling,
\be \label{eq:scaling}
 u_i \rightarrow \alpha u_i \, , \quad
\tilde{u}_i \rightarrow \alpha^{-1} \tilde{u}_i \, , \quad
w_i \rightarrow \alpha^{-1} w_i \, , \quad
\tilde{w}_i \rightarrow \alpha \tilde{w}_i \, ,
\ee
which is a reflection of scaling redundancy of $\xi$ and $\zeta$. 
Additionally, there is a shift redundancy in $w_{i}$,  
\be \label{eq:shifting}
w_{i}\rightarrow w_{i}+b_{i}u_{i} \, ,
\ee 
with $\sum_{i=1}^3 b_{i}=0$ corresponds to the redundancy $\zeta\rightarrow\zeta+b\,\xi$ in the defining condition $\langle \zeta\xi \rangle=1$. Let us now fix this $\text{SL}(2,\mathbb{C})$ redundancy by setting $\xi=(1,0)$ and $\zeta=(0,1)$. Then
\be
M_{i}=\left(\begin{array}{cc}
u_{i}^{+} & u_{i}^{-}\\
\sigma_{ij}\sigma_{ik}w_{i}^{+} & \sigma_{ij}\sigma_{ik}w_{i}^{-}
\end{array}\right),
\ee
and similarly for the conjugate. We will now focus on the chiral
piece $F_{3}$. Following \cite{Dennen:2009vk}, we define ${\bf w}_{i}=w_{i}^{a}\eta_{i,a}$ and ${\bf u}_{i}=u_{i}^{a}\eta_{i,a}$. Then we evaluate the fermionic integrals as follows
\begin{align} \label{eq:3pts_Fermion}
F_{3}^{(1,0)} & =  \frac{1}{V_3}\int d\chi^{+}_0 d\chi^{-}_0 dg \prod_{i=1}^{3}\delta(\sigma_{ij}\sigma_{ik}{\bf w}_{i}-\chi^{+}_0 -g\, \sigma_{i})\delta({\bf u}_{i}-\chi^{-}_0 )\nn\\
 & =  \frac{1}{V_3}({\bf u}_{1}-{\bf u}_{2})({\bf u}_{1}-{\bf u}_{3})\int d\chi^{+}_0 dg\prod_{i=1}^{3}\delta(\sigma_{i\,i+1}\sigma_{i\,i+2}{\bf w}_{i}-\chi^{+}_0-g \, \sigma_{i})\nn\\
& =  ({\bf u}_{1}{\bf u}_{2} + {\bf u}_{2} {\bf u}_{3}+{\bf u}_{3}{\bf u}_{1})({\bf w}_{1}+{\bf w}_{2}+{\bf w}_{3}),
\end{align}
where we have omitted the notation ``$\delta$" for fermionic delta functions. The final result is in precise agreement with the three-point superamplitude given, e.g., in \cite{Czech:2011dk}. For example, the three-gluon amplitude is:
\be
\mathcal{A}_3(A^{a\hat{a}}_1,A^{b\hat{b}}_2,A^{c\hat{c}}_3) = \left(u_1^a u_2^b w_3^c + u_1^a w_2^b u_3^c + w_1^a u_2^b u_3^c\right) 
 \left(\tilde u_1^{\hat a} \tilde u_2^{\hat b} \tilde w_3^{\hat c} 
+ \tilde u_1^{\hat a} \tilde w_2^{\hat b} \tilde u_3^{\hat c} 
+ \tilde w_1^{\hat a} \tilde u_2^{\hat b} \tilde u_3^{\hat c}\right)\,.
\ee

\section{Linear Form of the Maps}
\label{sec:LinearConstraints}
In this section, we present an alternative version of the connected formula for tree-level scattering amplitudes in 6D $\mathcal{N}=(1,1)$ SYM. We make use of ``linear" constraints involving $\lambda^A_a$ and $\eta_a$ directly, instead of the quadratic combinations $p^{AB} = \langle \lambda^A \lambda^B \rangle$ and $q^A = \langle \lambda^A \eta \rangle$. This form of the constraints is a natural generalization of the 4D Witten--RSV formula, in the form of \eqref{t-variables}. We have previously presented the linear constraints in \eqref{linearmeasure}. However, our conventions in this section differ from the previous formula by the change of variables $W_i = M_i^{-1}$. Since the $M_i$'s are $2\times 2$ matrices, the two formulations differ by where $W_i$ appears in the constraints as well as an overall Jacobian. For certain computations such as the soft limits, it may be preferable to use the previous version of the constraints.

One way in which the linear constraints differ from the quadratic constraints is that the on-shell conditions are no longer built in. Instead, they are enforced by the introduction of spinor-helicity variables. Another feature of the linear form is that it makes manifest more of the symmetries, including the $\text{SU}(2) \times \text{SU}(2)$ R symmetry. We will also give evidence that this representation may be a step towards a Grassmannian formulation of 6D theories \cite{ArkaniHamed:2012nw}.

As in the previous formulation of 6D theories, there are additional subtleties when the number of particles $n$ is odd. As before, the maps appropriate for odd $n$ require the $T$ symmetry, which acts as a redundancy of these maps. SYM amplitudes follow by pairing these constraints with the integrands found previously. 

Using the linear constraints for even- and odd-point SYM amplitudes, in Section \ref{sec:veronese} we obtain a version of these constraints that is even closer to the original Witten--RSV form. In the case of 4D, this version is sometimes known as the \emph{Veronese embedding} \cite{ArkaniHamed:2009dg}. This is achieved by evaluating the integral over the original rational maps $\rho^A_a(z)$, $\chi_a(z)$, $\tilde{\chi}_{\hat a}(z)$, leaving an integral over only the punctures and the $W_i$ variables. This allows one to view the linear constraints as those for a symplectic (or Lagrangian) Grassmannian acting on a vector built from the external kinematic data.

As an application of this formulation, we also present an alternative version of the tree-level amplitudes of the Abelian $(2,0)$ M5-brane theory. Since this theory does not have odd-point amplitudes, it is not a focus of the present work. Still, the linear version of the tree amplitudes of this theory have some advantages compared to the formula presented in \cite{Heydeman:2017yww}.

\subsection{\label{sec:linear-even}Linear Even-Point Measure}

The linear form of the 6D even-point measure is obtained by introducing an integration over $\text{GL}(2)$ matrices $(W_i)^b_a$ associated to each particle (or puncture):
\begin{align}
\label{Eq:LinearMeasure}
\int d\mu_{n \text{ even}}^{\text{6D}} &= \int \frac{\prod_{i=1}^n d\sigma_i\, \prod_{k=0}^{m} d^8 \rho_k}{\vol( \slc_\sigma \times \slc_\rho)} \frac{1}{V_n^2}\prod_{i=1}^n \delta^6 \left( p^{AB}_i - \frac{\langle \rho^{A}(\sigma_i)\, \rho^{B}(\sigma_i) \rangle }{\prod_{j\neq i} \sigma_{ij}}\right) \nonumber \\
& = \left(\prod_{i=1}^n \d(p_i^2)\right) \!\int\! \frac{\prod_{i=1}^n\, \prod_{k=0}^{m} d^8 \rho_k}{\vol( \slc_\sigma \times \slc_\rho)}  {\cal W} (\lambda, \rho, \s),
\end{align}
where
\be
{\cal W} (\lambda, \rho, \s) = \prod_{i=1}^n \int\!  d^{4}W_i  \, \d^8 \!\left ( \l^A_{ia} - (W_i)^b_a \r^A_{b}(\s_i) \right ) \d\!\left(|W_i| - {1 \over\prod_{j\neq i}\sigma_{ij} } \right) 
\ee
and $|W_i|={\rm det}\, W_i$. The total number of delta functions exceeds the number of integrations by $n+6$, accounting for the mass-shell and momentum-conservation delta functions. This step introduces $4n$ integrals in addition to the previous $5n-6$ that were previously present after accounting for the $\slc_\sigma \times \slc_\rho$ symmetry. It allowed us to extract the $n$ mass-shell constraints $\delta(p_i^2)$.
 
Before proceeding, let us comment on the $\slc$ indices of the matrix $(W_i)^b_a$.
Throughout this work, we have used the Latin indices $a = +, -$ to denote both the ``global" $ \slc_\rho$ indices as well as the little-group indices of the external particles. The latter was not visible when all the external data entered the formulas through the little-group invariant combinations $p^{AB}_i$, $q_i^A$, and $\tilde{q}_{iA}$. In passing to the linear form, we have introduced one matrix $(W_i)^b_a$ \emph{per particle}. We should view the upper index as global, because it contracts with the maps, whereas the lower index must transform under the little group of the $i$th external particle in order for the delta functions to be little-group invariant. So each $W_i$ transforms as a bi-fundamental under the global $\slc_{\r}$ and the $i$th $\slc$ little group. (The corresponding feature was also present in 4D when the $t_i$ and $\tilde{t}_i$ variables were introduced.) More explicitly, it is sometimes useful to solve for them in favor of the maps as follows: If we pick $\{A,B\}\subset \{1,2,3,4\}$, then 
\begin{equation}\label{eq:solW}
p_i^{AB} W^a_{i,b} =\frac{\rho^{[A,a}(\sigma_i) \, \lambda^{B]}_{i,b}}{\prod_{j\neq i}\sigma_{ij}},
\end{equation}
the above solution also makes clear the difference between these two $\slc$ indices. Despite this subtlety, we have elected not to use different notations for the different kinds of the $\slc$ indices, though it is always easy to distinguish them based on the context. 

The passage to linear constraints works analogously for the fermionic delta functions. The relevant identity is now:
\bea
\label{eq:linearfermions}
\D_F &=&  \prod^n_{i=1}  \delta^{4} \left(q^{A}_i -   { \langle \rho^A(\sigma_i)\, \chi (\sigma_i) \rangle \over \prod_{j \neq i} \sigma_{ij} }  \right) =  \prod_{i=1}^n \d^{2} \! \left (\tilde{\l}_{i A \hat{a}} q^{A}_i \right ) \d^{2}\! \left ( \eta^a_i - (W_i)^a_b \chi^{b}(\s_i) \right ) , \\
\widetilde{\D}_F &=&  \prod^n_{i=1} \delta^{4} \left( \tilde{q}_{i,A} -   { [\tilde{\rho}_{A}(\sigma_i) \, \tilde{\chi} (\sigma_i) ] \over \prod_{j \neq i} \sigma_{ij} }  \right) = \prod_{i=1}^n \d^{2} \! \left( \l^A_{i a} \tilde{q}_{iA} \right) \d^{2}\! \left ( \tilde{\eta}^{\hat a}_i - (\widetilde{W}_i)^{\hat{a}}_{\hat b} \tilde{\chi}^{\hat b}(\s_i) \right ).
\eea
These formulas are only valid on the support of the bosonic delta functions. Just like $\tilde{\rho}_k$, the conjugate set of matrices, $\widetilde{W}_i$, are not integrated over. Rather, they are solved for by the conjugate set of constraints, as in (\ref{eq:conjconstraints}). As before, this form allows us to explicitly extract the super-wave-function factors leaving linear fermionic delta functions in the $\eta$ and $\tilde{\eta}$ variables. 

For the case of 6D $\mathcal{N}=(1,1)$ SYM with $n$ even, the right-hand integrand, which is the Parke--Taylor factor, does not depend on this change of variables. So we can now assemble the even-point integrand, which in terms of the usual maps,
\be
\chi^{a} (z)   = \sum^{m}_{k=0} \chi^{ a}_k \, z^k, \qquad \quad \tilde{\chi}^{\hat{a}} (z)   = \sum^{m}_{k=0} \tilde{\chi}^{ \hat{a}}_k \, z^k,
\ee
is given by
\be 
\mathcal{I}_{n \text{ even}}^{\mathcal{N}=(1,1) \, {\rm SYM}}= {\rm PT}( \alpha ) \left(\,  V_n \,{\rm Pf'} A_n   \int  \left( \prod_{k=0}^{m}  d^2\chi_k\,  d^2\tilde{\chi}_k  \right)\, \D_F \widetilde{\D}_F, \right) .
\ee
Removing the mass-shell delta functions, the explicit formula for the linear form of the even-point scattering amplitudes of 6D $\mathcal{N}=(1,1)$ SYM is
\begin{empheq}[box=\fbox]{align}
\label{eq:(1,1)SYM-even-linear}
\mathcal{A}_{n\text{ even}}^{\mathcal{N}=(1,1)\text{ SYM}} (\alpha) =& \!\int \! \frac{\prod_{i=1}^n d\sigma_i\, \prod_{k=0}^{m} d^8 \rho_k \,  d^2\chi_k\,  d^2\tilde{\chi}_k}{\vol( \slc_\sigma \times \slc_\rho)} \;{\rm PT}( \alpha ) \; {\cal W} (\lambda, \rho, \s) \nn\\
& \times  V_n \,  {\rm Pf'} A_n \prod_{i=1}^{n} \, \d^2\! \left ( \eta^a_i - (W_i)^a_b \chi^{b}(\s_i) \right ) \d^2 \!\left ( \tilde{\eta}^{\hat a}_i - (\widetilde{W}_i)^{\hat{a}}_{\hat b} \tilde{\chi}^{\hat b}(\s_i) \right )   \,  .
\end{empheq}

So far we have used the fact that the kinematic data associated to a given particle in 6D can be encoded in two pairs of spinors, $\lambda^A_{ia}$ and $\tilde\lambda^{i{\hat a}}_A$ . However, using the overall scaling it is also possible to associate the chiral part, $\lambda^A_{ia}$, with a line in $\mathbb{CP}^3$ and two points on it, where the two components, $a = \pm$, label the points. The linear formula implements the transformation from one description to the other. $\rho^A_a(\sigma_i)$ can be taken to define a line in $\mathbb{CP}^3$, while each row of the $2\times 2$ matrix $W_i$ can be interpreted as defining the homogeneous coordinates for two points on this line. We believe that this new viewpoint would be useful in writing formulas in the 6D version of twistor space.

An added benefit of the linear form is that it makes parts of the non-linearly realized R symmetry generators manifest, as we mentioned previously. Recalling \eqref{eq:Rgenerators}, the generators are quadratic in the $\eta_i$, $\tilde{\eta}_i$ variables and their derivatives. In particular, let us consider the generators $R^+ = \sum_{i=1}^n \eta_{i,a} \eta_i^a$ and $\widetilde{R}^+ = \sum_{i=1}^n \tilde{\eta}_{i,\hat a} \tilde{\eta}_i^{\hat a}$.
One may verify that these are symmetry generators by first noting that under the support of the delta functions
\be
\eta_{ia} = (W_i)_{ab} \chi^{b}(\s_i), \quad \tilde{\eta}_{i\hat a} = (\widetilde{W}_i)_{\hat{a}\hat b}\tilde{\chi}^{\hat b}(\s_i)\,.
\ee
Similar to how one constructs the momenta $p^{AB}_i$ from antisymmetric combinations of the analogous bosonic delta functions for $\l^A_{ia}$, we can construct the combinations:
\be
R^+ = \sum_{i=1}^n \langle \eta_{i} \eta_i \rangle = \sum_{i=1}^n (W_i)_{ab}(W_i)^a_{c} \chi^{b}(\s_i) \chi^{c}(\s_i)  = \sum_{i=1}^n |W_i|\, \chi_{b}(\s_i) \chi^{b}(\s_i) \,,
\ee
and similarly for $\widetilde{R}^+$. Under the support of the bosonic delta functions the determinant $|W_i|$ can be replaced by $(\prod_{j\neq i}\sigma_{ij})^{-1}$, whereas $\chi_{b}(\s_i) \chi^{b}(\s_i)$ is a polynomial of degree $n{-}2$ in $\s_i$. Using the identity
\begin{align}
\label{eq:magicidentity}
\sum_{i=1}^n \frac{\sigma_i^k}{\prod_{j\neq i}\sigma_{ij}}=0 \, , \qquad {\rm for} \qquad k=0,1,\ldots, n-2 \, ,
\end{align}
which can be understood as a consequence of a residue theorem, we find that $R^+=0$. This means that the amplitude is supported on configurations such that $\sum_i \eta_{ia}\eta_{i}^a = 0$ and $\sum_i \tilde{\eta}_{i \hat a}\tilde{\eta}_{i}^{\hat a} =0$, which proves the conservation of this R charge. The vanishing of the final R symmetry generators, $R^-$ and $\tilde R^-$, which are second derivative operators, is still not made manifest in this formulation, but it is not hard to prove. For example, a Grassmann Fourier transform interchanges the role of $\eta$ and $\partial / \partial \eta$.

As a final application, we apply the formalism of linear constraints to the tree amplitudes of a single M5-brane in 11D Minkowski spacetime. This provides an example of a 6D theory with $(2,0)$ supersymmetry; the amplitudes in the rational maps formalism are given by \cite{Heydeman:2017yww}:
\be
\mathcal{A}_{n}^{\text{M5-brane}} = \int \frac{\prod_{i=1}^n d\sigma_i \, \prod_{k=0}^{m} d^8 \rho_k \,  d^4\chi_k}{\vol( \slc_\sigma \times \slc_\rho)} \D_B \, \D_F \, {\left( {\rm Pf}'A_n \right)^3 \over V_n} \, ,
\ee
where 
\be
\D_B = \prod^n_{i=1} \d^6 \! \left ( p^{AB}_i -  \frac{ \langle \r^{A}(\s_i) \r^{B}(\s_i) \rangle}{ \prod_{j\neq i} \sigma_{ij}} \right),
\ee
\be
\D_F =  \prod^n_{i=1}  \delta^{8} \left(q^{AI}_i -   { \langle \rho^A(\sigma_i)\, \chi^I (\sigma_i) \rangle \over \prod_{j \neq i} \sigma_{ij} }  \right),
\ee
and $I = 1,2$ denotes the two chiral supercharges. 

Since this theory has only even-point amplitudes, we do not need the machinery of odd-point rational maps in this case. Introducing the $W_i$ variables, the bosonic measure is identical to that of SYM. The fermionic delta functions with $\mathcal{N}=(2,0)$ supersymmetry become: 
\be
\D_F  =  \prod_{i=1}^n \d^{4} \! \left (\tilde{\l}_{i A \hat{a}} q^{AI}_i \right) \d^{4}\! \left ( \eta^{aI}_i - (W_i)^a_b \chi^{bI}(\s_i) \right) ,
\ee
so the amplitudes have the representation:
\begin{empheq}[box=\fbox]{align} \label{eq:(2,0)M5-linear}
&\mathcal{A}_{n}^{\text{M5-brane}} = \int \frac{\prod_{i=1}^n d\sigma_i\, \prod_{k=0}^{m} d^8 \rho_k \,  d^4\chi_k}{\vol( \slc_\sigma \times \slc_\rho)} \; {\cal W} (\lambda, \rho, \s) \nn \\
& \qquad \qquad \qquad \qquad \times ({\rm Pf'} A_n)^3 \, V_n\, \, \prod_{i=1}^n
\d^{4}\! \left ( \eta^{aI}_i - (W_i)^a_b \chi^{bI}(\s_i) \right)\, .\end{empheq}
It is worth noting that for chiral $\mathcal{N}=(2,0)$ supersymmetry there is no need to introduce $\tilde{\rho}$, $\widetilde{W}_i$, or $\tilde{\chi}$. In some sense, the 6D chiral theories appear more natural than their non-chiral counterparts. This theory has $\text{USp}(4)$ R symmetry, which can be verified
in the linear formulation by the technique described above.

By the same reasoning, the D5-brane formula \cite{Heydeman:2017yww}, which has $\N=(1,1)$ supersymmetry, can be recast in a similar form
with the same fermionic delta functions as in \eqref{eq:(1,1)SYM-even-linear}

\subsection{\label{sec:linear-odd}Linear Odd-Point Measure}

To complete the discussion for the $\N=(1,1)$ SYM odd-point measure and integrand in this formalism, we introduce the parametrization of the odd-point maps described in Section \ref{sec:odd-rational-maps}. As before, we define
\bea
\r^A_a(z) &=& \sum_{k=0}^{m-1} \r^A_{a,k}\,z^k + \omega^A \xi_a \,z^m , \\
\tilde{\r}_A^{\hat{a}}(z) &=& \sum_{k=0}^{m-1} \tilde{\r}_{A\,k}^{\hat{a}} \,z^k + \tilde{\omega}_A \tilde{\xi}^{\hat a} \,z^m ,
\eea
and similarly for the fermionic partners, where $m={(n-1)/ 2}$. In the case where we used constraints for $p^{AB}(z)$, the introduction of this parametrization of the maps included a new redundancy. This was because the polynomial $\langle \rho^A(z) \rho^B(z) \rangle$ has a shift symmetry of the form $\rho^{A}(z) \;\rightarrow\; \hat{\rho}^{A}(z)  =  (\mathbb{I}+z\,T)\rho^{A}(z)$, where $T^{a}_{b} = \alpha \xi^a \xi_b$ and $\alpha$ is a parameter. The invariance of the product is still required in the linear formalism, and there must be a redundancy that reduces the number of components of $\omega^A$ and $\xi_a$. As before, the integrations over the moduli and the Riemann sphere are completely localized by the bosonic delta functions, which requires five independent components.

We will find that the appropriate choice is to keep the general action of the T-shift on $\rho^A_a(\sigma_i)$ but now allowing the $W_i$ to transform at the same time. The linear constraint $\d^8 \left (\lambda^A_{ia} - (W_i)^b_a \rho^A_b(\s_i) \right )$ is left invariant under the T-shift, which now explicitly depends on each puncture $\s_i$:
\begin{align}
\rho^A_b(\sigma_i) &\rightarrow \rho^A_b(\sigma_i) + \alpha \s_i \xi^c \xi_b \rho^A_c (\s_i) \\
(W_i)_a^b &\rightarrow (W_i)_a^b - \alpha \s_i \xi^b \xi_c  (W_i)_a^c\, , \label{eq:TsymW}
\end{align}
or more abstractly,
\begin{align}
\rho^A(\sigma_i) &\rightarrow \left ( \mathbb{I}+\s_i \,T \right ) \rho^A (\s_i) \\
(W_i)_a &\rightarrow \left ( \mathbb{I} - \s_i \,T^\intercal \right ) (W_i)_a \, .
\end{align}
These transformations leave the product invariant by virtue of \eqref{eq:Tconditions}. Recall that the lower index $a$ on $(W_i)^b_a$ is the little-group index for the $i$th particle, and it does not participate in the shift.

With the maps and redundancy more or less the same as in Section \ref{sec:odd-pt}, we may now write down the measure associated to the linear constraints for odd $n$, which takes a similar form as the even-point one:
\begin{align}
\label{Eq:LinearMeasure-Odd}
\int d\mu_{n \,\text{odd}}^{\text{6D}} &= \int \frac{\left(\prod_{i=1}^{n} d\sigma_i \prod_{k=0}^{m-1}d^{8}\rho_{k}\right)\, d^{4}\omega\, \langle \xi d\xi\rangle }{\vol \left( \slc_{\sigma} , \slc_{\rho} , \text{T} \right)}\, \frac{1}{V^{2}_n} \prod_{i=1}^{n}\delta^6 \left(p_{i}^{AB}-\frac{p^{AB}(\sigma_{i})}{\prod_{j\neq i} \sigma_{ij}}\right) \nonumber \\
& = \left(\prod_{i=1}^n \d(p_i^2) \right) \int \frac{\left(\prod_{i=1}^{n} d\sigma_i \prod_{k=0}^{m-1}d^{8}\rho_{k}\right)\, d^{4}\omega\, \langle \xi d\xi\rangle }{\vol \left( \slc_{\sigma} , \slc_{\rho} , \text{T} \right)} \; {\cal W} (\lambda, \rho, \s).
\end{align}
We are free to fix the scaling and T-shift symmetry of this measure exactly as before, so all Jacobians will be the same as in previous sections. Therefore in terms of the linear maps, the superamplitudes of 6D $\mathcal{N}=(1,1)$ SYM can be expressed as
\begin{empheq}[box=\fbox]{align} \label{eq:(1,1)SYM-odd-linear}
&\mathcal{A}_{n \text{ odd}}^{\mathcal{N}=(1,1)\text{ SYM}} (\alpha) = \int \frac{\left(\prod_{i=1}^{n} d\sigma_i \prod_{k=0}^{m-1}d^{8}\rho_{k}\right)\, d^{4}\omega\, \langle \xi d\xi\rangle }{\vol \left( \slc_{\sigma} , \slc_{\rho} , \text{T} \right)} \; {\cal W} (\lambda, \rho, \s) \; {\rm PT}( \alpha )  \;{\rm Pf'} \widehat{A}_n \nn\\
&  \qquad\times   V_n \int \prod_{k=0}^{m-1} d^2 \chi_k d^2 \tilde{\chi}_k \, dg \, d\tilde{g} 
 \prod_{i=1}^n    \d^2 \!\left ( \eta^a_i - (W_i)^a_b \chi^{b}(\s_i) \right ) \d^2 \!\left ( \tilde{\eta}^{\hat a}_i - (\widetilde{W}_i)^{\hat{a}}_{\hat b} \tilde{\chi}^{\hat b}(\s_i) \right ) \!,
\end{empheq}
where, as before, the fermionic maps for $n$ odd are defined to be
\begin{align}
\chi^{a}(z)& =\sum_{k=0}^{m-1} \chi^{a}_k\, z^k  + g \xi^a z^m, \\
\tilde{\chi}^{\hat{a}}(z)& = \sum_{k=0}^{m-1} \tilde{\chi}^{\hat{a}}_k\, z^k + \tilde{g} \tilde{\xi}^{\hat{a}} z^m.
\end{align}

\subsection{Veronese Maps and Symplectic Grassmannian}
\label{sec:veronese}

The preceding results can be brought even closer to the original Witten--RSV formulation by integrating out the moduli $\rho^A_{a,k}$ of the maps, which leaves an integral over only the $\s_i$ and the $W_i$. This will allow us to show that these constraints apply to the elements of a \emph{symplectic} Grassmannian. Let us begin with the even-$n$ case and recast the bosonic measure:
\begin{align}
&\int d\mu_{n \text{ even}}^{\text{6D}} = 
 \int \frac{\left(\prod_{i=1}^n d\sigma_i\,  d^{4}W_i\right) \prod_{k=0}^{m} d^8 \rho_k}{\vol( \slc_\sigma \times \slc_\rho)} \prod_{i=1}^n    \, \d^8 \left ( \l^A_{ia} - (W_i)^b_a \r^A_{b}(\s_i) \right ) \d\!\left(|W_i| - {1 \over\prod_{j\neq i}\sigma_{ij} } \right) \nonumber \\
&= \int \frac{\prod_{i=1}^n d\sigma_i\, d^{4}W_i  }{\vol( \slc_\sigma \times \slc_W)} \prod^m_{k=0} \d^8 \left(\sum_{i=1}^n (W_i)^b_a \s^k_i \l^A_{ib}\right)\prod_{i=1}^n \d\left(|W_i| - {1 \over\prod_{j\neq i}\sigma_{ij} }\right) .
\end{align} 
This result can be obtained by using the following identity for each of the eight components separately \cite{Roiban:2004yf,Witten:2004cp,Cachazo:2013zc},
\be\label{eq:residueidentity}
\prod_{k=0}^m \delta\left(\sum_{i=1}^n \sigma_i^k X_i \right) = V_n  \int \left(\prod_{k=0}^{m} d {\rho}_k \right) \prod_{i=1}^n \delta\left({\rho}(\s_i) - X_i \prod_{j\neq i} \s_{ij} \right)\, ,
\ee
where ${\rho}(z) = \sum_{k=0}^{m} {\rho}_k \,z^k$ denotes any component of the polynomial map.
Starting with \eqref{eq:linearfermions}, one can obtain a similar result for the fermions. Specifically, 
\begin{align}
\int \left(\prod_{k=0}^m  d^2 \chi_k\right) \, \prod_{i=1}^n\delta^2 \left(\eta_{i,a} - (W_i)^b_a \chi_b(\s_i) \right)=\prod_{k=0}^{m} \delta^2 \left(\sum_{i=1}^n (W_i)^b_a \s_i^k \eta_{i,b} \right) \, .
\end{align} 

We now note that $(W_i)^b_a \s_i^k$ forms an $n\times 2n$ matrix:
\begin{equation}
C_{a,k; i,b}=(W_i)^b_a\, \s_i^k\, ,
\end{equation}
where we group the index $k$ with the global $\slc$ index $a$ and the index $i$ with the $i$th little-group $\slc$ index $b$. Interestingly, under the constraints $|W_i| - {1 \over\prod_{j\neq i}\sigma_{ij}} =0$, the matrix $C$ formed in this way is symplectic satisfying
\begin{align}
C\cdot \Omega \cdot C^T =0 \,,
\end{align}
which follows from the application of the identity \eqref{eq:magicidentity} to each block matrix of the product. Here $\Omega$ is a symplectic metric: an anti-symmetric $2n\times 2n$ matrix with non-zero entries at $\Omega_{i,i+1}=-\Omega_{i+1,i}=1$. Therefore $C$ is a symplectic Grassmannian, which was mentioned in \cite{ArkaniHamed:2012nw} for its possible applications to scattering amplitudes. Here we construct the sympletic Grassmannian explicitly in the spirit of the Veronese maps as discussed in \cite{ArkaniHamed:2009dg} to relate Witten--RSV formulas with Grassmannian formulations for 4D $\mathcal{N}=4$ SYM \cite{Bourjaily:2010kw}. Using the $n\times 2n$ matrix $C$, one may rewrite the constraints nicely as
\be
\sum_{i=1}^n (W_i)^b_a \s^k_i \l^A_{ib}:= (C\cdot \Omega \cdot \Lambda)_a{}^A =0 \, ,
\ee
where $\Lambda^A=\l^A_{i,b}$ is a $2n$-dimensional vector. The fermionic constraints take a similar form with the same Grassmannian. Geometrically, this is a 6D version of the orthogonality conditions of the 4D Grassmannian described in \cite{ArkaniHamed:2009dn}.

Similarly, when $n = 2m+1$ is odd, the identity \eqref{eq:residueidentity} leads to
\begin{align}
\int\! d\mu_{n \text{ odd}}^{\text{6D}} \!&= \!\!\!
 \int\! \frac{\prod_{i=1}^n d\sigma_i d^{4}W_i\, (\prod_{k=0}^{m-1} d^8 \rho_k) d^4 \omega \langle \xi d\xi \rangle}{\vol( \slc_\sigma \, , \slc_\rho \, , T)} \!\prod_{i=1}^n  \! \d^8 \!\left (\! \l^A_{ia} \!-\! (W_i)^b_a \r^A_{b}(\s_i) \!\right ) \!\d\!\left(\!|W_i| - {1 \over\prod_{j\neq i}\sigma_{ij} } \!\right) \nonumber \\
&= \!\!\int \frac{\left(\prod_{i=1}^n d\sigma_i\, d^{4}W_i\right) \langle \xi d\xi \rangle}{\vol( \slc_\sigma \, , \slc_W \, , T)} \prod^{m-1}_{k=0} \d^8 \left(\sum_{i=1}^n (W_i)^b_a \s^k_i \l^A_{ib}\right) \nonumber \\
& \qquad \qquad \qquad \qquad \times \, \d^4 \left(\sum_{i=1}^n \xi^a (W_i)^b_a \s^{m}_i \l^A_{ib}\right)\prod_{i=1}^n \d\left(|W_i| - {1 \over\prod_{j\neq i}\sigma_{ij} }\right) .
\end{align} 
For odd $n$, the form of the Grassmannian constraints is unmodified except for the highest degree, $\s_i^m$. The highest-degree terms must be modified so that the number of constraints decreases by $5$ when passing from even to odd, which is the case for this expression. Note that we have integrated out the Lorentz spinor $\omega^A$ but not the global little-group spinor $\xi^a$. One of the $\slc_\rho$ generators can be used to fix the only independent component in $\xi^a$, making it effectively arbitrary. This nontrivial relation leaves only four independent constraints for the highest-degree part of the Grassmannian.

For odd $n$, this Veronese form also has the T-shift symmetry inherited from that of the $W_i$'s, as shown in \eqref{eq:TsymW}. The T-shift acts on the Grassmannian as
\begin{align}
(W_i)^b_a\, \s_i^k\, &= C_{a,k; i,b} \rightarrow C_{a,k; i,b} - \alpha \xi_a \xi^c C_{c,k+1; i,b} \, ,\quad k=0,\dots, m-1 \, , \\
\xi^a (W_i)^b_a\, \s_i^m\, &= \xi^a C_{a,m; i,b} \rightarrow \xi^a C_{a,m; i,b} \, .
\end{align}
The term of highest degree is invariant under the shift due to $\langle \xi \,\xi \rangle =0$. This shift can be interpreted as a special kind of row operation on the Grassmannian in which the rows of degree $k$ are translated by the rows of degree $k+1$ with the exception of the highest-degree rows.

One must now fix the various redundancies of this description. In the end the number of integrals should equal the number of constraints after gauge fixing. There are $5n$ integrals before fixing the two $\slc$'s and $5n-6$ after fixing them. These choices can be used to fix three of the punctures $\s_i$ as well as two values of a $W_i$ and one component of $\xi_a$. Finally, the T-shift can be used to fix the last value of the chosen $W_i$.

The fermionic delta functions satisfy a similar identity, 
\begin{align}
\int \!dg\prod_{k=0}^{m-1}  d^2 \chi_k  \prod_{i=1}^n \delta^2 \left(\eta_{i,a} - (W_i)^b_a \chi_b(\s_i) \right)= \d \left(\sum_{i=1}^n \xi^a (W_i)^b_a \s^{m}_i \eta_{i,b}\right) \prod_{k=0}^{m-1} \delta^2 \left(\sum_{i=1}^n (W_i)^b_a \s_i^k \eta_{i,b} \right)\! .
\end{align} 
Now $(W_i)^b_a \s_i^k$ with $k=0, 1, \ldots, m{-}1$ combines with $\xi^a (W_i)^b_a \s^{m}_i$ to form an $n\times 2n$ symplectic matrix acting on the vector of external Grassmann variables, entirely analogous to the constraints for the external spinors. 

Using these relations, it is then straightforward to rewrite all of the superamplitudes given in previous sections in terms of the Veronese maps. In the case of 6D $\mathcal{N}=(1,1)$ SYM and odd $n$, in the integrand, the term ${\rm Pf'} \widehat{A}_n$ contains a special ``momentum'' vector, which we recall here:
\begin{align}
p_{\star}^{AB} = \dfrac{ 2\, q^{[A} p^{B]C}(\sigma_\star) \tilde{q}_C }{ q^D  [\tilde{\rho}_D(\sigma_\star )\, \tilde{\xi}] \langle \rho^E(\sigma_\star)\, \xi \rangle  \tilde{q}_E}\,.
\end{align}
This shows that $p_{\star}^{AB}$ is in general a function of the moduli $\rho_{a,k}^A$ and $\tilde{\rho}_{A,k}^{\hat a}$. Therefore, when we integrate out the moduli and express the amplitudes in the Veronese form, we should solve for $\rho_{a,k}^A$ and $\tilde{\rho}_{A,k}^{\hat a}$ in terms of the $W_i$'s and $\widetilde{W}_i$'s, as well as the $\s_i$'s. If we choose $\s_{\star}$ to be one of the $\s_i$'s, then it is trivial to express $p_{\star}^{AB}$ in terms of $W_i$ and $\s_i$ by using the relation, 
\begin{align} \label{eq:rho-lambda-relation}
\rho^A_a(\sigma_i) = (M_i)^{b}_a \, \lambda^A_{i,b}\, ,
\end{align}
and a similar relation for $\tilde{\rho}^{\hat a}_A(\sigma_i)$, and recalling that $M_i = W_i^{-1}$. If, instead, we choose $\s_{\star}$ to be arbitrary, $\rho^A_a(\sigma_{\star})$ can also be determined in terms of $M_i$ and $\s_i$ using the above relation (\ref{eq:rho-lambda-relation}), since $\rho^A_a(z)$ is a degree $m=\frac{n-1}{2}$ polynomial, and there are $n$ such relations.

\section{\label{sec:applications}Various Theories in $\text{D}\mathbf{\leq 6}$ and $\mathbf{\N=4}$ SYM on the Coulomb Branch}

This section describes some interesting applications and consistency checks of the 6D SYM formulas that we have obtained. We start by writing down a formula for 6D $\N=(2,2)$ supergravity amplitudes in Section~\ref{sec:SUGRA}, which follows from the double copy of the formula of 6D $\N=(1,1)$ SYM studied in previous sections. Then in Section~\ref{sec:mixed-amplitudes} we consider mixed amplitudes by coupling the 6D $\N=(1,1)$ SYM with a single D5-brane. We will also study the dimensional reduction of these theories. We begin with the reduction to five dimensions in Section~\ref{sec:5d-theories}, followed by $\mathcal{N}=4$ SYM on the Coulomb branch in Section~\ref{sec:Coulomb-branch}. We obtain new connected formulas for all tree-level scattering amplitudes of these theories. In Section~\ref{sec:massless-4D} we study dimensional reduction to 4D $\mathcal{N}=4$ SYM at the origin of the moduli space.

\subsection{\label{sec:SUGRA}$\N=(2,2)$ Supergravity in Six Dimensions}

In this section we consider the tree amplitudes of $\N=(2,2)$ supergravity in 6D. Even though 6D ${\cal N} =(2,2)$ SUGRA is nonrenormalizable, it has a well-known UV completion. This completion is given by Type IIB superstring theory compactified on $T^4$ or (equivalently) M theory compactified on $T^5$. In either case, the theory has an 
$\text{E}_{5,5}(\mathbb{Z}) = \text{SO}(5,5;\mathbb{Z})$ U-duality group. This is a discrete global symmetry. (It is believed that string theory does not give rise to continuous global symmetries \cite{Banks:2010zn}.)
The low-energy effective description of this theory, which is the 6D supergravity theory under consideration here, extends this symmetry to the continuous non-compact global symmetry $\text{Spin}(5,5)$. However, much of this symmetry is non-perturbative, and only the compact subgroup $\text{Spin}(5) \times \text{Spin}(5)$ is realized as a symmetry of the supergravity tree-level scattering amplitudes. (Recall that $\text{Spin}(5) = \text{USp}(4)$.) This symmetry is the relevant R symmetry group. This is called an R symmetry group because particles with
different spins belong to different representations of this group even though they form an irreducible supermultiplet. The UV complete theory and its low-energy supergravity effective description are both maximally supersymmetric. This means that there are 32 local supersymmetries, gauged by the gravitino fields. It also implies that the supergravity theory has a 6D Minkowski-space solution that has 32 unbroken global supersymmetries. When we discuss scattering amplitudes, this is the background geometry under consideration. If we further reduce to four dimensions, we get $\mathcal{N}=8$ supergravity, which has nonperturbative $\text{E}_{7(7)}$ symmetry. Again, only the compact subgroup, which is $\text{SU}(8)$ in this case, is the R symmetry of the tree amplitudes.

The 6D ${\cal N} =(2,2)$ supergravity multiplet contains 128 bosonic and 128 fermionic degrees of freedom, which can be elegantly combined into a scalar superparticle by introducing eight Grassmann coordinates in a manner that will be described below. This multiplet contains six different spins, i.e., little-group representations, which we will now enumerate. They are characterized by their $\text{SU}(2) \times \text{SU}(2)$ little-group representations and their $\text{USp}(4) \times \text{USp}(4)$ R symmetry representations. The graviton transforms as $(\mathbf{3,3;1,1})$ under these four groups. Similarly, the eight gravitinos belong to $(\mathbf{3,2;1,4})+(\mathbf{2,3;4,1})$. Also, the ten two-form particles belong to $(\mathbf{3,1;1,5})+(\mathbf{1,3;5,1})$.  The 16 vector particles belong to $(\mathbf{2,2;4,4})$, the spinors belong to $(\mathbf{2,1;4,5})+(\mathbf{1,2;5,4})$, and the scalars belong $(\mathbf{1,1;5,5})$. As in the case of the SYM theory, the amplitudes will be presented in a form that makes the helicity properties of the particles straightforward to read off, but only a subgroup of the R symmetry will be manifest. With some effort, one can prove that the entire $\text{USp}(4) \times \text{USp}(4)$ R symmetry is actually realized. Even though this is a non-chiral (left-right symmetric) theory, corresponding left- and right-handed
particles have their R symmetry factors interchanged. So this interchange should be understood to be part of the definition of the reflection symmetry.

The on-shell superfield description of the supergravity multiplet, analogous to the one for the SYM multiplet in (\ref{eq:spectrum11}), utilizes eight Grassmann coordinates denoted $\eta^{I,a}$ and $\tilde\eta^{\hat I, \hat a}$. It contains 128 bosonic and 128 fermionic modes with the spectrum enumerated above. It has the schematic form
\be
\Phi(\eta)
=  \phi + \ldots +  \eta_a^I \, \eta_{b,I} \, \tilde{\eta}^{\hat I}_{\hat{a}} \,\tilde\eta_{\hat b, \hat I}\, G^{ab;\hat{a}\hat b} + \ldots
+ (\eta)^4 (\tilde{\eta})^4 \bar\phi \,. 
\ee
Note that $I=1,2$ and ${\hat I}={\hat 1}, {\hat 2}$ label components of an $\text{SU}(2) \times \text{SU}(2)$ subgroup of the R symmetry group. Only this subgroup of the $\text{USp}(4) \times \text{USp}(4)$
R symmetry is manifest in this formulation. The on-shell field $G^{ab;{\hat a}{\hat b}}$ in the middle of the on-shell superfield is the 6D graviton. We have only displayed this field and two of the 25 scalar fields. 

The supergravity superamplitudes have total symmetry in the $n$ scattered particles. This is to be contrasted with the cyclic symmetry of the color-stripped SYM amplitudes.  For instance, the four-point superamplitude is given by
\be \label{eq:sugra-4pts}
{M}_4^{\N=(2,2)\text{ SUGRA}} \;=\; \d^6 \left( \sum_{i=1}^4 p_i^{AB} \right) { \delta^{8} \left( \sum_{i=1}^4 q^{A, I}_i \right)  \delta^{8} \left( \sum_{i=1}^4 {\tilde q}^{ {\hat I} }_{i,{\hat A}}\right)  \over s_{12}\, s_{23}\, s_{13} } \, ,
\ee
which has manifest permutation symmetry. Here the supercharges are defined as $q^{A, I}_i = \lambda^A_{i, a} \eta^{I, a}_i$, and $  {\tilde q}^{ {\hat I} }_{i,{\hat A}}= \tilde{\lambda}_{i, \hat{A},\hat{a} } \tilde{\eta}^{ {\hat I}, \hat {a}}_i$. As in the case of the SYM theory, these are half of the supercharges, and the other half involve $\eta$ derivatives. Conservation of these additional supercharges automatically follows from the first set together with the R symmetry.

Thanks to the separation of the $\N=(1,1)$ SYM formulas into the measure, left- and right-integrands, the formulas for $\N=(2,2)$ SUGRA amplitudes follow from the standard KLT argument \cite{Kawai:1985xq} in the context of CHY formulations \cite{Cachazo:2013iea}. One replaces the Parke--Taylor factor with a second copy of the remaining half-integrand. The resulting connected formula for amplitudes of all multiplicities can be written in a compact form:
\be \label{eq:(2,2)SUGRA}
\boxed{\mathcal{M}_n^{\N=(2,2)\text{ SUGRA}} =  \int d\mu_n^{\text{6D}} \left({\text{Pf}}\,' A_n\right)^2 \int d\Omega_F^{(2,2)} .}   
\ee
Here the fermionic measure $d\Omega_F^{(2,2)}$ that implements the 6D $\N=(2,2)$ supersymmetry is the double copy of the $\N=(1,1)$ version $d\Omega_F^{(1,1)}$, with
\be
\chi_k^a \rightarrow \chi_k^{I\,a}, \quad \tilde{\chi}_k^{\hat{a}} \rightarrow \tilde\chi_k^{\hat{I}\,\hat{a}} \, , \quad g \rightarrow g^I, \quad \tilde{g} \rightarrow \tilde{g}^{\hat{I}} .
\ee
Here $I=1,2$ and $\hat{I}=1,2$ are the $\text{SU}(2) \times \text{SU}(2)$ R symmetry indices. Explicitly, the measure is defined as
\be
d\Omega_F^{(2,2)}  = V^2_n \left( \prod_{k=0}^{m}  d^4\chi_k\,  d^4\tilde{\chi}_k  \right)\, \D^{(8)}_F \, \widetilde{\D}^{(8)}_F \, ,
\ee
for even $n=2m+2$, and 
\be
d{\widehat{\Omega}}_F^{(2,2)} = V^2_n\, d^2g\,d^2\tilde{g} \left(\prod_{k=0}^{m-1}d^4\chi_{k}\,d^4\tilde{\chi}_{k} \right) \, \D^{(8)}_F \, \widetilde{\D}^{(8)}_F \, , 
\ee
for odd $n=2m+1$. The fermionic delta functions are also a double copy of the $\mathcal{N}=(1,1)$ ones, and they are given by
\bea
\D^{(8)}_F&=& \prod^n_{i=1}  \delta^{8} \left(q^{I, A}_i -   { \rho^A_a(\sigma_i) \chi^{I,a} (\sigma_i)  \over \prod_{j \neq i} \sigma_{ij}}  \right) \, , \\
\widetilde{\D}^{(8)}_F &=& \prod^n_{i=1} \delta^{8} \left( \tilde{q}^{\hat{I}}_{i,A} -   { \tilde{\rho}_{A, \hat{a}}(\sigma_i) \tilde{\chi}^{\hat{I}, \hat{a}} (\sigma_i)  \over \prod_{j \neq i} \sigma_{ij} }  \right) \, .
\eea
Finally, it is understood that the reduced Pfaffian in the integrand refers to the matrix $A_n$ in \eqref{A-matrix-def} for the even-point case and the hatted matrix $\widehat{A}_n$ in \eqref{odd-matrix-A} for the odd-point case.

\subsection{$\N=(1,1)$ Super Yang--Mills Coupled to D5-branes} \label{sec:mixed-amplitudes}

Since we now have connected formulas for the scattering amplitudes in the effective field theories of the D5-brane and $\N=(1,1)$ SYM in 6D, we can consider mixed amplitudes involving both kinds of particles. It was proposed in \cite{Cachazo:2016njl} that these types of amplitudes admit a simple CHY formula, which interpolates between the Parke--Taylor factor $\text{PT}(\alpha)$ for the non-Abelian theory and $(\text{Pf}' A_n)^2$ for the Abelian one. Such a construction was used in \cite{Cachazo:2016njl} to write down amplitudes coupling Non-linear Sigma Model (NLSM) pions to bi-adjoint scalars, as well as their supersymmetrization in 4D involving Volkov--Akulov theory (effective theory on a D3-brane) \cite{Volkov:1973ix, Kallosh:1997aw} and $\N=4$ SYM. Related models were later written down in the context of string-theory amplitudes \cite{Carrasco:2016ygv}. These mixed amplitudes are also parts of the unifying relations for scattering amplitudes \cite{Cheung:2017ems, Cheung:2017yef}. In all of the above cases, the connected formula selects preferred couplings between the two theories. They were identified in \cite{Cheung:2016prv,Low:2017mlh} in the case of the NLSM coupled to bi-adjoint scalar theory.

Following the same approach allows us to write down a formula coupling the D5-brane effective theory to 6D SYM:
\be\label{D5-SYM-theory}
\boxed{\mathcal{A}^{\text{D5-brane } \oplus \text{ SYM}}_n (\alpha) = \int d\mu_n^{\text{6D}} \,\bigg( \text{PT}(\alpha)\, \left(\text{Pf} A_{\overline{\alpha}}\right)^2 \bigg) \left( \text{Pf}\,' A_n \int d\Omega_{\text{F}}^{(1,1)} \right),}
\ee
where $\alpha$ represents the states of SYM, which are color ordered, and the complement, $\overline{\alpha}$, represents states of the Abelian D5-brane theory. We have also used the fact that the D5-brane theory and 6D SYM have identical supermultiplets and the same supersymmetry. Here the right-integrand, which is common between the two theories, remains unchanged. Of course, whenever the total number of particles $n$ is odd, one should make use of the odd-multiplicity counterparts of the reduced Pfaffian and the bosonic and fermionic measures. In the left-integrand we have a Parke--Taylor factor constructed out of only the SYM states that enter the color ordering $\alpha$. The D5-brane states belonging to $\overline{\alpha}$ do not have color labels, and hence they appear in the formula through the permutation-invariant Pfaffian. The matrix $A_{\overline{\alpha}}$ is an $|\overline{\alpha}| \times |\overline{\alpha}|$ minor of $\left[\frac{p_i \cdot p_j}{\sigma_{ij}}\right]$ with columns and rows labeled by the D5-brane states. This implies that the above amplitude in non-vanishing only if the number of D5-brane particles $|\overline{\alpha}|$ is even. 

Note that whenever $|{\alpha}|=2$, i.e., only two states are SYM particles, the left integrand reduces to the square of a reduced Pfaffian, and the amplitude is equal to the D5-brane amplitude, though two particles carry color labels. Hence the first non-trivial amplitude in this mixed theory arises for $n=5$:
\be
\mathcal{A}^{\text{D5-brane } \oplus \text{ SYM}}_5 (345) = \frac{1}{4} s_{12} \left( s_{23}\, \mathcal{A}^{\N=(1,1) \text{ SYM}}_5 (12345) - s_{24}\, \mathcal{A}^{\N=(1,1) \text{ SYM}}_5 (12435) \right).
\ee
Here we used KLT to rewrite \eqref{D5-SYM-theory} in terms of the NLSM $\oplus$ $\phi^3$ amplitudes from \cite{Cachazo:2016njl} and 6D $\N=(1,1)$ SYM ones, and presented the final result in terms of the SYM amplitudes. Symmetry in labels $1,2$ and antisymmetry with respect to $3,4,5$ of the right-hand side follows from the BCJ relations \cite{Bern:2008qj}. Expressions for $5$-point SYM amplitudes can be found in \cite{Dennen:2009vk}.

The construction of these mixed amplitudes uniquely defines nontrivial interactions between the two sectors, as the amplitude given above illustrates.
It is a curious fact that these interactions have not yet been explored from a Lagrangian point of view. There are indications that the interactions implied by these amplitude constructions may have better soft behavior than any other possible interactions. This warrants further exploration.

\subsection{\label{sec:5d-theories}5D SYM and SUGRA}

Let us now consider 5D SYM and SUGRA with maximal supersymmetry. The spin of a massless particle in 5D is given by a $\text{Spin}(3) = \text{SU}(2)$ little-group representation. The appropriate spinor-helicity formalism can be conveniently obtained from the 6D one, with additional constraints, see for instance \cite{Wang:2015aua}. Concretely, a 5D massless momentum can be expressed
\be
p^{AB}= \lambda^A_a \lambda^B_b \epsilon^{ab}  \, .
\ee
This is identical to the 6D formula, but now there is only one kind of $\lambda^A_a$ due to the fact that the little-group consists of a single $\text{SU}(2)$, which can be identified with the diagonal subgroup of the  $\text{SU}(2)\times\text{SU}(2)$ little group in 6D. Of course, one still has to impose a further condition to restrict the momentum to 5D. The additional constraint that achieves this is 
\be \label{eq:addconstrain}
\Omega_{AB} \lambda^A_a \lambda^B_b \epsilon^{ab} =0 \, .
\ee
Here $\Omega_{AB}$ is the anti-symmetric invariant tensor of $\text{Spin}(4,1)$, which is a non-compact version of $\text{USp}(4)$. Here we choose $\Omega_{13}=\Omega_{24}=1$, and the other components of $\Omega_{AB}$ vanish for $A<B$.  Note that the antisymmetry of $\Omega_{AB}$ implies that $\Omega_{AB} \lambda^A_a \lambda^B_b = c\, \epsilon_{ab}$. Therefore (\ref{eq:addconstrain}) actually implies that $\Omega_{AB} \lambda^A_a \lambda^B_b=0$ for all $a,b=1,2$. This fact will be useful later. 

Having set up the kinematics, we are now ready to present the formulas for the scattering amplitudes of 5D theories. Let us begin with 5D maximal SYM theory. This theory has $\text{Spin}(5) = \text{USp}(4)$ R symmetry. The spectrum of an on-shell supermultiplet consists of a vector that transforms as $(\mathbf{3,1})$, spinors $(\mathbf{2,4})$, and scalars $(\mathbf{1,5})$. The bold-face integers label little-group and R symmetry representations. The on-shell superfield of the theory can be expressed, 
\be  \label{eq:spectrum_5DSYM}
\Phi(\eta)
=  \phi + \eta^I_a  \psi^{a }_I  + \epsilon_{IJ} \eta^I_a {\eta}^J_{b} A^{ab} + \epsilon^{ab} \eta^I_a {\eta}^J_{b} \phi_{IJ}  + (\eta^3)^I_a (\bar{\psi})^a_I + (\eta^4) \bar{\phi} \,. 
\ee
The index $I=1,2$ labels a doublet of an $\text{SU}(2)$ subgroup of the R symmetry group, whereas the entire little-group properties are manifest. This superfield is the dimensional reduction of the 6D on-shell superfield (\ref{eq:spectrum11}) obtained by removing all hats from 6D little-group indices. This works because the 5D $\text{SU}(2)$ little group corresponds to the diagonal subgroup of the 6D $\text{SU}(2)\times \text{SU}(2)$ little group. One consequence of this is that the 6D gluon reduces to the 5D gluon with three degrees of freedom and a scalar. Similarly, 5D amplitudes can be obtained directly from the 6D ones by making the substitution
\be \label{eq:6Dto5D}
\tilde{\l}^{\hat a}_{A} \rightarrow \Omega_{AB} \l^{a \, B} \, .
\ee
A 6D Lorentz contraction, such as ${V}^{A}\tilde{V}_{A}$, now is realized by the use of $\Omega_{AB}$, namely ${V}^{A}\tilde{V}_{A} \rightarrow \Omega_{AB}  {V}^{A} {V}^{B}$. For instance, the four-gluon amplitude is given by
\be
\mathcal{A}_4 (A_{a_1 b_1}, A_{a_2 b_2}, A_{a_3 b_3}, A_{a_4 b_4}) = \d^5 \left( \sum_{i=1}^4 p_i^{AB} \right) { \langle 1_{a_1} 2_{b_1} 3_{c_1} 4_{d_1}  \rangle \langle 1_{a_2} 2_{b_2} 3_{c_2} 4_{d_2}  \rangle \over s_{12}\, s_{23} } + {\rm sym.} \, ,
\ee
where the symmetrization is over the little-group indices of each gluon. 

This procedure gives the following color-ordered tree-level superamplitudes for 5D maximal SYM:
\be \label{eq:5DYM}
\boxed{ {\cal A}^{ {\rm 5D\,  SYM}}_n( \alpha ) = \int d\mu_n^{\text{5D}}  \; {\rm PT}( \alpha ) \; \left(
{ {\rm Pf}^{\prime} A_n  } \int d\Omega^{(8)}_F \right) .}
\ee
Here the 5D measure is defined as
\begin{align}
\int d\mu_{n \text{ even}}^{\text{5D}} = \int \frac{\prod_{i=1}^n d\sigma_i\, \prod_{k=0}^{m} d^8 \rho_k}{\vol( \slc_\sigma \times \slc_\rho)} \frac{1}{V_n^2}\; \D^{5D}_B ,
\end{align}
for even $n$, and
\begin{align}
\label{Eq:LinearMeasure-Odd-5D}
&\int d\mu_{n \,\text{odd}}^{\text{5D}} = \int \frac{\left(\prod_{i=1}^{n} d\sigma_i \prod_{k=0}^{m-1}d^{8}\rho_{k}\right)\, d^{4}\omega\, \langle \xi d\xi\rangle }{\vol \left( \slc_{\sigma} , \slc_{\rho} , \text{T} \right)}\, \frac{1}{V^{2}_n} \; \D^{5D}_B  \, ,
\end{align}
for odd $n$. The 5D delta-function constraints $\D^{\text{5D}}_B$ will be defined later. We see that the integration variables and symmetry groups are identical to those of 6D, and the same for the maps,
\be
\r^A_a(z) = \sum_{k=0}^m \r^A_{a\,k} \, z^k \, , \qquad 
\chi_a(z) = \sum_{k=0}^m \chi_{a\,k} \, z^k 
\ee
and similarly for conjugate ones. Here $m={n \over 2} -1$ or $m={n-1 \over 2} $ depending on whether $n$ is even or odd, and the highest coefficients factorize if $n$ is odd, namely $\r^A_{a\,m} = \omega^A \xi_a, \chi_{a\,m}= g\, \xi_a$ for $n=2m+1$. 

Let us now examine the 5D delta-function constraints $\D^{\text{5D}}_B$. We propose that the 5D conditions for the rational maps are given by
\be \label{eq:DB20}
\D^{\text{5D}}_B  = \prod^n_{i=1} \d^6 \left(p^{AB}_i
-  \frac{ \langle \r^{A} (\s_i) \r^{B} (\s_i) \rangle }{ \prod_{j\neq i} \s_{ij} } \right) 
 \prod^{n-1}_{j=1} \delta \left(  \frac{  \Omega_{AB} \langle\r^{A} (\s_j) \r^{B} (\s_j) \rangle }{ \prod_{l \neq j} \s_{jl} }  \right) \,, 
\ee 
where the first part is identical to the 6D version, and the second part imposes additional constraints to incorporate the 5D kinematic constraints (\ref{eq:addconstrain}). The constraints should only be imposed for $(n{-}1)$ particles, because the remaining one is then automatically satisfied due to momentum conservation. As in the case of 6D, momentum conservation and on-shell conditions are built into (\ref{eq:DB20}), so to compute the usual scattering amplitudes we should pull out the
corresponding delta functions, 
\be
\D^{\text{5D}}_B =\delta^5(\sum^n_{i=1} p_i^{AB}) \left( \prod^n_{i=1} \delta(p_i^2) \delta(\Omega_{AB} p_i^{AB}) \right)\, \hat{\D}^{5D}_B \, .
\ee
Note that besides the usual on-shell conditions $p_i^2=0$, there are additional conditions $\Omega_{AB}\, p_i^{AB}=0$ that one has to extract. 5D momentum conservation is now implemented by restricting, for instance, the Lorentz indices in the $\delta^5$-function to be $\{A,B\} \neq \{2,4\}$. Then the remaining constraints $\hat{\D}^{5D}_B$ are given by
\bea
\hat{\D}^{\text{5D}}_B &=& \prod^{n-1}_{j=1} \delta \left(  \frac{  \Omega_{AB} \langle\r^{A} (\s_j) \r^{B} (\s_j) \rangle }{ \prod_{l \neq j} \s_{jl} }  \right)   \prod^{n-2}_{i=1} \d^4 \left(p^{AB}_i
-  \frac{ \langle \r^{A} (\s_i)\, \r^{B} (\s_i) \rangle}{ \prod_{j\neq i} \s_{ij}  } \right)\nonumber \\
&\times& \d^3 \left(p^{AB}_n
-  \frac{\langle \r^{A} (\s_n)\, \r^{B} (\s_i)\rangle}{ \prod_{j\neq n} \s_{nj} } \right)  \prod^{n}_{i=1} p_i^{12} \, \left( { p^{14}_{n-1}
\over p_{n-1}^{12} } - { p^{14}_{n} \over p_{n}^{12} } \right) , 
\eea
where the $\d^4$-function has $\{A,B\}=\{1,2\},\{1,3\},\{1,4\},\{2,3\}$, and the $\d^3$-function has $\{A,B\}=\{1,2\},\{1,3\},\{1,4\}$. Of course, the final result is independent of the choices we make here. Altogether the number of independent of delta functions is $5n-6$, which matches with the number of integration variables (after modding out the symmetry factors). It is also straightforward to check that the formula has the correct power counting for the scattering amplitudes of 5D SYM. Finally, we remark that just like the rational maps in 6D, the 5D rational constraints also incorporate all $(n-3)!$ solutions because of the non-trivial summation over the little-group indices. 

The reduction of supersymmetry to lower dimensions is straightforward, and therefore the 5D fermionic measure, $d\Omega^{(8)}_F$, is almost identical to the 6D version, except that the fermionic maps $\chi^{a} (\s_i)$ and $\tilde{\chi}^{\hat{a}} (\s_i)$ now combine into $\chi^{Ia} (\s_i)$ (with $I=1,2$), just as the $\eta$'s and $\tilde{\eta}$'s combined to give $\eta^I$, as we discussed previously. The corresponding 5D fermionic delta functions are therefore given by
\be \label{eq:DF20}
\D^{(8)}_F  = \prod^n_{i=1} \d^{8}\left(q^{A I}_i
- \frac{ \langle \r^{A} (\s_i) \chi^{I} (\s_i) \rangle } { \prod_{j\neq i} \s_{ij} } \right)\, ,
\ee
whereas the fermionic on-shell conditions that have to be taken out for computing scattering amplitudes become, 
\be
\prod^n_{i=1} \delta^4 \left( \Omega_{AB} \lambda^A_{a, i} q^{BI}_i \right)\,.
\ee
As usual, these constraints allow one to introduce the Grassmann coordinates $\eta_{ia}^I$ by writing the supercharges in the form $\lambda \eta$.
Furthermore, the meaning of the factor denoted ${\rm Pf'}A_n$ in (\ref{eq:5DYM}) takes a different form depending on whether the number of particles is even or odd.  Recall that if $n$ is even, ${\rm Pf'}A_n$ is defined in (\ref{reduced-pfaffian}), whereas for odd $n$, it is given in (\ref{odd-matrix-A}). For both cases, the reduction to 5D is straightforward using (\ref{eq:6Dto5D}) and the discussion following it. We have carried out various checks of this formula by comparing it with explicit component amplitudes from Feynman diagrams; these analytically agree for $n=3,4$, and numerically agree up to $n=8$. 

Next we present the formula for the tree-level amplitudes of 5D maximal supergravity, which can be obtained either by a double copy of the 5D SYM formula or by a direct reduction of the 6D SUGRA formula. Either procedure gives the result
\be\label{eq:5DSUGRA}
\boxed{{\cal M}^{ {\rm 5D\,  SUGRA}}_n = \int d\mu_n^{\text{5D}}\, 
({ {\rm Pf}^{\prime} A_n  } )^2 \int d\Omega^{(16)}_F, }
\ee
with the fermionic measures and delta-functions all doubled up,
\be \label{eq:DF22}
\D^{(16)}_F  = \prod^n_{i=1} \d^{16}\left(q^{A I}_i
- \frac{ \langle \r^{A}(\s_i) \, \chi^{I} (\s_i) \rangle } { \prod_{j\neq i} \s_{ij} } \right)\, ,
\ee
where now $I=1,2,3,4$. This makes an $\text{SU}(4)$ subgroup of the $\text{USp}(8)$ R symmetry manifest. Again, the details of the formula depend on whether $n$ is even or odd. 

Finally, it is worth mentioning that there are analogous formulas for the superamplitudes of the world-volume theory of a D4-brane, which are nonzero only when $n$ is even. These can be obtained either as the dimensional reduction of a D5-brane world-volume theory or of an M5-brane world-volume theory. Using the 5D measures, the probe D4-brane amplitudes can be expressed as
\be
\mathcal{A}^{\text{D4{-}brane}}_n = \int d\mu^{\rm 5D}_n \,({\rm Pf}^{\prime} A_n)^2\,  \left( \, {\rm Pf}^{\prime} A_n \int d\Omega^{(8)}_F \right)  ,
\ee
where the number of particles, $n$, is always taken to be even.

\subsection{\label{sec:Coulomb-branch}$\mathcal{N}=4$ SYM on the Coulomb Branch}

A further application of our 6D formulas involves the embedding of 4D \emph{massive} kinematics into the 6D massless kinematics. In this approach, we view some components of the 6D spinors as 4D masses \cite{Bern:2010qa,Huang:2011um}. In the case of 6D $\mathcal{N}=(1,1)$ SYM, this procedure allows us to obtain amplitudes for 4D $\mathcal{N}=4$ SYM on the Coulomb branch.

\subsubsection{4D Massive Amplitudes from 6D Massless Ones}

Four-dimensional $\mathcal{N}=4$ SYM on the Coulomb branch can be achieved by giving vevs to scalar fields of the theory. For instance, in the simplest case, 
\be
\langle (\phi^{12})^I_J \rangle =\langle (\phi^{34})^I_J \rangle = v\,\delta^I_J \,,
\ee
other scalars have zero vev. Here ``$12$" and ``$34$" are $SU(4)$ R symmetry indices, whereas $I, J$ are color indices for the gauge group $\text{U}(M)$. So the vev spontaneously breaks the gauge group from $\text{U}(M{+}N)$ to $\text{U}(N)\times \text{U}(M)$, and the off-diagonal gauge bosons, which are bifundamentals of $\text{U}(N)\times \text{U}(M)$, denoted $W$ and $\overline{W}$, gain mass. In the simple example, given above, all of the masses are equal, with $m=g_{\rm YM}\,v$. One can consider more general situations with different masses, as our formulas will describe. There have been many interesting studies of $\mathcal{N}=4$ SYM on the Coulomb branch in the context of scattering amplitudes. For instance, the masses introduced by moving onto the Coulomb branch can be used as IR regulators \cite{Alday:2009zm, Henn:2010bk, Henn:2010ir}; one can also study the low-energy effective action by integrating out the masses, which has led to interesting supersymmetric non-renormalization theorems \cite{Chen:2015hpa, Bianchi:2016viy}. The subject we are interested in here is to study the tree-level massive amplitudes of $\mathcal{N}=4$ SYM on the Coulomb branch \cite{Craig:2011ws}.  

One can obtain 4D massive amplitudes from 6D massless ones via dimensional reduction. As discussed in \cite{Bern:2010qa}, 4D massive kinematics can be parametrized by choosing the 6D spinor-helicity coordinates to take the special form 
\be \label{eq:mass6D}
\l^A_a = \begin{pmatrix} 
- \kappa \mu_{\alpha} & \l_{\alpha} \\
\tilde{\l}^{\dot {\alpha} } & \tilde{\kappa} \tilde{\mu}^{\dot {\alpha}}
\end{pmatrix} \, , \quad \quad
\tilde{\l}_{A \hat {a}} = \begin{pmatrix} 
 \kappa' \mu^{\alpha} & \l^{\alpha} \\
-\tilde{\l}_{\dot {\alpha} } & \tilde{\kappa}' \tilde{\mu}_{\dot {\alpha}}
\end{pmatrix} \, ,
\ee
where 
\be \label{eq:kappa}
\kappa = {M \over \langle \l \mu \rangle} \, , \quad 
\tilde{\kappa} = { \widetilde {M} \over [ \l \mu ]} \, , \quad 
\kappa' = {  \widetilde {M} \over \langle \l \mu \rangle} \, , \quad 
\tilde{\kappa}'= {M \over [ \l \mu ]} \, ,
\ee
and $M \widetilde{M} = m^2$ is the mass squared. As usual, the indices $\alpha$ and $\dot\alpha$ label spinor representations of the 4D Lorentz group $\text{SL}(2,\C)$. With this setup, a 4D massive momentum is given by
\be
p_{\alpha \dot \alpha} = \l_{\alpha} \tilde{\l}_{\dot \alpha}  + \rho\, \mu_{\alpha}  \tilde{\mu}_{\dot \alpha} \, ,
\ee
with $\rho = \kappa \tilde{\kappa}  = \kappa' \tilde{\kappa}'$. We have decomposed a massive momentum into two light-like momenta, where $\mu_{\alpha}  \tilde{\mu}_{\dot \alpha}$ can be considered a reference momentum. 

$\mathcal{N}=4$ SYM on the Coulomb branch can be viewed as a dimensional reduction of 6D $\mathcal{N}=(1,1)$ SYM with massless particles. For instance, the four-point amplitude involving two massive conjugate $W$ bosons and two massless gluons, $A(W_1^+, \overline{W}_2^-, g_3^-, g_4^-)$ can be obtained from the 6D pure gluon amplitude, 
\be
A^{\text{6D YM}}_4(A^{+\widehat{+}}_1,A^{-\widehat{-}}_2, A^{-\widehat{-}}_3, A^{-\widehat{-}}_4) = { \langle 1^+ 2^- 3^- 4^-\rangle [1^{\widehat +} 2^{\widehat -} 3^{\widehat -} 4^{\widehat -} ] \over s_{12} \, s_{23} } \,.
\ee
Plugging in the massive spinors (\ref{eq:mass6D}), and using the identity, 
\be
 \langle 1_+ 2_- 3_- 4_-\rangle =- \tilde{\kappa}_2 [1 \mu ]  \langle 34\rangle \, , \quad
[1_{\widehat +} 2_{\widehat -} 3_{\widehat -} 4_{\widehat -} ]  =- \tilde{\kappa}'_2 [1 \mu ]  \langle 34\rangle \, ,
\ee
as well as the definition of $\kappa$ in \eqref{eq:kappa}, the result can be expressed as,
\be
A_4^{\text{6D}}(W_1^+, \overline{W}_2^-, g_3^-, g_4^-)=   { m^2 [1 \mu ]^2  \langle 34\rangle^2 \over  [2 \mu ]^2  s_{12} ( s_{23} - m^2 )}\, ,
\ee
which agrees with the result in \cite{Craig:2011ws}.  

Alternatively, one can choose a different way of parameterizing 4D massive kinematics, 
\be \label{eq:mass6D2}
\l^A_a = \begin{pmatrix} 
\l_{\alpha,1} & \l_{\alpha, 2} \\
\tilde{\l}^{\dot {\alpha}}_{\,1} & \tilde{\l}^{\dot {\alpha}}_{\,2}
\end{pmatrix} \, , \quad \quad
\tilde{\l}_{A \hat {a}} = \begin{pmatrix} 
\l^{\alpha}_{\,1} & \l^{\alpha}_{\,2} \\
\tilde{\l}_{\dot {\alpha}, 1} & \tilde{\l}_{\dot {\alpha}, 2}
\end{pmatrix} \, ,
\ee
where we split the Lorentz indices $A \Rightarrow \{\alpha, \dot{\alpha} \}$, and $1$ and $2$ are little-group indices of massive particles in 4D.  
The momentum and mass are given by
\be
p_{\alpha, \dot{\alpha} } = \l_{\alpha, a} \tilde{\l}_{\dot {\alpha}, b} \epsilon^{ab}\, , \quad 
 \l_{\alpha, a} {\l}_{{\beta}, b} \epsilon^{ab} = M \epsilon_{ \alpha \beta } \, , \quad 
\tilde {\l}_{\dot{\alpha}, a} \tilde{\l}_{ \dot{\beta}, b} \epsilon^{ab} = {M} \epsilon_{ \dot{\alpha} \dot{\beta} } \, ,
\ee
with $M^2 = m^2$. The advantage of this setup is that it makes the massive 4D little group $\text{Spin}(3) = \text{SU}(2)$ manifest. In fact, it actually leads to the massive spinor-helicity formalism of the recent work \cite{Arkani-Hamed:2017jhn}, which one can refer to for further details. In this formalism, for instance, 
\be
A^{\text{6D}}_4 (W_1^{ab}, \overline{W}_2^{cd}, g_3^-, g_4^-)=   {([1_{a}  2_{c} ][1_{b}  2_{d} ])  \langle 34\rangle^2 \over s_{12}\, ( s_{23} - m^2 )} + {\rm sym} \, ,
\ee
and
\be \label{eq:2W-2g}
A^{\text{6D}}_4 (W_1^{ab}, \overline{W}_2^{cd}, g_3^+, g_4^-)=   { (\langle 1_{a} 4 \rangle [2_{c} 3] -  \langle 2_{c} 4 \rangle [1_{a} 3])(\langle 1_{b} 4 \rangle [2_{d} 3] -  \langle 2_{d} 4 \rangle [1_{b} 3]) \over s_{12} \,( s_{23} - m^2 )} + {\rm sym}\, , 
\ee
where $a, b$ and $c,d$ are $\text{SU}(2)$ little-group indices of the massive particles $W_1^{ab} = W_1^{ba}$ and $\overline{W}_2^{cd} = \overline{W}_2^{dc}$, respectively. The notation ``$+\, {\rm sym}$'' means that one should symmetrize on the little-group indices of each massive W boson.  Here  we have also defined
\be
 [1_a  2_b ] = \tilde{\l}_{1, \dot{\alpha}, a } \tilde{\l}_{2, \dot{\beta}, b }  \epsilon^{\dot{\alpha} \dot{\beta}} \, , \quad 
 \langle 1_a  2_b \rangle = {\l}_{1, {\alpha}, a } {\l}_{2, {\beta}, b }  \epsilon^{{\alpha} {\beta}} \, ,
\ee
for massive spinors. Note if $a \neq b$, they vanish in the massless limit which sets $\l_{ \dot{\alpha}, + } = \tilde{\l}_{\dot{\alpha}, -}=0$. While if $a=b$, they reduce to the usual spinor brackets for 4D massless particles.
Clearly, this formalism is very convenient for massive amplitudes, as was emphasized in \cite{Arkani-Hamed:2017jhn}. 

\subsubsection{Massive SUSY}

Amplitudes for 4D $\mathcal{N}=4$ SYM on the Coulomb branch can be constructed using the massive spinor-helicity formalism. Recall that the 16 supercharges of 
a particle in 6D $(1,1)$ SYM can be expressed in the form 
\bea
q^{A} = \l^A_a \eta^a \, , \quad {\overline q}^{A} = \l^A_a {\partial \over \partial {\eta_a}} \, , \\
\tilde{q}_{A} = \tilde{\l}_{A {\hat a}} \tilde{\eta}^{\hat a} \, , \quad {\overline { \tilde{q} }}_{A} = \tilde{\l}_{A {\hat a}} {\partial \over \partial \tilde{\eta}_{\hat a} } .
\eea
The reduction to the supercharges of a 4D massive particle can be obtained using (\ref{eq:mass6D2}),  
\bea
q^{I \alpha} = \l^{\alpha}_- \eta^I_+  - \l^{\alpha}_+ \eta^I_- \, , \quad {\overline q}^{I \dot{\alpha} } = \tilde{\l}_+^{\dot{\alpha} }  {\partial \over \partial {\eta^I_+}}  + \tilde{\l}^{\dot{\alpha}}_- { \partial \over \partial {\eta^I_-}  } \, , \\
\tilde{q}^I_{\alpha}  = {\l}_{\alpha -} { \partial \over \partial {\eta^I_-}  }  + \l_{\alpha +} { \partial \over \partial {\eta^I_+}  } \, , \quad {\overline { \tilde{q} }}_{\dot \alpha} = \tilde{\l}_{\dot \alpha +} \eta^I_- -  \tilde{\l}_{\dot \alpha -} \eta_+^I \, ,
\eea
where we have identified $\{\eta, \tilde{\eta}\}$ as $\eta^I$ with $I=1,2$. Their anti-commutation relations are 
\bea
\{ q^{I \alpha}, \tilde{q}^{J \beta} \} =  M \e^{IJ} \e^{\alpha \beta} \, , \quad 
\{ {\overline q}^{I  {\dot \alpha}}, \overline{\tilde{q}}^{J  {\dot \beta} } \}  = {M} \e^{IJ} \e^{\alpha \beta},  \\
\{ q^{I \alpha}, \overline{q}^{J \dot{\alpha} } \} = \e^{IJ} p^{\alpha \dot{\alpha} } \, , \quad 
\{ {\tilde q}^{I  { \alpha}}, \overline{\tilde{q}}^{J  {\dot \beta} } \} = -\e^{IJ}p^{\alpha \dot{\alpha} }  .
\eea
The central charge $Z$ satisfies $Z^2= M^2 =m^2$, which reflects the fact that the $W$'s of $\mathcal{N}=4$ SYM on the Coulomb branch are BPS. To reduce to the massless case, one sets $ \l^{\alpha}_+= \tilde{\l}_{\dot \alpha -}  =0$ and identifies $ \l^{\alpha}_+= \l^{\alpha}$ and $\tilde{\l}_{\dot \alpha -} =\tilde{\l}_{\dot \alpha }$. That is, of course, the familiar (super) spinor-helicity formalism for $\mathcal{N}=4$ SYM at the origin of moduli space. With the introduction of supersymmetry, a massive supermultiplet of $\mathcal{N}=4$ SYM on the Coulomb branch can be neatly written as
\be
\Phi(\eta)
=  \phi + \eta^I_a  \psi^a_I + \epsilon_{ab} \eta^{Ia} \eta^{Jb} \phi_{IJ} +\epsilon_{IJ}  \eta^I_a {\eta}^J_{b} A^{a b}
+  ({\eta})^2 \eta^J_a  \bar{\psi}^a_{J}  + (\eta)^4 \bar{\phi} \, ,
\ee
which contains one vector, four fermions, and five scalars. One scalar has been eaten by the vector, compared to the massless case with six scalars. 

We can also express the massive amplitudes supersymmetrically. For instance, the superamplitude for the four-point amplitude with a pair of conjugate W-bosons, considered previously, can be written as 
\be
A_4 = {\delta^4_F \, \tilde{\delta}^4_F \over s_{12} (s_{23} - m^2)} \,,
\ee 
with the fermionic delta-functions given by
\bea
\delta^4_F &=& \delta^4( \l^{\a}_{1a} \eta^{Ia}_1+ \l^{\a}_{2a} \eta^{Ia}_2 + \l^{\a}_{3} \eta^{I,-}_3 + \l^{\a}_{4} \eta^{I,-}_4) \, , \\
\tilde{\delta}^4_F &=& \delta^4(  \tilde{\l}^{\dot \a}_{1a} {\eta}^{Ia}_1+  \tilde{\l}^{\dot \a}_{2a}  {\eta}^{Ia}_2 +  \tilde{\l}^{\dot \a}_{3}  {\eta}^{I,+}_3 +  \tilde{\l}^{\dot \a}_{4}  {\eta}^{I,+}_4) \, .
\eea
These delta functions make the conservation of eight supercharges manifest.

\subsubsection{Massive Amplitudes on the Coulomb Branch of $\mathcal{N}=4$ SYM}

Having set up the 4D massive kinematics and supersymmetry, we are ready to write down a general Witten--RSV formula for 4D scattering amplitudes of $\mathcal{N}=4$ SYM on the Coulomb branch by a simple dimensional reduction of 6D massless $\mathcal{N}=(1,1)$ SYM. The formula is 
\be \label{eq:N4SYMCol}
\boxed{\mathcal{A}^{\N=4 \text{ SYM CB}}_n (\alpha) = \int d\mu^{\text{CB}}_n \; {\rm PT}(\alpha) \left(\, { {\rm Pf'} A_n } \int d\Omega^{(4),\text{CB}}_{F} \right).}
\ee
The measure $d\mu^{\text{CB}}_n$ is obtained directly from the 6D massless one with the following replacement of the bosonic delta functions:
\be
\D_B \rightarrow \prod^n_{i=1} \delta^4 \left( p_i^{ \a \dot \a } - { \langle \r^{\a} (\s_i)\, \tilde{\r}^{\dot \a} (\s_i)\rangle \over \prod_{j\neq i} \s_{ij}  } \right)  \delta\left( M_i - { \langle \r^{1} (\s_i) {\r}^{2} (\s_i) \rangle \over \prod_{j\neq i} \s_{ij} } \right)  \delta\left( \widetilde{M}_i - { \langle \tilde{\r}^{\dot 1} (\s_i) \tilde{\r}^{{ \dot 2}} (\s_i) \rangle \over \prod_{j\neq i} \s_{ij} } \right) ,
\ee 
and using massive kinematics of (\ref{eq:mass6D2}), we set $\widetilde{M}_i={M}_i$ for $i=1,2,\ldots,n-1$, where ${M}_i^2=m_i^2$ is the mass squared of the $i$th particle ($\widetilde{M}_n = M_n$ is a consequence of 6D momentum conservation). The mass $m^2_i$, is $m^2_W$ or $0$, as appropriate, for the simple symmetry breaking pattern described previously. Similarly, for the fermionic part
\be
\D_F \rightarrow \prod^n_{i=1} \delta^4\left(q_i^{\a, I}  -  { \langle {\rho}^{\a}(\sigma_i)\, \chi^{I} (\sigma_i) \rangle  \over \prod_{j\neq i} \s_{ij} } \right) \, , \quad 
\tilde{\D}_F \rightarrow \prod^n_{i=1} \delta^4\left(\tilde{q}_i^{{\dot \a}, I}  -  { \langle \tilde{\rho}^{\dot \a}(\sigma_i) \,\chi^{I} (\sigma_i) \rangle  \over \prod_{j\neq i} \s_{ij} } \right) ,
\ee
where the supercharges $q_i^{\a, I}$ and $\tilde{q}_i^{{\dot \a}, I}$ are defined in the previous section. 

The polynomial maps are defined as usual, 
\be
{\rho}^{\a}_{a}(z)  = \sum_{k=0}^{m} {\rho}^{ \a}_{k, a} \, z^k \, ,\quad
\tilde{\rho}^{\dot \a}_{a}(z) = \sum_{k=0}^{m} \tilde{\rho}^{\dot \a}_{k,a} \, z^k \, , \quad
\chi^{I}_a (z) = \sum_{k=0}^{m} {\chi}^{I}_{k,a} \, z^k \, .
\ee
They can be understood as a reduction from the 6D maps, 
\be \label{eq:rho_reduction}
{\rho}^{A}_{a}(z) = \begin{pmatrix} 
{\rho}_{\alpha,1}(z) & {\rho}_{\alpha, 2}(z) \\
\tilde{{\rho}}^{\dot {\alpha}}_1(z) & \tilde{{\rho}}^{\dot {\alpha}}_2(z)
\end{pmatrix}\, , \quad \quad
\tilde{{\rho}}_{A \hat {a}}(z) = \begin{pmatrix} 
{\rho}^{\alpha}_{1}(z) & {\rho}^{\alpha}_{2}(z) \\
\tilde{{\rho}}_{\dot {\alpha}, 1}(z) & \tilde{{\rho}}_{\dot {\alpha}, 2}(z)
\end{pmatrix} \, .
\ee
Again, we have to treat amplitudes with $n$ even and $n$ odd differently. So $n = 2m+2$ or $n= 2m+1$ if $n$ is even or odd, and the highest coefficients in the maps take the factorized form if $n$ is odd. 

The factor ${\rm Pf'} A_n$ in the integrand is defined differently depending on whether $n$ is even or odd, but they are straightforward reductions from 6D ones. For instance, we find that the odd-point Pfaffian can be constructed with the additional vector
\begin{equation}
p^{\alpha \dot{\alpha}}_{*}=\frac{2\, r^{\alpha}\,\tilde{\rho}^{\dot{\alpha}}_a(\sigma_*)\langle\rho^a(\sigma_*),r\rangle}{(\xi_b \langle\rho^b(\sigma_*),r\rangle)^2}\,,\qquad m_{*}=0\,.
\end{equation}
This is obtained from \eqref{eq:pstar} by splitting the 6D spinor index $A$ into 4D ones $\alpha, \dot{\alpha}$ according to (\ref{eq:rho_reduction}), and choosing the reference spinors as $q^A=(r_{\alpha};0)$, $\tilde{q}_A=(r^{\alpha};0)$. The same manipulations are required for the description of the $n$ scattered particles, according to (\ref{eq:mass6D2}).

If the amplitudes involve massless external particles, we set $m_i =0$ for them. The massive particle masses should satisfy the conservation constraint $\sum_i m_i=0$, which is imposed by the rational maps automatically. Note that it is necessary to keep track of the signs of masses, even though the inertial mass is always $|m|$. This would be the only condition for a general 4D theory obtained by dimensional reduction. Specifying the particular Coulomb branch of $\mathcal{N} = 4$ SYM requires that we impose further conditions. In the simplest cases, where all of the massive particles have the same mass, we have $m_i^2=m^2_{W}$ for all $i$, but all the W bosons have mass $m_W$, whereas all the $\overline{W}$'s have mass $-m_W$. More generally, different masses can be assigned to different massive particles, but if we assign $m$ as the mass of a W boson, then we should then assign $-m$ to the corresponding conjugate $\overline{\rm W}$ boson. Therefore $\sum_i m_i=0$ is satisfied in pairs. Finally, due to the color structure, a W boson and its conjugate $\overline{\rm W}$ boson should appear in adjacent pairs with gluons sandwiched in between. For instance, there are nontrivial amplitudes of the type $A_n(W_1, g_2, \ldots, g_{i-1}, \overline{W}_i, \tilde{g}_{i+1}, \ldots, g_n)$, with gluons $g$ and $\tilde{g}$ belonging to the gauge groups $\text{U}(N)$ and $\text{U}(M)$, respectively. 

We checked that the formula produces correct four-point amplitudes in previous section. It also gives correct five- and six-point ones such as
\bea
A_5 (g^+_1,g^+_2,g^+_3, W^{ab}_4,\overline{W}^{cd}_5)
&=& { \langle 4_{a} 5_{c}\rangle \langle 4_{b} 5_{d}\rangle[1|p_5 (p_1+p_2)|3] \over \langle 12 \rangle \langle 23 \rangle (s_{51} - m^2)  ( s_{34} - m^2)} + {\rm sym} \, ,\\
A_6 (g^+_1,g^+_2,g^+_3,g^+_4, W^{ab}_5,\overline{W}^{cd}_6)
&=& { \langle 5_{a} 6_{c}\rangle \langle 5_{b} 6_{d}\rangle [1|p_6 (p_1+p_2)(p_3+p_4)p_5|4] \over \langle 12 \rangle \langle 23 \rangle \langle 34 \rangle  (s_{61} - m^2) ( s_{612} -m^2 ) ( s_{45} - m^2)} + {\rm sym} \, ,  \nonumber
\eea
or $\text{SU}(4)$ R symmetry-violating amplitudes that vanish in the massless limit, such as
\bea
A_5 (\phi^{34}_1,\phi^{34}_2,\phi^{34}_3,W^{ab}_4,\overline{W}^{cd}_5)
&=&-  { m \, \langle 4_a 5_c\rangle [4_b 5_d]  \over  (s_{51} - m^2) ( s_{34} - m^2)} + {\rm sym} \, , \\
A_6 (\phi^{34}_1,\phi^{34}_2,\phi^{34}_3,\phi^{34}_4, W^{ab}_5,\overline{W}^{cd}_6)
&=&-  { m^{2} \langle 5_a 6_c\rangle [5_b 6_d]  \over  (s_{61} - m^2) ( s_{612} -m^2 ) ( s_{45} - m^2)} + {\rm sym} \, .
\eea
When restricted to W bosons with helicity $\pm 1$, they are all in agreement with the results in \cite{Craig:2011ws}, but now in a form with manifest $\text{SU}(2)$ little-group symmetry for the massive particles. One can also consider cases in which the massive particles are not adjacent, for instance
\begin{align}
A_4(W^{ab}_1, g_2^+, \overline{W}^{cd}_3, \tilde{g}^+_4) &= { \langle 1_a  3_c \rangle \langle 1_b  3_d \rangle [2 4]^2 \over (s_{12}-m^2)(s_{23} - m^2)  } + {\rm sym} \,, \\
A_4(W^{ab}_1, g_2^-, \overline{W}^{cd}_3, \tilde{g}^+_4) &= { (\langle 1_a  2 \rangle [3_c 4]-\langle 3_c  2 \rangle [1_a 4]) (\langle 1_b  2 \rangle [3_d 4]-\langle 3_d  2 \rangle [1_b 4]) \over (s_{12}-m^2)\, (s_{23} - m^2)  }+ {\rm sym} \,.
\end{align}

\subsection{Reduction to Four Dimensions: Special Sectors} \label{sec:massless-4D}

One can further reduce our 6D formulas down to 4D massless kinematics. It is interesting that 4D kinematics induces a separation into sectors, as reviewed in Section~\ref{sec:4d_rational}, whereas there is no natural separation into sectors in higher dimensions. In fact, one of the motivations for developing formulas in 6D is to unify all of the 4D sectors. Here we will explain how to naturally obtain the integrand of 4D theories from 6D via dimensional reduction in the middle ($d=\tilde{d}$) and ``next to middle'' ($d=\tilde{d}\pm1$) sectors for even and odd multiplicity, respectively. However, the emergence of the other sectors is more difficult to see via dimensional reduction, even though all sectors are present. We will comment on this at the end of this subsection. 

For the first case, it was already argued in \cite{Heydeman:2017yww} that the 6D measure for rational maps reduces to the corresponding 4D measure
provided the maps behave regularly under the dimensional reduction, i.e.,
they reduce to the ones appearing in the Witten--RSV formula. After reviewing the reduction for $n$ even, we will generalize the
argument to odd $n$ for the near-to-middle sectors, i.e., $d=\tilde{d}\pm1$.\footnote{One can input kinematics $\{p_{i}^{\text{\ensuremath{\mu}}}\}$ in $D=4+\epsilon$ dimensions and study the behaviour of the 6D maps
as $\epsilon\rightarrow0$. We find that when the solution corresponds
to the aforementioned sectors the maps are regular. This implies that
the measure is finite and reproduces the CHY measure of Section~\ref{sec:rational-maps}, valid for both 6D and 4D. For other sectors the maps become divergent and additional care is needed to define the limit of the measure.} 

Let us first consider the even-point case, $n=2m +2$. For the solutions corresponding to the middle sector $d=\tilde{d}=m$, the maps behave as follows \cite{Heydeman:2017yww}:
\be\label{4dbeh}
\rho_{a}^{A}(z)\rightarrow\left(\begin{array}{cc}
0 & \rho_{\alpha}(z)\\
\tilde{\rho}^{\dot{\alpha}}(z) & 0
\end{array}\right)\,,\qquad\tilde{\rho}_{A \hat a}(z)\rightarrow\left(\begin{array}{cc}
0 & \rho^{\alpha}(z)\\
\tilde{\rho}_{\dot{\alpha}}(z) & 0
\end{array}\right),
\ee
where ${\rm deg}\, \rho_{\alpha}(z) = {\rm deg}\, \tilde{\rho}_{\dot{\alpha}}(z)=d$.
Here we have used the 4D embedding described in \cite{Cheung:2009dc},
with the analogous behaviour for the kinematic data $\lambda_{a}^{A}$
and $\tilde{\lambda}_{A\hat{a}}$. This corresponds to setting $p_{i}^{AB}=0$
for $\{A,B\}=\{1,2\},\{3,4\}$. Note further that the action of the
subgroup $\text{GL(1,\ensuremath{\mathbb{C}})}\subset\text{SL}(2,\mathbb{C})\times\text{SL}(2,\mathbb{C})$
is manifest and given by $\rho_{-}^{A}\rightarrow\ell\rho_{-}^{A}$,
$\rho_{+}^{A}\rightarrow\frac{1}{\ell}\rho_{+}^{A}$, etc. Consider
now the fermionic piece of the SYM integrand in \eqref{eq:(1,1)SYM-even}:
\be
V_{n}\int\prod_{k=0}^{d}d^{2}\chi_{k}d^{2}\tilde{\chi}_{k}\prod_{i=1}^{n}\delta^{4}\left(q_{i}^{A}-\frac{\rho_{a}^{A}(\sigma_{i})\chi^{a}(\sigma_{i})}{\prod_{j\neq i}\sigma_{ij}}\right)\delta^{4}\left(\tilde{q}_{i,A}-\frac{\tilde{\rho}_{A,\hat{a}}(\sigma_{i})\tilde{\chi}^{\hat{a}}(\sigma_{i})}{\prod_{j\neq i}\sigma_{ij}}\right).
\ee
Under the embedding \eqref{4dbeh} this becomes 
\begin{align}
V_{n}\int\prod_{k=0}^{d}d\chi_{k}^{+}d\chi_{k}^{-}d\tilde{\chi}_{k-}d\tilde{\chi}_{k+}\times\prod_{i=1}^{n}\delta^{2}\left(q_{i\alpha}^{1}-\frac{\rho_{\alpha}(\sigma_{i})\chi^{-}(\sigma_{i})}{\prod_{j\neq i}\sigma_{ij}}\right)\delta^{2}\left(\tilde{q}_{i}^{\dot{\alpha}1}-\frac{\tilde{\rho}^{\dot{\alpha}}(\sigma_{i})\chi^{+}(\sigma_{i})}{\prod_{j\neq i}\sigma_{ij}}\right)\nn\\
\qquad\times \, \delta^{2}\left(\tilde{q}_{i\dot{\alpha}}^{2}-\frac{\tilde{\rho}_{\dot{\alpha}}(\sigma_{i})\tilde{\chi}_{-}(\sigma_{i})}{\prod_{j\neq i}\sigma_{ij}}\right)\delta^{2}\left(q_{i}^{\alpha2}-\frac{\rho^{\alpha}(\sigma_{i})\tilde{\chi}_{+}(\sigma_{i})}{\prod_{j\neq i}\sigma_{ij}}\right),
\end{align}
where we have labeled $q^{A}=(q_{\alpha}^{1};\tilde{q}^{\dot{\alpha}1})$
and $\tilde{q}_{A}=(\tilde{q}_{\dot{\alpha}}^{2};q^{\alpha2})$. We
can now identify the 4D fermionic degrees of freedom as 
\be
\tilde{\chi}^{\hat{I}}=(\chi^{-},\tilde{\chi}_{+})\,,\qquad\chi^{I}=(\chi^{+},\tilde{\chi}_{-}),
\ee
with $I=1,2$ and $\hat{I}=1,2$ transforming under the manifest $\text{SU}(2)\times\text{SU}(2)\subset\text{SU}(4)$ R symmetry group in 4D. Hence, the fermionic piece is
\be
\int d\Omega_{F}=V_{n}\int\prod_{k=0}^{d}d^{2}\chi_{k}^{I}d^{2}\tilde{\chi}_{k}^{\hat{I}}\, \times\, \prod_{i=1}^{n}\delta^{4}\left(q_{i}^{\alpha\hat{I}}-\frac{\rho^{\alpha}(\sigma_{i})\tilde{\chi}^{\hat{I}}(\sigma_{i})}{\prod_{j\neq i}\sigma_{ij}}\right)\delta^{4}\left(\tilde{q}_{i}^{\dot{\alpha}I}-\frac{\tilde{\rho}^{\dot{\alpha}}(\sigma_{i})\chi^{I}(\sigma_{i})}{\prod_{j\neq i}\sigma_{ij}}\right)\, .
\ee

The remaining part of the even-multiplicity integrand is trivially
reduced to four dimensions, since the matrix $[A_n]_{ij}=\frac{p_i \cdot p_j}{\sigma_{ij}}$ is not sensitive to any specific dimension. Alternatively, it can be seen that under the embedding \eqref{4dbeh} and the support of the bosonic delta functions \cite{Heydeman:2017yww}
\be
V_{n} \, \text{Pf}^{\prime}A_n\rightarrow R^{d}(\rho) \, R^{\tilde{d}}(\tilde{\rho}) \, .
\ee
Let us now derive the analogous statement for $n=2m+1$. We assume $\tilde{d}=d-1$ (with the case $\tilde{d}=d+1$ being completely analogous). The embedding \eqref{4dbeh} can then be obtained by fixing the components $\xi=\tilde{\xi}=(1,0)$ and $\zeta=\tilde{\zeta}=(0,1)$ for the odd-point maps (recall that we defined $\{\xi,\zeta\}$ as an $\slc_{\rho}$ basis). For the fermionic part we again introduce two polynomials $\chi^{I}(\sigma)$ and $\tilde{\chi}^{\hat{I}}(\sigma)$ of degrees $d$ and $\tilde{d}$. The top components of the polynomial $\chi^{I}$ can be identified as
\be
(\chi_{d}^{1},\chi_{d}^{2})=(\chi_{d}^{+},\tilde{\chi}_{-})=(g,\tilde{g})
\ee
according to \eqref{fermaps1} and \eqref{fermaps2}. The bosonic part of the integrand becomes
\begin{align}
\text{Pf}'\widehat{A}_{n} & = \frac{1}{\sigma_{n1}}\sum_{i=2}^{n-1}\frac{(-1)^{i}}{\sigma_{ni}}\frac{[q|P(\sigma_{i})|\tilde{\rho}_{\hat{a}}(\sigma_{n})]\zeta^{\d \hat{a}}}{[q|\rho_{a}(\sigma_{n})\rangle\xi^{a}}\text{Pf}\,A^{[1in]}\\
 & = \frac{1}{\sigma_{n1}}\sum_{i=2}^{n-1}\frac{(-1)^{i}}{\sigma_{ni}}\frac{[qi]\langle i\mbox{\ensuremath{\rho}}(\sigma_{n})\rangle}{[q\tilde{\rho}(\sigma_{n})]}\text{Pf}\,A^{[1,i,n]}.
\end{align}
We have checked numerically up to $n=7$ that this expression
coincides with $V_{n}^{-1}R^{d}(\rho)R^{\tilde{d}}(\tilde{\rho})$
for $\tilde{d}=d-1$ on the support of the 4D equations \eqref{4d-scattering-maps}. Hence for this sector ($d=\tilde{d}$ for even $n$ or  $d=\tilde{d} \pm 1 $ for odd $n$) the integrand can be recast into the non-chiral form
of the Witten--RSV formula, and the amplitude is given by \cite{Cachazo:2013iaa}:
\bea
\mathcal{A}^{\N=4 \text{ SYM}}_{n,d} &=& \int \mu_{n, d}^{\rm 4D} \, R^{d}(\rho)R^{\tilde{d}}(\tilde{\rho})\int\prod_{k=0}^{d}d^{2}\chi_{k}^{I}\prod_{k=0}^{\tilde{d}}d^{2}\tilde{\chi}_{k}^{\hat{I}} \nonumber \\ && \times \, \prod_{i=1}^{n}\delta^{4}\left(q_{i}^{\alpha\hat{I}}-\frac{\rho^{\alpha}(\sigma_{i})\tilde{\chi}^{\hat{I}}(\sigma_{i})}{\prod_{j\neq i}\sigma_{ij}}\right)\delta^{4}\left(\tilde{q}_{i}^{\dot{\alpha}I}-\frac{\tilde{\rho}^{\dot{\alpha}}(\sigma_{i})\chi^{I}(\sigma_{i})}{\prod_{j\neq i}\sigma_{ij}}\right).
\eea

Let us finally comment on other sectors. First of all, given the fact that the 6D rational maps contain all $(n-3)!$ solutions, it is clear that all the sectors are there. One can see it by considering completely integrating out all the moduli $\rho$'s, then reducing the 6D formulas to 4D will not be different from the dimensional reduction of the original CHY formulations. However, from the procedure outlined above, it is subtle to see how other sectors emerge directly by dimensional reduction. As we will discuss in Section~\ref{sec:degenerate_kinematics}, this subtlety is closely related to the fact that both ${\rm Pf}^{\prime}A_n$ (for even $n$) and ${\rm Pf}^{\prime}\widehat{A}_n$ (for odd $n$) vanish for the kinematics of the non-middle sectors (for even $n$) and the non next-to-middle sectors (for odd $n$).

\section{\label{sec:outlook}Conclusion and Discussion}

In this work we presented new connected formulas for tree-level scattering amplitudes of 6D $\N=(1,1)$ SYM theory as well as for $\N=(2,2)$ SUGRA via the KLT double-copy procedure. Due to the peculiar properties of 6D spinor-helicity variables, scattering amplitudes of even and odd number of particles must be treated differently. In the case of even multiplicity, our formulas are direct extensions of the results for the world-volume theory of a probe D5-brane \cite{Heydeman:2017yww}. By considering a soft limit of the even-point formulas we obtained the rational maps and the integrands for odd multiplicity, with many interesting features and novelties. In particular, a new redundancy, which we call T-shift symmetry, emerges for the odd-point worldsheet formulas. Interestingly, the T shift intertwines with the original M\"obius $\slc_{\sigma}$ and $\slc_{\rho}$ redundancies. Another new feature is the generalized Pfaffian ${\rm Pf}^{\prime} \widehat{A}_n$ in the integrand. Besides the original $n$ punctures, it contains an additional reference puncture, which can be set to an arbitrary value. Associated to the new puncture there is a special ``momentum" vector. The special vector is used to increase the size of the original matrix ${A}_n$ to $(n{+}1)\times (n{+}1)$ such that it has a non-vanishing reduced Pfaffian for odd $n$. Moreover, since the special null vector $p_\star$ has zero mass dimension, ${\rm Pf}^{\prime} \widehat{A}_n$ has the correct mass dimension for Yang--Mills amplitudes. It would be of great interest to better understand the physical origin of the additional puncture and the additional special vector. One clear future direction is to obtain an ambitwistor model that realizes all of these new features of the odd-multiplicity connected formulas. 

We also presented the 6D formulas in alternative forms, with constraints linearly in terms of the 6D external helicity spinors. They are a direct analog of the Witten--RSV formulations for 4D $\mathcal{N}=4$ SYM. By integrating out the moduli of maps, the linear maps can be further recast into a form with a symplectic Grassmannian structure. The symplectic Grassmannian is realized in terms of 6D version of the Veronese maps. 

Having obtained formulas for 6D theories, we also considered their dimensional reduction to 5D and 4D leading to various new connected formulas. By reducing to 5D for massless kinematics and utilizing the 5D spinor-helicity formalism, we obtained new formulas for 5D SYM and SUGRA theories. Reduction to 4D reproduced the original Witten--RSV formula for $\mathcal{N}=4$ SYM in 4D for the middle helicity sector for even $n$ and the next-to-middle sector for odd $n$. The appearance of other disconnected sectors for 4D kinematics is rather subtle, and we leave it for future investigations. On the other hand, it is very nice that reduction to 4D massive kinematics turns out to be more straightforward without such subtleties. By doing so, we deduced a connected formula for massive amplitudes of 4D $\mathcal{N}=4$ SYM on the Coulomb branch. 

Another natural future application of our 6D formulas would be to use the procedure of forward limits in \cite{He:2015yua, Cachazo:2015aol} to obtain the one-loop integrand of 4D $\mathcal{N}=4$ SYM. Since now we have manifestly supersymmetric formulas for amplitudes in 5D and 6D, we are in a good position to apply the forward limit procedure of \cite{Cachazo:2015aol} supersymmetrically. This procedure might lead to an analog of the Witten--RSV formulas at loop level, which might be genuinely different from previous formulations \cite{Geyer:2015bja,Geyer:2015jch,Cachazo:2015aol,Geyer:2016wjx,Geyer:2018xwu}. We leave this as a future research direction. 

Even though 6D $\mathcal{N}=(1,1)$ SYM is not a conformal theory, its planar scattering amplitudes enjoy a dual conformal symmetry \cite{Dennen:2010dh} just like $\mathcal{N}=4$ SYM in 4D \cite{Drummond:2006rz, Drummond:2008vq}. Such hidden symmetries are often obscured in traditional ways of representing the amplitudes (such as Feynman diagrams), and become more transparent in modern formulations, such as the Grassmannian \cite{ArkaniHamed:2009dn, ArkaniHamed:2012nw}, as shown in \cite{ArkaniHamed:2009vw}. It would be of interest to investigate whether our 6D $\mathcal{N}=(1,1)$ SYM formulas, especially the version in terms of the Veronese maps or its ultimate symplectic Grassmannian form, can make dual conformal symmetry manifest.   

Having succeeded in using the spinor-helicity formalism to study supersymmetric theories in 6D, it is tempting to try to carry out analogous constructions in even higher dimensions where supersymmetric theories still exist, such as ten or eleven. The main challenge is that in $D>6$ one has to impose non-linear constraints on the spinors. Not long after the 6D spinor-helicity formalism was developed, a proposal for a 10D version was introduced \cite{CaronHuot:2010rj}, also see recent work \cite{Bandos:2016tsm, Bandos:2017eof} for 10D and 11D theories. It would be interesting to pursue this line of research further.    

Finally, there are two issues that are very natural open questions and deserve a detailed discussion. The first has to do with a mysterious but natural object that has 6D $\mathcal{N}=(2,0)$ symmetry and a non-abelian structure similar to that of Yang--Mills. The second is related to the mathematical characterization of the moduli space of maps from $\mathbb{CP}^1$ to the null-cone in six dimensions. 

\subsection*{Non-abelian $\N=(2,0)$ Formula}
\label{sec:nonab(2,0)}

As discussed in Section~\ref{sec:even_points}, the 6D $\mathcal{N}=(1,1)$ non-abelian SYM amplitudes for even $n$ can be obtained from those of the D5-brane theory by replacing $({\rm Pf}^{\prime} A_n)^2$ with the Parke--Taylor factor ${\rm PT}(\a)$. It is natural to ask what happens if we apply the same replacement to the M5-brane formula \cite{Heydeman:2017yww}, at least for an even number $n$ of particles.
This procedure leads to a formula with a non-abelian structure and $\N=(2,0)$ supersymmetry, 
\begin{align}\label{myst}
\mathcal{A}^{\N=(2,0)}_n(\alpha) = \int &\frac{\prod_{i=1}^n d\sigma_i\, \prod_{k=0}^{m} d^8 \rho_k \,  d^4\chi_k}{\vol( \slc_\sigma \times \slc_\rho)} \;\prod^n_{i=1} \d^6 \! \left ( p^{AB}_i -  \frac{ \langle \r^{A}(\s_i)\, \r^{B}(\s_i) \rangle}{ \prod_{j\neq i} \sigma_{ij}} \right) \, \nn \\
&\times \, \d^{8} \! \left(q^{A I}_i - \frac{ \langle \r^{A}(\s_i)\, \chi^{I} (\s_i) \rangle} { \prod_{j\neq i} \sigma_{ij} } \right)\, { {\rm Pf}'A_n  \over V_n}\, {\mathrm{PT}}(\alpha)\, .
\end{align}
One would be tempted to speculate that these new formulas compute some observable in the mysterious $\N=(2,0)$ theory that arises in the world-volume of multiple coincident M5-branes. Of course, this would be too naive based on what it is currently known about the $\N=(2,0)$ theory; simple dimensional arguments suggest that the $\N=(2,0)$ theory does not have a perturbative parameter and hence a perturbative S matrix. Moreover, explicit no-go results have been obtained preventing the existence of three-particle amplitudes with all the necessary symmetries \cite{Huang:2010rn, Czech:2011dk}. 

Here we take the viewpoint that since \eqref{myst} is well defined as an integral, i.e., it has all correct redundancies, $\text{SL}(2,\mathbb{C})_{\s} \times \text{SL}(2,\mathbb{C})_{\r}$, it is worth exploring in its own right. Moreover, the new non-abelian $\N=(2,0)$ formulas can be combined with non-abelian $\N=(0,2)$ formulas using the KLT procedure in order to compute $\N=(2,2)$ supergravity amplitudes. Given that the non-abelian $\N=(2,0)$ formulas are purely chiral, they have some computational advantages over their $\N=(1,1)$ Yang--Mills cousins, which are traditionally used in KLT. 

A natural step in the study of any connected formula based on rational maps is to consider its behavior under factorization. 
Any physical amplitude must satisfy locality and unitarity: a tree-level amplitude should only have simple poles when non-overlapping Mandelstam variables approach to zero, and the corresponding residues should be products of lower-point ones. 

Let us investigate these physical properties of the non-abelian $\N=(2,0)$ formula. Already for $n=4$ we find a peculiar behavior under factorization. As we discussed in Section~\ref{sec:even_points}, the net effect of changing from $({\rm Pf}^{\prime} A_4)^2$ to the Parke--Taylor factor ${\rm PT}(1234)$ is to introduce an additional factor of ${1/(s_{12}\, s_{23})}$. Therefore, for $n=4$ the non-abelian $(2,0)$ formula gives \cite{Congkao-UCLA-talk}:
\be\label{frt}
\mathcal{A}_4^{\N=(2,0)}(1234) = \d^6\left(\sum_{i=1}^{4} p^{AB}\right) { \d^8(\sum_{i=1}^{4} q_i^{A,I}) \over s_{12} \, s_{23}  } \, ,
\ee 
which is related to that of 6D $\N=(1,1)$ SYM by a simple change to the fermionic delta functions. Comparing with the four-point amplitude of the theory of a probe M5-brane, the new feature is that $\mathcal{A}_4^{\mathcal{N}=(2,0)}(1234)$ has simple poles at $s_{12} \rightarrow 0$ and $s_{23} \rightarrow 0$, and the question is what the corresponding residues are. In order to explore the singularity in the $s_{12}$-channel, let us define the following two objects at $s_{12}=0$:
\begin{equation}
x_{23}=w_2^a\langle 2_a | 3_{\hat{b}}]\tilde{u}_3^{\hat{b}}\,, \qquad \tilde{x}_{23}=\tilde{w}_2^{\hat{a}} [2_{\hat{a}} | 3_b \rangle u_3^{b}.
\end{equation}
It is easy to check that $s_{23}=x_{23}\tilde{x}_{23}$. One can then show that the residue is given by
\begin{equation} \label{20res}
\lim_{s_{12}\rightarrow 0}s_{12}\, \mathcal{A}_4^{\mathcal{N}=(2,0)}(1234) = \delta^6\left(\sum_{i=1}^{4} p^{AB}_i\right) \frac{x_{23}^2}{s_{23}}\int d^4\eta^I_P\, F_3^{(2,0)}(1,2,P) F_3^{(2,0)}(-P,3,4),
\end{equation}
where $F_3^{(2,0)}$ is obtained from $\mathcal{A}_3^{\mathcal{N}=(1,1)\,{\rm SYM}}$ by the replacement of fermionic delta functions \eqref{eq:3pts_Fermion} to make it $\mathcal{N}=(2,0)$ supersymmetric. Note that the left-hand side still diverges as $s_{23}\rightarrow 0$. These three-point objects, $F_3^{(2,0)}$, are ambiguous since they are not invariant under $\alpha$-scaling of (\ref{eq:scaling}) as we discussed in Section~\ref{sec:3pts}, which is a redundancy of the three-particle special kinematics \cite{Huang:2010rn, Czech:2011dk}. However, equation \eqref{20res} is well-defined, because the prefactor on the right-hand side precisely cancels out the ambiguity. Moreover, the scaling acts by sending $x_{23}\rightarrow \alpha x_{23}$, $\tilde{x}_{23}\rightarrow\alpha^{-1} \tilde{x}_{23}$, so it is clear that there is a choice of $\alpha=\alpha(w_2,\tilde{u}_3)$ that sets $x^2_{23}=s_{23}$. For this choice the four-particle residue can in fact be written as a product of the three-point objects $F_3^{(2,0)}$ summed over internal states. Note, however, that the two $F_3^{(2,0)}$ factors cannot be regarded as independent amplitudes, i.e., they are non-local, since they are defined only in the frame $\frac{x^2_{23}}{s_{23}}=1$, which in turn depends on all four particles involved. A similar decomposition can be achieved by implementing an unfixed $\alpha$-scale, but using the shift redundancy (\ref{eq:shifting}), $w_i\rightarrow w_i + b_i u_i$, to set
\begin{equation}\label{eq:hsframe}
w_2^a\langle 2_a | 3_{\hat{b}}]\tilde{w}_3^{\hat{b}} + w_1^a\langle 1_a | 3_{\hat{b}}]\tilde{w}_3^{\hat{b}}=0.
\end{equation}  
In this frame we find $\frac{x^2_{23}}{s_{23}}=[ \tilde{u}_P\tilde{u}_{-P}] \langle w_P w_{-P}\rangle$, and we can write\footnote{We thank Yu-tin Huang for this observation.}
\begin{equation}
\lim_{s_{12}\rightarrow 0}s_{12}\, \mathcal{A}_4^{\mathcal{N}=(2,0)}(1234) = \delta^6\left(\sum_{i=1}^{4} p^{AB}_i\right) \int d^4\eta^I_P\, F_3^{a\hat{a}}(1,2,P) F_{3,a\hat{a}}(-P,3,4)\,,
\end{equation}
with
\begin{equation}
F_3^{a\hat{a}}(1,2,P) :=F_3^{(2,0)}(1,2,P) w_P^a \tilde{u}_P^{\hat{a}}\,,
\end{equation}
which now resembles the three-particle amplitude of higher spin states with $\mathcal{N}=(2,0)$ supersymmetry, as described in \cite{Czech:2011dk}. Non-locality is now present because the objects are not $b$-shift invariant. In fact, the defining frame given by  \eqref{eq:hsframe} again depends on the kinematics of all the particles involved. We hence recognize two different ``frames" in which the residue of $\mathcal{A}_4^{\mathcal{N}=(2,0)}(1234)$ is given by a sum over exchanges between three-point $\mathcal{N}=(2,0)$ objects.

Since the residue is not given by local quantities we expect that the non-abelian $\mathcal{N}=(2,0)$ formulas give rise to a generalization of physical scattering amplitudes whose meaning might be interesting to explore. Note that the same computation for the 6D $\mathcal{N}=(1,1)$ SYM theory yields no prefactor, and therefore the residue of a four-point amplitude is precisely a product of two three-point amplitudes summed over the exchange of all allowed on-shell states in the theory, as required by unitarity. 

We have further checked that the naive non-abelian $(2,0)$ integral formula for odd multiplicity does not have the required $(\text{SL}(2,\mathbb{C})_{\s} , \text{SL}(2,\mathbb{C})_{\r} , T)$ redundancies anymore, i.e., it depends on the ``fixing'' of $\sigma$'s and $\rho$'s. In the case of three particles this is a reflection of the $\alpha$-scaling ambiguity and is again in agreement with the analysis of \cite{Huang:2010rn, Czech:2011dk}.

Along the same line of thought, one may further construct 6D $\mathcal{N}=(4,0)$ ``supergravity'' formulas by the  double copy of two non-abelian $\mathcal{N}=(2,0)$ formulas discussed previously and $\mathcal{N}=(3,1)$ ``supergravity'' formulas by the  double copy of the non-abelian $\mathcal{N}=(2,0)$ formulas with $\mathcal{N}=(1,1)$ SYM. The possible existence of a 6D $\mathcal{N}=(4,0)$ theory and its relation to supergravity theories have been discussed in \cite{Hull:2000zn}; also see the recent works \cite{Henneaux:2017xsb, Henneaux:2018rub} on constructing the actions of 6D free theories with $\mathcal{N}=(4,0)$ or $\mathcal{N}=(3,1)$ supersymmetry.\footnote{The double copy of the $(2,0)$ spectrum to produce the $(4,0)$ one was discussed in \cite{Chiodaroli:2011pp}, and more recently in \cite{Borsten:2017jpt}.} These constructions clearly will lead to well-defined integral formulas as far as the $\text{SL}(2,\mathbb{C})_{\s} \times \text{SL}(2,\mathbb{C})_{\r}$ redundancies are concerned. For instance, the four-point formulas should be given by (\ref{eq:sugra-4pts}), but with a change of the fermionic delta functions in the numerator such that they implement $\mathcal{N}=(4,0)$  or $\mathcal{N}=(3,1)$  supersymmetry. However, as we can see already at four points, the formulas contain kinematics poles, and the residues do not have clear physical interpretations, just like the $\mathcal{N}=(2,0)$ non-abelian formulas above.

It is worth mentioning that, even though all these formulas are pathological in 6D, upon dimensional reduction to lower dimensions the non-abelian $\N=(2,0)$ formulas or the $\mathcal{N}=(4,0)$ and $\mathcal{N}=(3,1)$ ``supergravity'' formulas actually behave as well as 6D $\mathcal{N}=(1,1)$ SYM or 6D $\mathcal{N}=(2,2)$ supergravity. In fact, they give the same results. This phenomenon has already been observed for branes, where the  $\N=(2,0)$ M5-brane formulas and the $\N=(1,1)$ D5-brane formulas both reduce to the same D4-brane amplitudes in 5D.

\subsection*{Degenerate Kinematics in 6D} \label{sec:degenerate_kinematics}

The last topic we address has to do with a very important assumption made in the construction of our formulas. Up to this point we have been using maps of degree $n-2$ from $\mathbb{CP}^1$ into the null cone defined by
\be
p^{AB}(z) = \langle \rho^A(z)\,\rho^B(z)\rangle = \rho^{A,+}(z)\rho^{B,-}(z) - \rho^{A,-}(z)\rho^{B,+}(z),
\ee
with $\rho^{A,+}(z)$ and $\rho^{A,-}(z)$ both polynomials of degree $(n-2)/2$ for even $n$ and $(n-1)/2$ for odd $n$. The assumption made so far is that these maps are sufficient to cover the entire relevant moduli space for the computation of Yang--Mills amplitudes. In particular, a natural question is what happens when $d_+ = {\rm deg}~\rho^{A,+}(z)$ and $d_-={\rm deg}~\rho^{A,-}(z)$ are allowed to be distinct and whether such maps are needed to cover regions of the moduli space when the external kinematics takes special values. 

Let us start the discussion with $n$ even. Considering maps of general degrees $d_+$ and $d_-$, subject to the constraint  $d_++d_-=n-2$, we may require $\Delta := d_+-d_-\geq 0$ without loss of generality. While for generic kinematics $\Delta=0$ maps exist for all $(n-3)!$ solutions of the scattering equations, we find that there are codimension one or higher subspaces for which some solutions escape the ``coordinate patch" covered by $\Delta=0$ maps.

There are three matrices that control all connected formulas presented in this work. They are $K_n$, $A_n$ and $\Phi_n$. The first and the last one only appeared implicitly. For reader's convenience we list below the definition of all three even though $A_n$ has been previously defined:
\be
(K_n)_{ij} = \left\{
        \begin{array}{ll}
          p_{i} \cdot p_{j} &\; \hbox{\quad $i\neq j$,} \\
          0 &\; \hbox{\quad $i=j$,}
        \end{array}
      \right.  \, \, (A_n)_{ij} = \left\{
        \begin{array}{ll}
          \frac{p_i \cdot p_j}{\sigma_{ij}} & \hbox{\quad $i\neq j$,} \\
          0 & \hbox{\quad $i=j$,}
        \end{array}
      \right.  \, \, (\Phi_n)_{ij} = \left\{
        \begin{array}{ll}
          \frac{p_i \cdot p_j}{\sigma_{ij}^2}, & \hbox{\quad$i\neq j$,} \\
          -\sum_{k\neq i} \frac{p_i \cdot p_k}{\sigma_{ik}^2} & \hbox{\quad$i=j$.}
        \end{array}
      \right.\nn
\ee
The physical meaning of the first one is clear: It is the matrix of kinematic invariants. The second is the familiar $A_n$ matrix whose reduced Pfaffian enters in all the formulas we have presented. Finally, $\Phi_n$ is the Jacobian matrix of the scattering equations.

In dimensions $D\geq n-1$ the number of independent Mandelstam invariants is $n(n-3)/2$, and therefore the matrix $K_n$ has rank $n-1$. When $D<n-1$ the space of Mandelstam invariants has the lower dimension $(D{-}1)n-D(D{+}1)/2$, and therefore the matrix $K_n$ has a lower rank. This is easy to understand as the momentum vectors in $p_i\cdot p_j$ start to satisfy linear dependencies. In general, if the dimension is $D$, then so is the rank of the $K_n$ matrix. The rank of $K_n$ is therefore a measure of the minimal spacetime dimension where a given set of kinematic invariants $p_{i} \cdot p_{j}$ can be realized as physical momentum vectors. By contrast, in general the matrices $A_n$ and $\Phi_n$ have ranks $n-2$ and $n-3$, respectively, for any spacetime dimension $D$.

At this point we have numerical evidence up to $n=10$ to support the following picture: There exist subspaces in the space of 6D kinematic invariants where some solutions to the scattering equations lower the rank of $A_n$ while keeping the rank of $K_n$ and $\Phi_n$ the same as is expected for generic kinematics. 

Since $A_n$ is antisymmetric, its rank decreases by multiples of two. Moreover, we find that when the rank has decreased by $2r$, i.e., its new rank is $n-2(r+1)$, the maps that cover such solutions of the scattering equations are those for which $\Delta  = 2r$. From the definition $\Delta = d_+-d_-$ it is clear that the maps needed to cover these new regions are of degree $d_+=n/2+(r-1)$ and $d_-=n/2-(r+1)$.

The extreme case $r=n/2-2$, i.e., when $d_- =1$, is never reached while keeping the rank of $K_n$ equal to six. In fact, it is only when the rank of $K_n$ becomes four that such maps are needed. Note that decreasing the rank of $K_n$ to four implies that such kinematic points can be realized by momenta embedded in 4D spacetime. In 4D it is well-known that the solutions to the scattering equations split into sectors, as discussed in Section~\ref{sec:rational-maps}, and maps of different degrees are needed to cover all solutions.


For odd $n$ the preceding statement needs to be refined. To see why, recall that in Section~\ref{subsec:maptransf} we introduced the notion of $z$-dependent $\slc$ transformations \eqref{gen}. In particular, for $d_{+}=\Delta + d_{-}$ one has the following redundancy of the maps:
\begin{equation}\label{odd-point-redundancy}
\rho^{A,+}(z) \rightarrow \rho^{A,+}(z)+ u(z)\rho^{A,-}(z)\,,
\end{equation}
where $\text{deg}\, u(z) = \Delta$. This is an intrinsic redundancy of each sector, since the maps (and hence the matrix $A_{ij}$) are invariant under such transformation. However, when $d_{+}<\Delta + d_{-}$ this transformation will ``shift'' between sectors, leading to maps that satisfy $d_{+}=\Delta + d_{-}$, but are equivalent to those with lower degree. We see that these points of the moduli space should be modded out in order to define sector decomposition. The way to recognize them is to notice that when $d_{+}<\Delta + d_{-}$ the transformation \eqref{odd-point-redundancy} will determine coefficients of $\rho^{+}(z)$ in terms of those of $\rho^{-}(z)$. Hence, for $n$ even, the natural way of modding out such cases is to consider the moduli space with completely independent coefficients of $\rho^{+}(z)$ and $\rho^{-}(z)$, which is what we did so far. For $n$ odd, this is not the case since in general the top coefficients of $\rho^{+}(z)$ and $\rho^{-}(z)$ are related, i.e., $\rho^+_{d_{+}}=\xi\rho^-_{d_{-}}$. A natural choice is then to set $\xi=0$, which effectively removes linear dependences within coefficients. Hence we define the $\Delta=1$ sector as the one with $d_{+}=d_{-}+1$ and all the coefficients being independent. The transformations \eqref{odd-point-redundancy} are now the standard $\slc_{\rho}$ shift and the T shift, which we leave as the redundancies of the sector. We further note that since the degree of the map $p^{AB}(z)=\rho^{[A}_{+}(z)\rho^{B]}_{-}(z)$ is odd, for even $\Delta=d_{+}-d_{-}$ there will be trivial linear relations among coefficients. This motivates us to label the sectors as
\begin{equation}
\Delta=d_{+}-d_{-}= 2r+1.
\end{equation}
The maps that we have used so far correspond to $r=0$. We find that for $r>0$ it is the odd-point analog of the reduced Pfaffian, ${\rm Pf}^{\prime}\widehat{A}_n$, defined in Section~\ref{sec:odd-pt}, that vanishes for the troublesome solutions supported by these maps.

Finally, we comment that, even though integrands of all the theories we consider in this paper contain ${\rm Pf}{A}_n$ or ${\rm Pf}^{\prime}\widehat{A}_n$, the full integrand does not necessarily vanish for the missing solutions in the degenerate kinematics sector (as we approach it via analytic continuation). In fact, depending on the projected components of the supermultiplet, the fermionic integrations may generate singularities for these solutions such that they contribute finitely. This can happen in all the theories considered so far except in the case of M5 brane and D-branes, where there are enough powers of ${\rm Pf}{A}_n$ to generate a zero for the degenerate solutions. These facts can also be seen by considering purely bosonic amplitudes and directly using CHY formulas. This means that at degenerate kinematic points there are solutions to the scattering equations that require maps with $|\Delta| >0$. However, this is of course not a problem for our formulas: As we mentioned earlier the degenerate regions of kinematic space are of codimension one or higher, so we can define the amplitudes by analytic continuation
of the $\Delta =0$ formulas. In practice, the integral over the maps moduli space can be first performed in a generic configuration close to the degenerate kinematics, after which the degenerate configuration can be easily reached. We leave a complete exploration of the moduli space of maps for all values of $\Delta$, together with the related topic of 4D dimensional reduction, for future research.

\acknowledgments
We thank Yvonne Geyer, Song He, Yu-tin Huang, Lionel Mason, and Kai Roehrig for useful discussions, and we are grateful to Song He, Yu-tin Huang, Zhengwen Liu, and
Yong Zhang for comments on early drafts. This research was supported in part by Perimeter Institute for Theoretical Physics. Research at Perimeter Institute is supported by the Government of Canada through the Department of Innovation, Science and Economic Development Canada and by the Province of Ontario through the Ministry of Research, Innovation and Science. A.G. thanks CONICYT for financial support. C.W. is supported by a Royal Society University Research Fellowship no. UF160350. This work was supported in part by the Walter Burke Institute for Theoretical Physics at Caltech and by U.S. DOE Grant DE-SC0011632. M.H. would like to thank Perimeter Institute for their hospitality.

\appendix

\section{\label{app:symmetry-algebra}Symmetry Algebra}

This appendix examines the group of redundancies of the odd-point scattering maps that preserves their polynomial form. This consists of a five-dimensional subalgebra of the full Lie algebra. We will examine this five-dimensional algebra now, and leave the analysis of the full algebra for the future. More concretely, we first fix two generators of $\slc_\rho$ corresponding to dilations and special conformal transformations in a suitable way, and then show that the algebra of residual symmetries corresponds to the semidirect product $\slc_\sigma \ltimes\mathbb{C}^{2}$.

It is instructive to start by analyzing the even-point symmetry group
$\slc_{\sigma}\times \slc_{\rho}$ in this setup. For $n=2m+2$ let us call the polynomials $\rho^{A,+}(z)=\varpi^{A}(z)$ and $\rho^{A,-}(z)=\vartheta^{A}(z)$, both of degree $m$. We consider transformations
$(z,\sigma_{i},\rho^{A,a})\rightarrow(\hat{z},\hat{\sigma}_{i},\hat{\rho}^{A,a})$ such that
\be
\frac{\hat{\varpi}^{[A}(\hat{z})\hat{\vartheta}^{B]}(\hat{z})}{\prod_{i=1}^{n}(\hat{z}-\hat{\sigma}_{i})}d\hat{z} \;=\; \frac{\varpi^{[A}(z)\vartheta^{B]}(z)}{\prod_{i=1}^{n}(z-\sigma_{i})} dz.
\ee
This contains the $\slc_{\rho}$ transformations, which can be defined as the stability subgroup satisfying $\hat{z}=z$. Among these, let us consider only the shift:\footnote{In this section we will mostly suppress the $\text{SU}^*(4)$ index, since it is not relevant to what follows.}
\be
J=\left(\begin{array}{cc}
0 & 1\\
0 & 0
\end{array}\right)\in \slc_{\rho}\,,\qquad e^{\alpha J}:\quad\hat{\varpi}(z)=\varpi(z)+\alpha\vartheta(z)\,,\qquad \hat{\vartheta}(z)=\vartheta(z).
\ee
The other
two generators should be thought as fixed. For instance, consider the
$\slc_{\sigma}$ scaling $\hat{z}=e^{\alpha} z$.
This induces the following transformation on the polynomials:
\be
e^{\alpha\ell_{0}}:\qquad\hat{\varpi}(z)=e^{p\alpha}\varpi(e^{-\alpha} z)\,,\qquad\hat{\vartheta}(z)=e^{q\alpha}\vartheta(e^{-\alpha} z),
\ee
with $p+q=n-1$. Since the generator of $\slc_{\rho}$
scaling is fixed, so are the values of $p$ and $q$, which will be
determined below. Similarly, for the shift $\hat{z}=z+\beta$
we find
\be
e^{\beta\ell_{-1}}:\qquad\hat{\varpi}(z)=\varpi(z-\beta)\,,\,\qquad\hat{\vartheta}(z)=\vartheta(z-\beta).
\ee
The last generator is defined by $\ell_{+1}=\mathcal{I}\, \ell_{-1}\, \mathcal{I}$, where inversion $\mathcal{I}$ acts in the following way. Consider
the transformation $\hat{z}=-1/z$. The polynomials
should then transform as
\be
\mathcal{I}:\qquad\hat{\varpi}(z)=z^{m}\,Y^{\frac{1}{2}}\,\varpi\left(-1/z\right)\,\text{,}\qquad\hat{\vartheta}(z)=z^{m}\,Y^{\frac{1}{2}}\,\vartheta\left(-1/z\right),
\ee
where $Y=\prod_{i=1}^{n}\sigma_{i}$. It is straightforward to check
that $\mathcal{I}^{2}=(-1)^{m}\mathds{1}$. The minus sign can be
neglected since we are only interested in a representation of $\text{PSL}(2,\mathbb{C})_\sigma$,
which corresponds to the M\"obius transformations acting on the punctures,
for which we have the $\mathbb{Z}_{2}$ identification $-\mathds{1}\cong\mathds{1}$.
Let us consider the action of the following composition
\begin{align}\label{eq:inv}
\mathbb{\mathcal{I}}\, e^{\alpha\ell_{0}}\, \mathcal{\mathcal{I}}(\varpi(z)) & =  \mathbb{\mathcal{I}}\, e^{\alpha\ell_{0}}\left(z^{m}\, Y^{\frac{1}{2}}(\sigma_{i})\,\varpi\left(-1/z\right)\right)\nonumber \\
& =  \mathcal{I}\left(e^{p\alpha}e^{-\alpha m}e^{\frac{-\alpha n}{2}}z^{m}\,Y^{\frac{1}{2}}(\sigma_{i})\,\varpi\left(-e^{\alpha}/z\right)\right)\nonumber \\
& \cong  e^{p\alpha}e^{-\alpha m}e^{\frac{-\alpha n}{2}}\varpi(e^{\alpha}z),
\end{align}
where the symbol $\cong$ indicates we have used the $\mathbb{Z}_{2}$
identification. Imposing $\mathbb{\mathcal{I}}\,\ell_{0}\,\mathcal{\mathcal{I}}=-\ell_{0}$ we find:
\be
-p+m+\frac{2m+2}{2}=p \quad\implies\quad p=q=m+\frac{1}{2},
\ee
which coincides with the choice of \cite{Heydeman:2017yww}.
The analysis for $\vartheta(z)$ is identical. It then follows that 
\be
e^{\alpha J}e^{\beta\ell_{0}}\left(\begin{array}{c}
\varpi(z)\\
\vartheta(z)
\end{array}\right)=e^{\beta\ell_{0}}e^{\alpha J}\left(\begin{array}{c}
\varpi(z)\\
\vartheta(z)
\end{array}\right),
\ee
or equivalently, $[J,\ell_{0}]=0$. We also have
\begin{align}
\mathbb{\mathcal{I}}\,e^{\alpha J}\,\mathcal{\mathcal{I}}(\varpi(z)) & =
 \mathbb{\mathcal{I}}\left[z^{m}\,Y^{\frac{1}{2}}\left(\varpi\left(-1/z\right)+\alpha\vartheta\left(-1/z\right)\right)\right]\nn\\
 & \cong  \varpi(z)+\alpha\vartheta(z)\nn\\
 & =  e^{\alpha J}\varpi(z),
\end{align}
which gives $\mathbb{\mathcal{I}}\,J\,\mathcal{\mathcal{I}}=J$ or $[\mathcal{I}, J] =0$. This analysis is consistent with the fact that we are considering the subalgebra $\slc_{\sigma}\times J$ of the direct product $\slc_{\sigma}\times \slc_{\rho}$ and $\mathcal{I}$ belongs to the first group. 

Let us now examine how this situation changes when considering
the odd-point maps with $n = 2m+1$. Now, we fix the generators of $\slc_{\rho}$ such that $\text{deg } \varpi(z) = m$ and $\text{\ensuremath{\deg} }\vartheta(z) =m-1$. Note that this is consistent with the fact that $J$ is a residual symmetry. In fact, the actions of $J$, $\ell_{0}$ and $\ell_{-1}$
are not modified, even though the values of $p,q$ differ, as we will
show below. The inversion $\mathcal{I}$ now acts as 
\be
\mathcal{I}:\qquad\hat{\varpi}(z)=z^{m}\,Y^{\frac{1}{2}}\,\varpi\left(-1/z\right)\,\text{,}\qquad\hat{\vartheta}(z)=z^{m-1}Y^{\frac{1}{2}}\,\vartheta\left(-1/z\right).
\ee
Repeating the computation in \eqref{eq:inv} we find that
\begin{eqnarray}
-p+m+\frac{2m+1}{2}=p & \quad\implies\quad \begin{cases} \;p=m+\frac{1}{4},\\
\;q= m-\frac{1}{4}.\end{cases}
\end{eqnarray}
Furthermore, we have:
\begin{align}
e^{\alpha J}e^{\beta\ell_{0}}\left(\begin{array}{c}
\varpi(z)\nn\\
\vartheta(z)
\end{array}\right) & =
 \left(\begin{array}{c}
e^{(m+1/4)\beta}\varpi(e^{-\beta}z)+\alpha e^{(m-1/4)\beta}\vartheta(e^{-\beta}z)\\
e^{(m-1/4)\beta}\vartheta(e^{-\beta}z)
\end{array}\right)\nn\\
 & =  \left(\begin{array}{c}
e^{(m+1/4)\beta}\left(\varpi(e^{-\beta}z)+\tilde{\alpha}\vartheta(e^{-\beta}z)\right)\\
e^{(m-1/4)\beta}\vartheta(e^{-\beta}z)
\end{array}\right)\nn\\
 & =  e^{\beta\ell_{0}}e^{\tilde{\alpha}J}\left(\begin{array}{c}
\varpi(z)\\
\vartheta(z)
\end{array}\right),
\end{align}
where $\tilde{\alpha}:=\alpha e^{-\beta/2}$. This means that $[J,\ell_{0}]=-\frac{1}{2}J\label{eq:Jweight}$.

In contrast to the case of even $n$, we have shown that for odd $n$ the group structure is a semidirect extension of $\slc_{\sigma}$
by an (Abelian) shift factor $J$. In other words, the Riemann sphere
symmetry group $\slc_{\sigma}$ and the group $\slc_{\rho}$
are intertwined. Moreover, we will now show that the $J$
extension of $\slc_{\sigma}$ is not enough to achieve closure
of the group. In fact, consider
\begin{align}
\mathbb{\mathcal{I}}\,e^{\alpha J}\,\mathcal{\mathcal{I}}(\varpi(z)) &= \mathbb{\mathcal{I}}\left[z^{m}\,Y^{\frac{1}{2}}\,\varpi\left(-1/z\right)+\alpha z^{m-1}Y^{\frac{1}{2}}\vartheta\left(-1/z\right)\right]\nn\\
 & = \mathbb{\mathcal{I}}\left[z^{m}\,Y^{\frac{1}{2}}\,\varpi_{(\alpha)}\!\left(-1/z\right)\right]\nn\\
 & \cong \varpi_{(\alpha)}(z),
\end{align}
where we have defined the polynomial
\be
\varpi_{(\alpha)}(z) := \varpi(z)-\alpha z\vartheta(z) = e^{\alpha T}\varpi(z).
\ee
This shows that conjugating the shift $J$ by an inversion
leads to a new shift symmetry not present in the even-$n$ case: $\mathcal{I}J\mathcal{I}= - T$. This precisely corresponds to the T-shift symmetry, introduced
previously, acting on the fixed frame with $\xi=(1,0)$. Conjugating the equation $[J,\ell_{0}]=-\frac{1}{2}J$ we find:
\be
[T,\ell_{0}]=\frac{1}{2}T.
\ee
Because $J$ and $T$ are Abelian shifts it follows that $[J,T]=0$, i.e., they generate the translation group $\mathbb{C}^{2}$ and transform as a doublet under $\slc_\sigma$. The rest of the $\slc_\sigma \ltimes\mathbb{C}^{2}$ algebra is
\be
[\ell_{1},T]  = [\ell_{-1},J]=0, \qquad [\ell_{-1},T]= -J, \qquad [\ell_{1},J]=T, \qquad [\ell_{i},\ell_{j}]  = (i-j)\ell_{i+j}.
\ee
More succinctly, if we define $(J,T) = (T_{-1/2}, T_{1/2})$, then we have $[T_r,T_s]=0$ and
\be
[\ell_i , T_r] = \left( \frac{i}{2} - r \right) T_{i+r}, \quad i=-1,0,1 \quad r= \pm 1/2,
\ee
as well as
\be
\mathcal{I}\, l_i\, \mathcal{I}^{-1} = - l_{-i} \quad i=-1,0,1 \quad
{\rm and} \quad \mathcal{I}\, T_r \, \mathcal{I}^{-1} = - T_{-r}
\quad r= \pm 1/2.
\ee

Finally, one can directly check that the remaining generators of $\slc_{\rho}$ do not preserve the polynomial form of the maps. Hence we claim that this is the maximal subalgebra associated to polynomial maps.

\section{\label{app:soft-limits}Details of the Soft-Limit Calculations}

In this appendix we analyze the soft limit of the connected formula, treating the measure and the integrand separately. Because of its simplicity, we start in \ref{4dm} with the soft limit of the CHY measure and the deformation of the maps in 4D. In \ref{app:even-to-odd} we turn to the even-$n$ measure for 6D, where several new technical ingredients appear due to the $\slc$ little-group structure. This analysis allows us to recover the form of the odd-point maps and measure as well as the emergent symmetry T
discussed in appendix A.  In \ref{app:soft-2} the odd-point integrand is derived from the even-point one for the case of $\N=(1,1)$ SYM. Finally, in \ref{app:softeven} we obtain the even-$n$ measure from the odd-$n$ one, and use it to prove that the constraints have $(n-3)!$ solutions.

\subsection{Four Dimensions \label{4dm}}

Let us illustrate the use of the soft limit by considering the simpler 4D case first. Here we will focus on the CHY measure in the Witten--RSV form and show that it has the form given in \eqref{soft-behavior}. In particular, we consider the measure associated to the $d$th sector and show that in the soft limit, where $p_{n+1}=\tau\,\hat{p}_{n+1}$ with $\tau \rightarrow0$, we have
\begin{equation}
\int d\mu_{n+1,d}^{\text{4D}}=\delta(p_{n+1}^{2})\int d\mu_{n,d-1}^{\text{4D}} \frac{1}{2\pi i}\oint_{[\tilde{\lambda}_{n+1},\tilde{\rho}(\sigma_{n+1})]=0}\frac{d\sigma_{n+1}}{E_{n+1}}+\text{conj.}+\mathcal{O}(\tau^0),\label{eq:4ds}
\end{equation}
where the scattering equation for the soft particle takes the form
\begin{equation}\label{scat4dsoft}
E_{n+1}=\sum_{i=1}^{n}\frac{p_{n+1}\cdot p_{i}}{\sigma_{n+1,i}}=\frac{\langle\lambda_{n+1}\,\rho(\sigma_{n+1})\rangle[\tilde{\lambda}_{n+1}\,\tilde{\rho}(\sigma_{n+1})]}{\prod_{i=1}^{n}\sigma_{n+1,i}}
\end{equation}
In \eqref{eq:4ds} ``conj.'' means to consider the first term with the conjugated
contour $[\tilde{\lambda}_{n+1}\,\tilde{\rho}(\sigma_{n+1})]\rightarrow\langle\lambda_{n+1}\,\rho(\sigma_{n+1})\rangle$
as well as conjugated sector $d\rightarrow \tilde d = n-2-d$. By summing \eqref{eq:4ds} over all sectors we obtain \eqref{soft-behavior}. 

Let us now consider the soft limit of
\begin{equation}
\int d\mu_{n+1,d}^{\text{4D}}=\int\frac{\prod_{i=1}^{n+1}d\sigma_{i}\prod_{k=0}^{d}d^{2}\rho_{k}\prod_{k=0}^{\tilde{d}}d^{2}\tilde{\rho}_{k}}{\text{vol SL}(2,\mathbb{C})\times{\rm GL}(1,\mathbb{C})}\frac{1}{R^{d}(\rho)R^{\tilde{d}}(\tilde{\rho})}\prod_{i=1}^{n+1}\delta^{4}\left(p_{i}^{\alpha\dot{\alpha}}-\frac{\rho^{\alpha}(\sigma_{i})\tilde{\rho}^{\dot{\alpha}}(\sigma_{i})}{\prod_{j\neq i}^{n+1}\sigma_{ij}}\right).\label{eq:dsec}
\end{equation}
The strategy is to first isolate the leading ${1}/{\tau}$ factor, which in this case comes from the resultants. As we will show, in the soft limit $R^{d}(\rho)R^{\tilde{d}}(\tilde{\rho})\sim\tau$, which allows us to evaluate the rest of the measure for $\tau=0$ (except for the factor $\delta(p_{n+1}^{2})$). What makes the case of 4D simple is that $p_{n+1}\rightarrow0$ has only two solutions: $\lambda_{n+1}\rightarrow0$ or $\tilde{\lambda}_{n+1}\rightarrow0$,
which account for the two terms in \eqref{eq:4ds}. Choosing $\lambda_{n+1} \to 0$, the delta function for the last particle in \eqref{eq:dsec} takes the form:
\begin{align}
\hspace{-.4em}\delta^{4}\!\left(p_{n+1}^{\alpha\dot{\alpha}}{-}\frac{\rho^{\alpha}(\sigma_{n+1})\tilde{\rho}^{\dot{\alpha}}(\sigma_{n+1})}{\prod_{i=1}^{n}\sigma_{n+1,i}}\right) &\rightarrow \int \!dt\, d\tilde{t}\, \delta^{2}(\tilde{\lambda}_{n+1}{-}\tilde{t}\tilde{\rho}(\sigma_{n+1}))\,\delta^{2}(t\rho(\sigma_{n+1}))\,\delta\!\left(\!t\,\tilde{t}{-}\frac{1}{\prod_{i=1}^{n}\sigma_{n+1,i}}\!\right)\nn\\
&=\left(\prod_{i=1}^{n}\sigma_{n+1,i}\!\right)^{2}\!\!\!\int\tilde{t}d\tilde{t}\,\delta^{2}(\tilde{\lambda}_{n+1}{-}\tilde{t}\tilde{\rho}(\sigma_{n+1}))\delta^{2}(\rho(\sigma_{n+1})),
\end{align}
where we have used \eqref{t-variables} and dropped the factor $\delta(p_{n+1}^{2})$. If we now introduce a reference spinor $|q]$, we can recast the result in the form
\begin{align}
\left(\prod_{i=1}^{n}\sigma_{n+1,i}\right)^{2}\int\tilde{t}d\tilde{t}\,\delta\left(\tilde{t}-\frac{\tilde{[\lambda}_{n+1}\,q]}{[\tilde{\rho}(\sigma_{n+1})\,q]}\right)\delta\left([\tilde{\lambda}_{n+1}\,\tilde{\rho}(\sigma_{n+1})]\right)\delta^{2}(\rho(\sigma_{n+1}))\,\,\qquad\quad\nn\\
=\,\,\left(\prod_{i=1}^{n}\sigma_{n+1,i}\right)\frac{1}{t}\,\delta\left([\tilde{\lambda}_{n+1}\,\tilde{\rho}(\sigma_{n+1})]\right)\delta^{2}(\rho(\sigma_{n+1})),\label{eq:4dcons} 
\end{align}
where now 
\be
t=\frac{1}{\prod_{i=1}^{n}\sigma_{n+1,i}}\frac{[\tilde{\rho}(\sigma_{n+1})\,q]}{[\tilde{\lambda}_{n+1}\,q]}.
\ee

The first constraint is a polynomial equation of degree $n-d$ in
$\sigma_{n+1}$, which we used for the contour in \eqref{eq:4ds}.
To manipulate the second constraint let us reparametrize the polynomial
as
\be
\rho^{\alpha}(z)=\hat{\rho}^{\alpha}(z)(z-\sigma_{n+1})+r^{\alpha}.
\ee
Here $\hat{\rho}^{\alpha}(z)=\sum_{k=0}^{d-1}\hat{\rho}_{k}^{\alpha} z^{k}$ is a polynomial of degree $d-1$, whose coefficients are shifted from those of $\rho^{\alpha}(z)$. Therefore the Jacobian is one, i.e.,
\be
\prod_{k=0}^{d}d^{2}\rho_{k}=d^{2}r\prod_{k=0}^{d-1}d^{2}\hat{\rho}_{k}
\ee
Integration over $r^{\alpha}$ eliminates the second delta function in \eqref{eq:4dcons}, since
\be
\int d^{2}r\delta^{2}(\rho(\sigma_{n+1}))=1\,,
\ee
setting $r=0$, i.e., $\rho^{\alpha}(z)=\hat{\rho}^{\alpha}(z)(z-\sigma_{n+1})$.
Putting everything together, \eqref{eq:dsec} becomes
\begin{align}
&\int\frac{\prod_{i}^{n}d\sigma_{i}\prod_{k=0}^{d-1}d^{2}\hat{\rho}_{k}\prod_{k=0}^{\tilde{d}}d^{2}\tilde{\rho}_{k}}{\text{vol SL}(2,\mathbb{C})\times{\rm GL}(1,\mathbb{C})} \prod_{i=1}^{n}\delta^{4}\left(p_{i}^{\alpha\dot{\alpha}}-\frac{\hat{\rho}^{\alpha}(\sigma_{i})\, \tilde{\rho}^{\dot{\alpha}}(\sigma_{i})}{\prod_{j\neq i}^{n}\sigma_{ij}}\right)\nn\\
&\qquad\qquad\qquad\times   \frac{1}{2\pi i}\oint_{[\tilde{\lambda}_{n+1}\,\, \tilde{\rho}(\sigma_{n+1})]=0}\frac{d\sigma_{n+1}}{\left(\frac{[\tilde{\lambda}_{n+1}\,\, \tilde{\rho}(\sigma_{n+1})]}{\prod_{i=1}^{n}\sigma_{n+1,i}}\right)}\left(\frac{1}{t\,R^{d}(\rho)R^{\tilde{d}}(\tilde{\rho})}\right)+\mathcal{O}(\tau^0).
\end{align}
Note that in the bosonic delta functions the puncture $\sigma_{n+1}$
has completely dropped thanks to the definition of $\hat{\rho}$.
We will not prove it here, but using the definition \eqref{detp} in terms
of the matrices $\Phi_{d}$ and $\tilde{\Phi}_{\tilde{d}}$ one can
show that in the soft limit the resultants behave as
\begin{equation}
t\,R^{d}(\rho)R^{\tilde{d}}(\tilde{\rho})=\langle\lambda_{n+1}\,\rho(\sigma_{n+1})\rangle R^{d-1}(\hat{\rho})R^{\tilde{d}}(\tilde{\rho})+\mathcal{O}(\tau^{2})
\end{equation}
where $\lambda_{n+1}=\mathcal{O}(\tau)$ is responsible for the leading behaviour,
as anticipated. This concludes the proof of \eqref{eq:4ds}. The extension of this procedure to the integrand in \eqref{sect}, including the redefinition of the fermionic maps, is straightforward in 4D, but we do not present it here. After including the integrand one can deform the contour for $\sigma_{n+1}$ such that it encloses two of the other punctures, i.e., at $\sigma_{n+1}=\sigma_i$. This leads to the soft limit of the $\mathcal{N}=4$ SYM amplitude.

\subsection{\label{app:even-to-odd}From Even to Odd Multiplicity in 6D}

Let us now consider the case of $n$ odd in 6D. We show that the expression \eqref{eq:prop} can be obtained from the soft limit of the $n+1=2m+2$ measure after extracting the corresponding wave function and scattering equation. That is,
\be
\int d\mu_{2m+2}^{\text{6D}} = \delta(p_{n+1}^{2})\int d\mu_{2m+1}^{\text{6D}} \frac{1}{2\pi i} \oint_{|\hat{E}_{n+1}| = \varepsilon} \frac{d\sigma_{n+1}}{E_{n+1}}+\mathcal{O}(\tau^{0}),\label{eq:soft-main}
\ee
where
\be
d\mu_{2m+1}^{\text{6D}} = \frac{\left(\prod_{i=1}^{n} d\sigma_i \right) \left(\prod_{k=0}^{m-1}d^{8}\rho_{k}\right)\, d^{4}\omega\, \langle \xi d\xi\rangle }{\vol \left( \slc_{\sigma}, \slc_{\rho} , \text{T} \right)}\, \frac{1}{V^{2}_n} \,\Delta_B .
\ee
The maps entering the bosonic delta functions $\Delta_{B}$ are defined in \eqref{bosonic-delta}. As in 4D, the strategy is to first isolate the $\tau^{-1}$ piece and then manipulate the delta function for particle $n+1$ to get the corresponding scattering equation. In Section \ref{app:softM} we achieve the first goal by proving that if $\hat{p}_{n+1}^{AB}=v^{[A}q^{B]}$
is the direction of the soft momentum, where $p_{n+1}=\tau\,\hat{p}_{n+1}$, then
\begin{align}\label{eq:soft-1}
\int d\mu_{2m+2}^{\text{6D}} & =\frac{1}{\tau} \delta(p_{n+1}^2) \int  \frac{d\sigma_{n+1} \,\left(\prod_{i=1}^n d\sigma_{i}\right) \left( \prod_{k=0}^m d^8\rho_k\right)}{{\text{vol} \left({\slc_\s} \times {\slc_\r}\right)}\,  V^2_n } \Delta_{B}^{(n)} \\
 &  \times\left(\prod_{i=1}^n \sigma_{n+1,i} \right)\int\,dx\,d^{2}\Xi\,\,\delta^{8}\left(\rho_{a}^{A}(\sigma_{n+1})-\Xi_{a}(q^{A}+x\,v^{A})\right)\,\,+\mathcal{O}(\tau^{0}). \nonumber
\end{align} 
Here  ${\Delta}_{B}^{(n)}$ contains the bosonic delta functions for the $n$ hard particles, but still depends on $\sigma_{n+1}$ and the even-multiplicity maps. Since the leading power of $\tau$ has been extracted in this expression, the integral can be evaluated for $\tau=0.$ Note that this expression is invariant under little-group transformations of the soft particle. In fact, the $\slc_\rho$ transformation
\begin{eqnarray}
q & \rightarrow & Dq+Bv\\
v & \rightarrow & Cq+Av
\end{eqnarray}
with $AD-BC=1$ is equivalent to the following change of variables
\begin{eqnarray}
x \;\rightarrow\; \hat{x} & = & \frac{Ax+B}{Cx+D},\,\\
\Xi_a \;\rightarrow\; \hat{\Xi}_{a} & = & \Xi_{a}(Cx+D),
\end{eqnarray}
which leaves the measure invariant, i.e., $dx\, d^{2}\Xi=d\hat{x}\, d^{2}\hat{\Xi}$. The reason for introducing the variables $x$ and $\Xi$ will become clear in the following section. In Section \ref{devb9} we redefine the maps and isolate the scattering equation as a contour prescription for the puncture $\sigma_{n+1}$ associated to the soft particle, leading to \eqref{eq:soft-main}.

\subsubsection{Derivation of \eqref{eq:soft-1} \label{app:softM}}

We start with the following identity
\begin{equation}
\Delta_{B}^{(n+1)} = \Delta_{B}^{(n)}\, \delta(p_{n+1}^{2})\int d^{4}M |M|^{3}\delta^{8} \left(\rho^{A a}(\sigma_{n+1})-M_{b}^{a}\lambda_{n}^{A b}\right) \delta\!\left(|M|- \prod_{i=1}^n \s_{n+1,i} \right) ,\label{eq:idd}
\end{equation}
where we have utilized the linear constraints in (\ref{linearmeasure}), and denoted $M=M_{n+1}$. 
Now, up to linear order in $\tau$, the most general form of the soft momenta
can be written as
\begin{equation}
\lambda_{n+1}^{A a}=\beta^{a}v^{A}+\tau q^{Aa}\label{eq:gensoft}
\end{equation}
which gives $p_{n+1}^{AB}=\tau v^{[A}q^{B]}+\mathcal{O}(\tau^{2})$, once we set $q^{A}:=\beta_{a}q^{Aa}.$ Unlike 4D, where the soft condition $p_{n+1}\rightarrow 0$ has only two branches (the holomorphic and antiholomorphic soft limits,) here we have a family of solutions due to the less trivial $\slc$ structure.
Let us now assume that as $\tau\rightarrow0$ all the components of the maps $\rho^{Aa}(z)$ and the $\sigma_i$'s stay finite, as determined by the delta functions $\Delta_{B}$, since they should be localized by the equations
of the hard particles. 

In the limit $\tau\rightarrow 0$, the matrix $M$ has a singular piece:
\be
M=\frac{\overline{M}}{\tau}+M_{0}+\mathcal{O}(\tau^{1}).
\ee
The strategy is to input this ansatz into the delta functions and
evaluate the result power by power in $\tau$ leaving only four components of $M$
to be integrated. That is, impose
\begin{align}
\rho^{A b}(\sigma_{n+1}) & = \left(\frac{\overline{M}_{a}^{b}}{\tau}+M_{0,a}^{b}\right)\left(\beta^{a}v^{A}+\tau q^{Aa}\right)\\
&= \frac{\overline{M}_{a}^{b} \, \beta^{a}}{\tau}v^{A}+M_{0,a}^{b}\beta^{a}v^{A}+\overline{M}_{a}^{b}q^{Aa}+\mathcal{O}(\tau^1)\,,\label{eq:contrac}\\
\prod_{i=1}^n \s_{n+1,i} & = |M|=\frac{1}{\tau^{2}}|\overline{M}|+\frac{\langle\overline{M}^{+}\,M_{0}^{-}\rangle-\langle\overline{M}^{-}\,M_{0}^{+}\rangle}{\tau}+|M_{0}| \label{eq:contrac2}\, .
\end{align}
Here $\overline{M}^{+},\overline{M}^{-},{M}^{+}_0, {M}^{-}_0$ denote the respective
columns of the matrices $\overline{M}$ and $M_{0}$. From the finiteness of the LHS of (\ref{eq:contrac}) and (\ref{eq:contrac2}), we see that $\overline{M}$
is degenerate and $\beta$ is a null eigenvector, that is
\be
\overline{M}_{a}^{b}=\Xi^{b}\,\beta_{a} \, .
\ee
Equating terms at order $\tau^{-1}$,
\be
0=\langle\overline{M}^{+}\,M_{0}^{-}\rangle-\langle\overline{M}^{-}\,M_{0}^{+}\rangle=\langle\beta\,\Xi_{a}M_{0}^{a}\rangle\Longrightarrow\Xi_{a}M_{0,b}^{a}\beta^{b}=0\, .
\ee
This result allows to introduce variables $x$ and $\overline{x}$ defined
by
\be
M_{0,b}^{a}\,\beta^{b}=x\,\Xi^{a} , \qquad\Xi_{a}\,M_{0,b}^{a}=\overline{x}\,\beta_{b} \, .
\ee
The general solution of these equations for $M_0$ can be expressed in the basis of spinors $\beta$ and $\Xi$ as
\be
M_{0,b}^{a} = \frac{\overline{x}\,\beta^{a}\beta_{b}+x\,\Xi^{a}\Xi_{b}}{\langle\Xi\beta\rangle}+\gamma\,\Xi^{a}\beta_{b}, 
\ee
and thus
\be
M_{b}^{a} = \frac{\overline{x}\,\beta^{a}\beta_{b}+x\,\Xi^{a}\Xi_{b}}{\langle\Xi\beta\rangle}+\left(\gamma+\frac{1}{\tau}\right)\,\Xi^{a}\beta_{b}\,. \label{eq:softm}
\ee
The component $\gamma$ is a fixed constant, which can only be determined
by considering subleading orders in $\tau$. This is consistent since
it only contributes to the result at order $O(\tau^{1})$. In
fact, choosing the change of variables $\{M_{b}^{a}\}\rightarrow\{x,\overline{x},\Xi^{+},\Xi^{-}\}$, we find
\be
d^{4}M=x\left(\frac{1+\gamma\tau}{\tau}\right)dx\,d\overline{x}\,d^{2}\Xi\sim 
\frac{x}{\tau}dx\,d\overline{x}\,d^{2}\Xi .
\ee

Having identified the singular dependence on $\tau$, we can now select the leading pieces of the arguments inside the delta functions, yielding
\begin{align}
\delta\left(|M|- \prod_{i=1}^n \s_{ n+1,i}\right) & = \delta\left(x\overline{x}- \prod_{i=1}^n \s_{n+1,i} \right)\, , \\
\delta^{8}\left(\rho^{A b}(\sigma_{n})-M_{a}^{b}\lambda_{n}^{A a}\right) & = \delta^{8}\left(\rho^{A b}(\sigma_{n})-\Xi^{b}(x\,v^{A}+q^{A})\right).
\end{align}
Integrating out $\overline{x}$, writing $V_{n+1}^{2}={V}_n^{2}\prod_{i=1}^n \s_{i,\, n+1}^{2}$, and substituting in the identity (\ref{eq:idd}), we finally arrive
at the desired result \eqref{eq:soft-1}.

\subsubsection{Derivation of \eqref{eq:soft-main}\label{devb9}}

In this section we consider the expression \eqref{eq:soft-1} without the integration over $\Xi^a$, i.e., taking $\Xi^a$ to be a fixed spinor. We will also introduce
an auxiliary spinor $\xi$ such that $\langle\Xi\,\xi\rangle=1$. Note
that $\xi$ still has one free component, which we choose to be $\xi^{+}=1$.
The integration over $\Xi^a$ will be restored later. 

We start by expanding the polynomial maps in basis vectors as
\begin{equation}
\rho^{A,a}(z)=\Xi^{a}\,\omega^{A}(z)+\xi^{a}\,\pi^{A}(z) ,\label{eq:bas}
\end{equation}
the delta functions of \eqref{eq:soft-1} as
\begin{align}
\delta^{8}\left(\rho^{A b}(\sigma_{n+1})-\Xi^{b}(x\,v^{A}+q^{A})\right) & = \delta^{4}\left(\pi^{A}(\sigma_{n+1})\right) \delta^{4}\left(\omega^{A}(\sigma_{n+1})-x\, v^{A}-q^{A}\right) \, ,\\
\Delta_{B}^{(n)} & = \prod_{i=1}^{n} \delta^{6}\left(p_{i}^{AB}-\frac{\omega^{[A}(\sigma_{i})\pi^{B]}(\sigma_{i})}{\prod_{j\neq i}^{n+1} \sigma_{ji}}\right)\, ,
\end{align}
and the integration measure as
\be
\prod^{m}_{k=0}d^{8}\rho_{k} =  \prod^{m}_{k=0} d^{4}\omega_{k}\, d^{4}\pi_{k}\, .
\ee
As in 4D, we now parametrize $\pi^{A}(z)=(z-\sigma_{n+1})\hat{\pi}^{A}(z) + r^{A}$, so that the first term vanishes at the last puncture. This change of variables gives, 
\begin{align}
\prod^{m}_{k=0}d^{4}\pi_k & = d^{4}r \prod^{m-1}_{k=0}d^{4}\hat{\pi}_k\, ,\\
\delta^{4}(\pi^{A}(\sigma_{n+1})) & = \delta^{4}(r^{A}) \label{eq:rrr}\, .
\end{align}
On the support of the first delta function,
\be
\Delta_{B}^{(n)}\Big|_{r^A = 0} = \prod_{i=1}^n \delta^{6}\left(p_{i}^{AB}-\frac{{\omega}^{[A}(\sigma_{i})\hat{\pi}^{B]}(\sigma_{i})}{\prod_{j\neq i}^n \sigma_{ij}}\right) =: \Delta_{B}^{(n)}(\omega,\hat{\pi}).
\ee
Note that this result does not depend on $\sigma_{n+1}$. 

The leading-order term in \eqref{eq:soft-1} can be rewritten in the form  
\begin{align}
& \frac{\delta(p_{n+1}^{2})}{\text{\ensuremath{\tau}}} {\int d^{2}\Xi} \int\frac{\prod^{m}_{k=0} d^{4}\omega_k \prod^{m-1}_{k=0} d^{4}\hat{\pi}_k \prod_{i=1}^n d\sigma_{i}}{{\text{vol} \left({\slc_\s} \times {\slc_\r} \right) }\, {V_n}^{2}}
\times {\Delta}_{B}^{(n)}(\omega,\hat{\pi}) \cr
& 
\qquad\qquad\qquad\qquad\times \int d\sigma_{n+1}\,dx\,\delta^{4} \left(\omega^{A}(\sigma_{n+1})-x\, v^{A}-q^{A}\right) \left(\prod_{i=1}^n\sigma_{n+1,i}\right).
\end{align}
The integration over $\int d^{2}\Xi$ has effectively
dropped out of the integral. In principle we could use it to cancel
two of the integrations over $\slc^{2}$ in the denominator. However,
this would fix part of the $\slc^{2}$ invariance, which we want to be present in the odd version of the measure. Instead, let us reintroduce the integration to get a manifestly
symmetric answer. To achieve this we revert to the change of basis (\ref{eq:bas}), i.e.,
for fixed $\{\Xi,\xi\}$ we define 
\begin{equation}
\hat{\rho}^{A,a}(z)=\xi^{a}\,\omega^{A}(z)-\Xi^{a}\,\hat{\pi}^{A}(z).\label{eq:oddpoly}
\end{equation}
This transformation is defined coefficient by coefficient as an $\slc$ transformation except for the top one, which is not invertible. In fact,
\be
d^{4}\omega_{m}\prod_{k=0}^{m-1}d^{4}\omega_{k}\, d^{4}\hat{\pi}_{k} = d^{4}\omega_{m}\prod_{k=0}^{m-1}d^{8}\hat{\rho}_{k}\, 
\ee
and
\be
{\Delta}_{B}^{(n)}(\omega,\hat{\pi})= {\Delta}_{B}^{(n)}(\hat{\rho}) = \prod_{i=1}^n \delta^{6}\left(p_{i}^{AB}-\frac{\langle \hat{\rho}^A(\sigma_i)\,\hat{\rho}^B(\sigma_i)\rangle}{\prod_{j\neq i}^n \sigma_{ij}}\right)\, ,
\ee
where the highest coefficient of the map is given by
\be
\hat{\rho}_{m}^{A,a}=\xi^a\,\omega_{m}^{A}=\left(\begin{array}{c}
1\\
\xi
\end{array}\right)\omega_{m}^{A},
\ee
with $\Xi^{+}\xi-\Xi^{-}=1$. Noting that
\be
\omega^{A}(\sigma_{n+1})=\langle\Xi\,\hat{\rho}^{A}(\sigma_{n+1})\rangle=\Xi^{+}\hat{\rho}^{A,-}(\sigma_{n+1})-\Xi^{-}\hat{\rho}^{A,+}(\sigma_{n+1})\,,
\ee
the integral becomes
\be
 \frac{\delta(p_{n+1}^{2})}{\text{\ensuremath{\tau}}} \int\frac{d^{4}\omega_{m}
\prod^{m-1}_{k=0} d^{8}\hat{\rho}_{k}
\prod_{i=1}^n d\sigma_{i}} {\text{vol} \left({\slc_\s} \times {\slc_\r} \right) V^2_n} {\Delta}_B^{(n)}(\hat{\rho})
  \int d\sigma_{n+1}\,d^2\Xi\,dx\,\delta^{4} \left(D^A\right)\left(\prod_{i=1}^n \sigma_{n+1,i}\right). 
\ee
where
\be
D^A = \Xi^{+}\hat{\rho}^{A,-}(\sigma_{n+1})-\Xi^{-}\hat{\rho}^{A,+}(\sigma_{n+1})-x\,v^{A}-q^{A} .
\ee
Now we note that
\be\label{softscateq}
 \left(\prod_{i=1}^n\sigma_{n+1,\,i} \right)\int d^2\Xi dx\,\delta^{4}\left(D^A\right)
= \left(\prod_{i=1}^n \sigma_{n+1,\,i}\right)\, \delta \left(\langle \hat{\rho}^{+}(\sigma_{n+1})\hat{\rho}^{-}(\sigma_{n+1})\,v\,q \rangle \right) = \delta(\hat{E}_{n+1}).
\ee
In the last line we recognize the scattering equation for the soft particle (in a form analogous to \eqref{scat4dsoft}), which we now implement as a contour integral for $\sigma_{n+1}$. This gives
\begin{equation}
\frac{\delta(p_{n+1}^{2})}{\text{\ensuremath{\tau}}} 
\int\frac{d^{4}\omega_{m} \prod_{k=0}^{m-1}d^{8}\hat{\rho}_{k} \prod_{i=1}^n d\sigma_{i}}{\text{vol} \left({\slc_\s} \times {\slc_\r} \right) \, {V}^{2}_n} \, \frac{1}{2\pi i}\oint_{|\hat{E}_{n+1}| = \varepsilon}\frac{d\sigma_{n+1}}{\hat{E}_{n+1}}\, {\Delta}_{B}^{(n)}(\hat{\rho}).\label{eq:tfixed}
\end{equation}

We have arrived at a compact expression. However, there is subtle but essential caveat. Recall that ${\Delta}_{B}^{(n)}(\hat{\rho})$ contains the variable
$\xi=\frac{1+\Xi^{-}}{\Xi^{+}}$ in the top component of the polynomial, 
$\hat{\rho}_{m}$. This variable still depends on the soft puncture
$\sigma_{n+1}$. In fact, it is implicitly defined through the relation 
\begin{equation}
\langle\Xi\,\hat{\rho}^{A}(\sigma_{n+1})\rangle=x\,v^{A}+q^{A}\,.\label{eq:softe}
\end{equation}
In order to decouple $\xi$ from this soft equation, we introduce
a new redundancy that will enable us to turn $\xi$ into an integration
variable (which will be fixed by the hard data). Since $v^{A}$ and $q^{A}$ are only defined through $\hat{p}_{n+1}^{AB}=v^{[A}q^{B]}$,
the formula must be invariant under $v\rightarrow\frac{v}{\alpha}$,
$q\rightarrow\alpha q$. According to (\ref{eq:softe}), such a transformation can be absorbed into a transformation of $(\Xi^{a}, x, \xi)$ as follows:
\be
x \rightarrow \frac{x}{\alpha^{2}} \, , \qquad
\Xi^{a} \rightarrow \frac{\Xi^{b}}{\alpha}\, , \qquad
\xi  \rightarrow \xi+\frac{\alpha-1}{\Xi^{+}}=\frac{\alpha+\Xi^{-}}{\Xi^{+}}\,. \label{eq:alphashift}
\ee
Since $\alpha$ is arbitrary, we add an additional integration in the form
\be
1=\frac{\text{\ensuremath{\int}}\frac{d\alpha}{\Xi^{+}}}{\text{vol(T)}}=\frac{\text{\ensuremath{\int}}d\xi}{\text{vol(T)}} \, ,
\ee
which should be regarded as a formal definition of the T-shift measure.
Note that this is not $\slc^2$ covariant, signaling that the Jacobian
is sensitive to the $\slc^2$ frame. Using this, we recast the formula as promised
\begin{equation}
\int d\mu_{2m+2}^{\rm CHY}\rightarrow\delta(p_{n+1}^{2})\int  \frac{d\xi d^{4}\omega \prod_{k=0}^{m-1}d^{8}\hat{\rho}_{k} \prod_{i=1}^n d\sigma_{i}}{\text{ vol}\left({\slc_\s} , {\slc_\r},\text{T} \right)  } \frac{\Delta_{B}^{(n)}(\hat{\rho})}{{V}^{2}_n }  \frac{1}{2\pi i}\oint_{|\hat{E}_{n+1}| = \varepsilon}\frac{d\sigma_{n+1}}{E_{n+1}}.
\end{equation}

Some comments are in order. We have used the little-group scaling
of the soft particle to introduce a new redundancy in the hard equations. As the notation makes clear, this redundancy can be identified with the shift transformation explored in Section~\ref{sec:odd-pt}. Note that this symmetry was absent in (\ref{eq:tfixed}), which can be regarded as a T-fixed version of the final measure. The reason is that while $\Delta_{B}^{(n)}(\hat{\rho})$ is invariant under the shift $\hat{\rho}(z )\rightarrow\hat{\rho}(z)+z \beta \xi\langle\xi,\hat{\rho}\rangle$,
equation \eqref{eq:softe} is not, meaning that the shift parameter $\beta$ can be determined in terms of $v$ and $q$. By averaging over the little group, i.e., over different choices of $v$ and $q$, we unfix this redundancy.

\subsection{\label{app:soft-2}Integrand of $\N=(1,1)$ SYM for Odd Multiplicity}

Let us now apply the prescription given in the previous section, this time at the level of the $\N=(1,1)$ integrand. For $n+1=2m+2$, this integrand can be broken down as follows:
\begin{align}
\mathcal{I}_{2m+2} & = \text{PT}(\I_{n+1})\, \text{Pf}'A_{n+1} \, V_{n+1}\int\prod^{m}_{k=0} d^2\chi_k d^2\tilde{\chi}_k\, \Delta^{(n+1)}_{F}\widetilde{\Delta}^{(n+1)}_{F}\nn\\
& = \delta^{2}\!\left(Q_{n+1}^{A}\tilde{\lambda}_{n+1,A,\hat{a}}\right)\delta^{2}\!\left(\lambda_{n+1,a}^{A}\tilde{Q}_{n+1,A}\right)\,V_{n}\,\text{PT}(\I_{n})\,\text{Pf}'A_{n+1} \left(\prod_{i=1}^{n}\sigma_{n+1,i}\right)\frac{\sigma_{1n}}{\sigma_{1,n+1}\sigma_{n+1,n}}\nn\\
& \times\!\!\int \prod^{m}_{k=0}d^2\chi_k d^2\tilde{\chi}_k \,{\Delta}_{F}^{(n)}\widetilde{\Delta}_{F}^{(n)}\delta^{2}\!\left(\eta_{n+1}^{a}- W_{b}^{a}\chi^{b}(\sigma_{n+1})\right)\delta^{2}\!\left(\tilde{\eta}_{n+1}^{\hat{a}}- \widetilde{W}_{\hat{b}}^{\hat{a}}\tilde{\chi}^{\hat{b}}(\sigma_{n+1})\right)\!.
\end{align}
Here $W=W_{n+1}=M_{n+1}^{-1}$, as defined in Section \ref{sec:LinearConstraints}. The fermionic delta functions are defined in \eqref{fermdelt}, from  which we have extracted the on-shell conditions of the soft particle
(recall that $Q^{A}=\lambda_{a}^{A}\eta^{a}$, etc.). We will first project out the $(n+1)$th gluon and then take the corresponding momentum to be soft. For a given polarization this will generate Weinberg's soft factor for the even point amplitude. In Section~\ref{sec:contour-deformation} we extract it to obtain the odd-point integrand.

A simple choice of polarization is $(a,\hat{a})=(+,\hat{+})$, where the spinor in (\ref{eq:gensoft}) and its conjugate are set
to 
\be
\beta=\tilde{\beta}=\left(\begin{array}{c}
0\\
1
\end{array}\right).
\ee
We will proceed with this special choice, but the answer for a general polarization $(a,\hat{a})$ will be deduced at the end. For now, note that the soft factor \eqref{soft-factor} for this choice is
\be\label{softfactint}
S^{+\hat{+}}=\frac{\tau^{2}\, [\tilde{q}|p_{1}\tilde{p}_{n}|q\rangle}{\tau^{2}\, \hat{s}_{n+1,1}\hat{s}_{n+1,n}},
\ee
where we have explicitly exhibited the powers of $\tau$. Since they
cancel, and the measure in \eqref{eq:soft-main} contributes
a power of $\tau^{-1}$, we expect the integrand to be of order $\tau^{1}$.
In fact, the factor of $\tau$ comes from the expansion of the
Pfaffian, i.e., $\text{Pf}'A_{n+1} = \tau\widehat{\text{Pf}'} A_{n+1}$. Now, to extract the aforementioned polarization from the amplitude we perform the following fermionic integration
\begin{align}
\mathcal{I}_{2m+1}^{+\hat{+}} &:= \int d^{4}\eta_{n+1}d^{4}\tilde{\eta}_{n+1}\,\eta_{n+1}^{1}\tilde{\eta}_{n+1}^{1}\, \widehat{\mathcal{I}}_{2m+2}\nn\\
&\,=\tau\, V_n \,\text{PT}(\I_n)\,\frac{\widehat{\text{Pf}'} A_{n+1}}{\prod_{i=1}^n \sigma_{n+1,i}}\frac{\sigma_{1n}}{\sigma_{1,n+1}\sigma_{n+1,n}}\nn\\
&\qquad\qquad\times \int \prod_{k=0}^{m} d^2\chi_k d^2\tilde{\chi}_k \Delta_{F}^{(n)} \widetilde{\Delta}_{F}^{(n)}\delta\left(\overline{W}_{a}^{+}\chi^{a}(\sigma_{n+1})\right)\delta\left(\widetilde{\overline{W}}_{\hat{a}}^{\hat{+}}\tilde{\chi}^{\hat{a}}(\sigma_{n+1})\right),
\end{align}
where $\widehat{\mathcal{I}}_{2m+2}$ corresponds to $\mathcal{I}_{2m+2}$ stripped of its on-shell delta functions. We also have $W=\overline{W}/(\prod_{i=1}^n \sigma_{n+1,i})$
with 
\begin{align}
\overline{W}_{b}^{a} & = \epsilon^{ac}\epsilon_{bd}M_{c}^{d}=\frac{\overline{x}\,\beta^{a}\beta_{b}+x\,\Xi^{a}\Xi_{b}}{\langle\Xi\beta\rangle}+\left(\gamma+\frac{1}{\tau}\right)\,\beta^{a}\Xi_{b}\label{eq:conjugatem}\\
\Rightarrow\overline{W}_{a}^{+} & = \frac{x\,\Xi^{+}\Xi_{a}}{\langle\Xi\beta\rangle}=x\,\Xi_{a}, 
\end{align}
using \eqref{eq:softm}. Here we have implicitly followed all of the
steps that were used in Section~\ref{app:even-to-odd} to simplify the form of the $W$ variables in the soft limit. The antichiral piece works in the same way.
Even though $\widetilde{M}$ is not integrated, its behaviour in the soft
limit allows us to define the antichiral counterparts $\tilde{\Xi}$
and $\tilde{x}$:
\be
\widetilde{\overline{W}}_{\hat{a}}^{\hat{+}}=\epsilon_{\hat{a}\hat{b}}\,\,\widetilde{M}_{\hat{-}}^{\hat{b}}=\tilde{x}\,\tilde{\Xi}_{\hat{a}}.
\ee
In direct correspondence to the bosonic case of Section \ref{app:softM}, we have managed to make explicit the $\tau$ dependence in the integrand,
and therefore we can evaluate the delta functions $\Delta_{F}^{(n)}\widetilde{\Delta}_{F}^{(n)}$ for $\tau=0$. 

We follow now Section \ref{devb9}, in which the basis element $\xi$ was
defined such that $\langle\xi\Xi\rangle=1$ for a given $\Xi^{a}$.
Then the polynomials are expanded as
\begin{eqnarray}
\chi^{a}(z) & = & \xi^{a}l(z)+\Xi^{a}r(z)\\
\tilde{\chi}^{\hat{a}}(z) & = & \tilde{\xi}^{\hat{a}}\tilde{l}(z)+\tilde{\Xi}^{\hat{a}}\tilde{r}(z),
\end{eqnarray}
where $l(z)$ and $r(z)$ are degree-$m$ polynomials with Grassmann
coefficients. Dropping the powers of $\tau$, we obtain
\begin{align}
\mathbb{\mathcal{I}}_{2m+1}^{+\hat{+}} =&\; V_n\,\text{PT}(\I_n)\,\frac{\widehat{\text{Pf}'} A}{\prod_{i=1}^n \sigma_{n+1,i}}\frac{\sigma_{1n}}{\sigma_{1,n+1}\sigma_{n+1,n}}\nn\\
& \times \int \prod^{m}_{k=0}\,dl_k\,dr_k\,d\tilde{l}_{k}\,d\tilde{r}_k\, \Delta_{F}^{(n)}\widetilde{\Delta}_{F}^{(n)} \delta\Big(l(\sigma_{n+1})\Big)\delta\left(\tilde{l}(\sigma_{n+1})\right)\,x\,\tilde{x}\,.
\end{align}
All of the following expressions for the integrand should be thought
as multiplied by the measure, as we continue to parallel the manipulations of
Section \ref{devb9}. Now we put $l(z)=(z-\sigma_{n+1})\hat{l}(z)+b$,
and we note that the fermionic delta functions fix $b=0$ in the
same way as the bosonic delta functions fixed $r^{A}=0$ in
\eqref{eq:rrr}. Using \eqref{eq:bas} we have
\begin{align}
\Delta_{F}^{(n)} & = \prod_{i=1}^{n}\delta^{4}\left(Q_{i}^{A}-\frac{\omega^{A}(\sigma_{i})l(\sigma_{i})-\pi^{A}(\sigma_{i})r(\sigma_{i})}{\prod_{j\neq i}^{n+1}\sigma_{ij}}\right)\nn\\
 & = \prod_{i=1}^{n}\delta^{4}\left(Q_{i}^{A}-\frac{\omega^{A}(\sigma_{i})\hat{l}(\sigma_{i})-\hat{\pi}^{A}(\sigma_{i})r(\sigma_{i})}{\prod_{j\neq i}^{n}\sigma_{ij}}\right)
  = \prod_{i=1}^{n}\delta^{4}\left(Q_{i}^{A}-\frac{\langle\hat{\rho}^{A}(\sigma_{i})\,\hat{\chi}(\sigma_{i})\rangle}{\prod_{j\neq i}^{n}\sigma_{ij}}\right),
\end{align}
where we have defined
\be
\hat{\chi}^{a}(z)=\xi^{a}r(z)- \Xi^{a}\hat{l}(z),
\ee
and $\hat{\rho}^{A}(\sigma)$ is given by (\ref{eq:oddpoly}).
The top component of this fermionic map is given by $\hat{\chi}_{m}^{a}=\xi^{a}r_{m}$. We identify $r_m=g$, hence agreeing with the fermionic maps introduced in Section \ref{sec:odd-pt}. We now have
\be
\mathcal{I}_{2m+1}^{+\hat{+}} = V_n\,\text{PT}(\I_{n})\,\frac{\widehat{\text{Pf}'} A}{\prod_{i=1}^{n}\sigma_{n+1,i}}\frac{\sigma_{1n}}{\sigma_{1,n+1}\sigma_{n+1,n}}x\,\tilde{x}\int dg\,d\tilde{g}\prod_{k=0}^{m-1}d^{2}\chi_{k}^{a}d^{2}\tilde{\chi}_{k}^{\hat{a}}\, \Delta_{F}^{(n)} \widetilde{\Delta}_{F}^{(n)}.
\ee

Recall that at this stage the map component $\xi=\frac{1+\Xi^{-}}{\Xi^{+}}$
is determined implicitly by \eqref{eq:softe}, which in turn depends on $\sigma_{n+1}$. Therefore the $\sigma_{n+1}$ dependence cannot be isolated yet. The final step is to turn $\xi$ into an extra variable, which is equivalent
to unfixing the T-shift symmetry, as explained at the end of Section \ref{devb9}. This is done by performing the transformation \eqref{eq:alphashift}. However,
as $\mathcal{I}_{2m+1}^{+\hat{+}}$ will be divided by $S^{+\hat{+}}$, given in \eqref{softfactint}, we also need to consider the scaling of the soft spinors $q\rightarrow{q}/{\alpha}$.  Doing the corresponding scaling for the antichiral piece, $\tilde{q}\rightarrow{\tilde{q}}/{\tilde{\alpha}}$, we effectively promote $\xi$ and $\tilde{\text{\ensuremath{\xi}}}$ into integration variables to be fixed by the bosonic equations. The relationship between the variables $\alpha$, $\tilde{\alpha}$ and the components $\xi$, $\tilde{\xi}$ can be read off from (\ref{eq:alphashift}):
\be
\alpha=\langle\Xi\,\xi\rangle\,,\quad\tilde{\alpha}=[\tilde{\Xi}\,\tilde{\xi}].
\ee

Including the scaling of the soft factor $S^{+\hat{+}}\rightarrow\alpha\tilde{\alpha}S^{+\hat{+}}$
and putting everything together, we find the following formula for the
$\mathcal{N}=(1,1)$ integrand:
\be
\frac{1}{2\pi i} \oint_{|\hat{E}_{n+1}| = \varepsilon} \frac{d\sigma_{n+1}}{E_{n+1}}\frac{\mathcal{I}_{2m+1}^{+\hat{+}}}{S^{+\hat{+}}}=\mathcal{J}_{2m+1}\times\int d\widehat{\Omega}_{\text{F}}^{(1,1)}.
\ee
The Vandermonde factor $V_n$ has been absorbed into the fermionic measure $d\widehat{\Omega}_{\text{F}}^{(1,1)}$, which is defined as:
\be
d\widehat{\Omega}_F^{(1,1)} = V_n\, dg\,d\tilde{g}\prod_{k=0}^{m-1}d^2\chi_{k}\,d^2\tilde{\chi}_{k} \prod_{i=1}^{n}\delta^{4}\bigg(q_i^A - \frac{\langle \rho^A(\sigma_i) \, \chi (\sigma_i)\rangle}{\prod_{j\neq i}\sigma_{ij}} \bigg) \delta^{4}\bigg(\tilde{q}_{i,A} - \frac{[ \tilde{\rho}_A(\sigma_i) \, \tilde{\chi} (\sigma_i)]}{\prod_{j\neq i}\sigma_{ij}} \bigg).\nn
\ee
The bosonic part of the integrand $\mathcal{J}_{2m+1}$ is given by
\be
\mathcal{J}_{2m+1}= \text{PT}(\mathbb{I}_n)\frac{\sigma_{1n}}{2\pi i}\oint_{|\hat{E}_{n+1}|= \varepsilon} \frac{d\sigma_{n+1}}{E_{n+1}} \frac{1}{S^{+\hat{+}}}\frac{x\,\tilde{x}}{\langle\Xi\xi\rangle[\tilde{\Xi}\tilde{\xi}]}\,\frac{1}{\prod_{i=1}^n\sigma_{n+1,i}}\left(\frac{\text{Pf}'\hat{A}}{\sigma_{1,n+1}\sigma_{n+1,n}}\right),
\ee
which encodes the complete $\sigma_{n+1}$ and $\hat{p}_{n+1}$ dependence. It is now straightforward to repeat these steps for other polarizations $(a,\hat{a})$.
In fact, from (\ref{eq:conjugatem}) we see that for the choice $a=-$,
the $\tau^{-1}$ contribution will dominate, yielding no factor of $x$
in the numerator. At the same time, the different $\tau$ dependence
of this integrand will be compensated by the different $\tau$ behaviour
of the soft factor $S^{a\hat{a}}$. For a general polarization we have: 
\be
\frac{x \tilde{x}}{ S^{+\hat{+}}} \;\to\; \frac{x^a \tilde{x}^{\hat{a}}}{ S^{a\hat{a}}},
\ee
where we have defined $x^{a}=(x,-1)$ and $\tilde{x}^{\hat{a}}=(\tilde{x},-1)$. Setting $\sigma_{n+1}=z$ and removing the fermionic delta functions, the integrand becomes
\be
\mathcal{J}_{2m+1} = \frac{1}{S^{a\hat{a}}} \text{PT}(\I_n) \, \frac{\sigma_{1n}}{2\pi i} \oint_{|\hat{\mathcal{E}}_{n+1}|=\varepsilon}\frac{dz}{\mathcal{E}_{n+1}} \frac{\text{Pf}'A_{n+1}}{(z-\sigma_{1})(z-\sigma_{n})}\,\frac{x^{a}}{\langle\xi\,\Xi\rangle}\,\frac{\tilde{x}^{\hat{a}}}{[\tilde{\xi}\,\tilde{\Xi}]},
\ee
where $\mathcal{E}_{n+1} = \tau \hat{\mathcal{E}}_{n+1} =p(z)\cdot p_{n+1}$ is the scattering equation for the
$(n+1)$th particle, valid on the support of the equations associated to hard particles. In this form the $\tau$ dependence cancels between the soft factor and the scattering equation. This form is taken as the starting point in Section~\ref{sec:odd-integrand}.

\subsection{From Odd to Even Multiplicity and the Number of Solutions \label{app:softeven}}

Here we consider taking a soft limit of the odd-point measure. The goal is to prove that the relation
\be
\int d\mu_{n+1}^{\text{6D}} = \delta(p_{n+1}^{2})\int d\mu_{n}^{\text{6D}} \frac{1}{2\pi i} \oint_{|\hat{E}_{n+1}| = \varepsilon} \frac{d\sigma_{n+1}}{E_{n+1}}+\mathcal{O}(\tau^{0}),\label{generalsoft}
\ee
holds for any $n$, whether it is even or odd. (The corresponding measures were defined in Sections \ref{sec:rational-maps} and \ref{sec:odd-rational-maps}). This result can be used to prove that the equations for the maps and the punctures of $n$ particles have $(n-3)!$ solutions,\footnote{This assumes generic kinematics in the sense of the discussion we give in Section \ref{sec:outlook}.} as claimed in Section \ref{sec:rational-maps}. Since we have already shown that integrating out the coefficients of the maps $\rho^{A,a}_{k}$ leaves delta functions localizing the $\sigma_i$'s, this implies that this measure should correspond to the CHY measure \eqref{CHY-measure} up to a trivial Jacobian. Such a Jacobian must not carry a nontrivial $\slc$ weight or mass dimension. This has been checked numerically.

The reasoning used to find the number of solutions follows closely the inductive proof in \cite{Cachazo:2013iaa}. For $n=3$ one can analytically check that there is one solution for the moduli $\{\rho,\sigma\}$. We then assume that the lower-point measure $d\mu_n$ in \eqref{generalsoft} has support on exactly $(n-3)!$ solutions. Then, we use the fact that in the soft limit $d\mu_{n+1}$ decouples into the lower-point measure and $\delta(E_{n+1})$. In the previous section we recognized $E_{n+1}$ as the soft limit of the scattering equation for $\sigma_{n+1}$, which has been shown to yield $n-2$ solutions for given hard data \cite{Cachazo:2013iaa}. This can also be seen directly from \eqref{softscateq}. Since the number of solutions cannot change in the soft limit, we conclude that $d\mu_{n+1}$ has support on $(n-2)!$ solutions, which completes the argument.

In order to show the validity of \eqref{generalsoft} for odd $n$ we begin with the same identity used in the previous section for $n$ odd:
\begin{align}
\Delta_B^{(n+1)} &=  \Delta_B^{(n)} 
\delta(p_{n+1}^2) \int d^4 M_{n+1} |M_{n+1}|^3 \nonumber \\ 
& ~~~~~~~~~~ \times \delta^8 \left (\rho^{A,a}(\sigma_{n+1}) - (M_{n+1})^a_b \lambda^{A,b}_{{n+1}} \right ) \delta \left(|M_{n+1}| - \prod_{i=1}^n \s_{n+1\,i} \right),
\end{align}
where we have used the odd-point parametrization of the rational maps,
\begin{equation}
\rho^{A,a}(z) = \sum_{k=0}^{m-1} \rho^{A,a}_{k} z^k + \omega'^A \xi'^a z^m \; ,
\end{equation}
and $m = ({n-1})/{2}$. To avoid confusion we have labeled the highest-degree coefficient using primed variables. As before, we parametrize the $(n+1)$th soft particle for $\tau \rightarrow 0$ using a 6D spinor of the form $\lambda^{A,a}_{n+1} = \xi^a v^A + \tau q^{A,a}$, which gives $p^{AB}_{n+1} \sim O(\tau)$. We also define $q^{A,a} \xi_a = q^A$. For the odd-point parametrization of the maps, the symmetry group $G$ includes the T-shift redundancy parametrized by the $\text{GL}(1,\mathbb{C})$ parameter $\alpha$. $\rho(z)$ and $M_i$ both transform under the T shift, as shown in \eqref{eq:TsymW} for $W_i = M_i^{-1}$.

Much of the soft-limit analysis for $n$ odd is similar to the case of $n$ even; the coefficients of the rational maps are fixed by the data of the hard particles while $M_{n+1}$ is allowed to have a singular piece in the soft limit. We may repeat the steps of Section \ref{app:softM}, inserting an ansatz for $M_{n+1}$ and decomposing it in a basis of spinors $\Xi^a$ and a modulus $x$. The dependence of the measure on the $(n+1)$th particle can we written in the soft limit as
\begin{align}
&\frac{1}{\tau} \delta(p_{n+1}^2) \int \frac{\prod_{k=0}^{m-1} d^8\rho_{k} \, d^4 \omega' \, \langle \xi' d \xi' \rangle \, d \sigma_{n+1}}{\mathrm{vol(\slc_{\s}, \slc_{\r}, T)}} \frac{\prod_{i=1}^n \s_{n+1,\, i} }{V_{n}^2} {\Delta}_B^{(n)} \nonumber \\ 
& ~~~~~~~~~~~~~~~~~~ \times \int dx \, d^2 \Xi \, \delta^8 \left(\rho^{A,a}(\sigma_{n+1}) - \Xi^a (q^A + x v^A)\right) \, .
\end{align}
After decomposing $M_{n+1}$ in the soft limit as done here, the transformation rule for $M_{n+1}$ becomes one for $\Xi^a$:
\begin{equation}
\delta \Xi^a = \alpha \, \sigma_{n+1} \, \xi'^a \langle \xi' \Xi \rangle \, .
\end{equation}

Having isolated the singular $\tau$ dependence in the soft limit, let us now examine the behavior of the even-point rational maps arising from the soft limit of odd-point amplitudes. At each point in the $d^2 \Xi$ integration, we expand the odd-point map in a special basis, the one determined by the two preferred spinors $\Xi^a$ and $\xi'^a$. This basis is not orthonormal, and $\langle \Xi \xi' \rangle \neq 1$. Changing variables to ($\pi^A$, $\omega^A$) spinor coordinates, the odd-point map $\omega'^A$ becomes the last component of the latter:
\begin{equation}
\rho^{A,a}(z) =  \Xi^a \pi^A(z) + \xi'^a \omega^A(z),
\end{equation}
or more explicitly
\begin{equation}
\rho^{A,a}(z) = \Xi^a \, \sum_{k=0}^{m-1} \pi^A_k z^k  + \xi'^a \left(\sum_{k=0}^{m-1} \omega^A_k z^k + \omega'^A z^{m} \right) .
\end{equation}

By taking linear combinations of the eight-dimensional constraint equations for $\rho^{A,a}$, we arrive at a split form involving the basis:
\begin{equation}
\delta^8 \left(\rho^{A,a}(\sigma_{n+1}) - \Xi^a (q^A + x v^A)\right) = \frac{1}{\langle \Xi \,\xi' \rangle^4 } \delta^4 \left(\omega^A(\sigma_{n+1})\right)\, \delta^4 \left(\pi^A(\sigma_{n+1}) - (q^A + x v^A)\right).
\end{equation}
Additionally, the remaining bosonic delta functions also change under this basis transformation:
\begin{equation}
{\Delta}_B^{(n)} = \prod_{i=1}^{n} \delta^6 \left ( p_i^{AB} - \langle \Xi \xi' \rangle \frac{\pi^{[A}(\sigma_i) \omega^{B]}(\sigma_i)}{\prod_{j=1}^n \s_{n+1,\, j}} \right ) \; ,
\end{equation}
along with the integration measure, which acquires a Jacobian
\begin{equation}
\left ( \prod_{k=0}^{m-1} d^8 \rho_{k} \right )\, d^4 \omega' \rightarrow \langle \Xi\, \xi' \rangle^{4m} \left ( \prod_{k=0}^{m-1} d^4 \pi_{k} \, d^4 \omega_{k} \right )\, d^4 \omega' .
\end{equation}
As in the case of taking a soft limit from even $n$ to odd $n$, we may now use the delta functions to reduce the degree of the map. To see this, we parametrize the map evaluated at the $(n+1)$th puncture as:
\begin{align}
&\omega^A(z) = (z - \sigma_{n+1}) \hat{\omega}^A(z) + r^A, \\
&\prod_{k=0}^{m-1} d^4 \omega_k \, d^4 \omega' \, \delta^4 \left(\omega^A(\sigma_{n+1})\right) \rightarrow d^4 r^A \prod_{k=0}^{m-1} d^4 \hat{\omega}_k \, \delta^4(r^A)\, .
\end{align}
The $r^A$ integrations are trivial, and now the $\omega'$ component has dropped out of the problem in favor of the $\hat{\omega}$ variables. This means we may now use the hatted variables in the remaining bosonic delta functions.

Having reduced the degree of the map, we may now switch back to the $\rho$ variables through another change of basis:
\begin{align}
\hat{\rho}^{A,a}(z) &= \Xi^a \pi^A(z) + \xi'^a \hat{\omega}^A(z), \\
\hat{\rho}^{A,a}(z) \Xi_a &=  \langle \Xi \xi' \rangle \, \hat{\omega}^A(z),\\
\hat{\rho}^{A,a}(z) \xi'_a &=  \langle \xi' \Xi \rangle \, \pi^A(z).
\end{align}
This has the effect of undoing several of the Jacobians acquired earlier, and the relevant piece of the measure and integrand becomes
\begin{align}
\frac{\prod_{i=1}^n \s_{n+1\,i}}{V_{n}^2} \int d\sigma_{n+1} \,  &\frac{\prod_{k=0}^{m-1} d^8\hat{\rho}_{k} \langle \xi' d \xi' \rangle \, d^2 \Xi\, dx }{\mathrm{vol(\slc_{\s}, \slc_{\r}, T)}} \prod_{i=1}^{n} \delta^6 \left ( p_i^{AB} -  \frac{\langle \hat{\rho}^{A}(\sigma_i) \hat{\rho}^{B} (\sigma_i) \rangle}{\prod_{j=1}^n \s_{n+1\, j}} \right ) \nonumber \\
&\times \frac{1}{\langle \Xi \xi' \rangle^4} \delta^4 \left(\frac{ \langle \hat{\rho}^{A}(\sigma_{n+1}) \, \xi' \rangle }{\langle \Xi \, \xi' \rangle} -q^A - x v^A \right ) .
\end{align}

The freedom to projectively scale $\xi'$ allows us to set the first component to 1 and define the second as $\xi'$ so that $\langle \xi' d\xi' \rangle = d\xi'$. Now we may focus on the last piece, which can be written as
\begin{equation}
\prod_{i=1}^n \s_{n+1\,i} \int d\sigma_{n+1} \, d \xi' \, d^2 \Xi\, dx \, \delta^4 \! \left(\hat{\rho}^{A,a}(\sigma_{n+1}) \xi'_a - \langle \Xi \xi' \rangle(q^A - x v^A) \right ) .
\end{equation}
There are now five integrations, four delta functions, and the $T$ redundancy to cancel. The strategy is to isolate the scattering equation for the last particle, integrate out the other delta functions, and cancel the T-shift symmetry. The scattering equation for the soft particle is supported on the solution of $E_{n+1} = \epsilon_{ABCD}\hat{\rho}^{A,+}(\sigma_{n+1}) \hat{\rho}^{B,-}(\sigma_{n+1}) v^C q^D = 0$. To get this, we first make the change of variables
\begin{align}
\langle \xi' \Xi \rangle = \Xi^- - \xi' \Xi^+ &\rightarrow u, \nonumber \\
x &\rightarrow \frac{x'}{u}, \nonumber \\
d \xi' \, d\Xi^+ \, d \Xi^- \, d x & \rightarrow \frac{d\Xi^+}{u} \, d \xi' \, du \, d x', \nonumber \\
\delta^4 \! \left(\hat{\rho}^{A,a}(\sigma_{n+1}) \xi'_a - \langle \Xi \xi' \rangle(q^A - x v^A) \right ) &\rightarrow \delta^4 \! \left(\hat{\rho}^{A,+}(\sigma_{n+1}) \xi' {-} \hat{\rho}^{A,-}(\sigma_{n+1}) {-} u q^A {-} x' v^A \right ).
\end{align}

Now we would like to evaluate the integrals over $u$, $x'$, and $\xi'$. As in the even-point case, we observe that these integrations give the scattering equation for the last particle after taking the appropriate linear combinations:
\begin{align}
&\frac{\prod_{i=1}^n \s_{n+1\,i}}{\tau} \int d\sigma_{n+1} \, d \xi' \, du \, dx' \, \delta^4 \! \left(\hat{\rho}^{A,+}(\sigma_{n+1}) \xi' - \hat{\rho}^{A,-}(\sigma_{n+1}) - u q^A - x' v^A \right )\nonumber \\
&= \frac{\prod_{i=1}^n \s_{n+1\,i}}{\tau} \int d\sigma_{n+1} \, \delta \! \left(\epsilon_{ABCD}\hat{\rho}^{A,+}(\sigma_{n+1}) \hat{\rho}^{B,-}(\sigma_{n+1}) v^C q^D \right ) = \int d\sigma_{n+1} \, \delta(E_{n+1}).
\end{align}
So we are left with
\begin{align}
&\delta(p_{n+1}^2) V_{n}^{-2} \int \frac{\prod_{k=0}^{m-1}  d^8\hat{\rho}_{k}}{\mathrm{vol(\slc_{\s} , \slc_{\rho} , T)}}\frac{d \Xi^+}{u}  \nonumber \\
&~~~~~~~~~ \times \prod_{i=1}^{n}\delta^6 \left ( p_i^{AB} - \frac{\langle \hat{\rho}^{A}(\sigma_i) \hat{\rho}^{B} (\sigma_i) \rangle}{\prod_{j=1}^n \s_{n+1\,j}} \right )
\int d\sigma_{n+1} \, \delta(E_{n+1}).
\end{align}
In this expression $u= \langle \xi' \Xi \rangle$ has a value determined by the constraints after doing the integral. Since T acts as a $\text{GL}(1,\mathbb{C})$ shift on the components of $\Xi$, we can absorb $u$ and cancel the symmetry. The result is the expected measure for $n$ even:
\begin{align}
\int \frac{\prod_{i=1}^n d\sigma_i \, \prod_{k=0}^{m-1} d^8\hat{\rho}_{k}}{\mathrm{vol(\slc_{\s} \times \slc_{\rho})}} \prod_{i=1}^{n} \delta^6 \left ( p_i^{AB} -  \frac{\langle \hat{\rho}^{A}(\sigma_i) \hat{\rho}^{B} (\sigma_i) \rangle}{\prod_{j=1}^n \s_{n+1\,j}} \right )
\delta(p_{n+1}^2) V_{n}^{-2} \int d\sigma_{n+1} \, \delta(E_{n+1}) .
\end{align}

\bibliographystyle{JHEP}
\bibliography{references}

\end{document}